\documentclass[aps,twocolumn,preprintnumbers,footinbib,superscriptaddress,longbibliography]{revtex4-1}

\usepackage[utf8]{inputenc}

\usepackage{graphicx,amssymb,amsmath,amsthm,amsfonts}
\usepackage[usenames,dvipsnames,svgnames,table]{xcolor}
\definecolor{darkred}{rgb}{0.5,0,0}
\usepackage{bm}
\usepackage{dcolumn}
\usepackage{latexsym}
\usepackage{rotating}

\usepackage[normalem]{ulem}

\usepackage{enumerate}
\usepackage{tensor,multirow}
\usepackage{url}
\usepackage[unicode,linktocpage]{hyperref}
\hypersetup{
    pdftitle={Instabilities of scalar fields around oscillating stars},
}

\usepackage{braket}

\def\be{\begin{equation}}
\def\ee{\end{equation}}
\newcommand{\beq}{\begin{eqnarray}}
\newcommand{\eeq}{\end{eqnarray}}

\def\ba{\begin{align}}
\def\ea{\end{align}}

\begin{document}

\title{Instabilities of scalar fields around oscillating stars}

\author{Taishi Ikeda}
\affiliation{Dipartimento di Fisica, 
``Sapienza'' Universit\'{a}
di Roma, Piazzale Aldo Moro 5, 00185, Roma, Italy}

\author{Vitor Cardoso}
\affiliation{CENTRA, Departamento de F\'{\i}sica, Instituto Superior T\'ecnico -- IST, Universidade de Lisboa -- UL,
Avenida Rovisco Pais 1, 1049-001 Lisboa, Portugal}
\affiliation{Niels Bohr International Academy, Niels Bohr Institute, Blegdamsvej 17, 2100 Copenhagen, Denmark}

\author{Miguel Zilh\~ao}
\affiliation{CENTRA, Departamento de F\'{\i}sica, Instituto Superior T\'ecnico -- IST, Universidade de Lisboa -- UL,
Avenida Rovisco Pais 1, 1049-001 Lisboa, Portugal}

\begin{abstract}
The behavior of fundamental fields in strong gravity or nontrivial environments is important for our understanding of nature. This problem has interesting applications in the context of dark matter, of dark energy physics or of quantum field theory. The dynamics of fundamental fields has been studied mainly in static or stationary backgrounds, whereas most of our Universe is dynamic.
In this paper we investigate ``blueshift'' and parametric instabilities of scalar
fields in dynamical backgrounds, which can be triggered (for instance) by
oscillating stars in scalar-tensor theories of gravity.
We discuss possible implications of our results, which include constraints on an otherwise hard-to-access parameter space of scalar-tensor theories.
\end{abstract}

\maketitle

% \tableofcontents

%%%%%%%%%%%%%%%%%%%%%%%%%%%%%%%%%%%%%%%%%%%%%%%%%%%%%%%%%%%%%%%%%%%%%%%%%%
\section{Introduction}

% \noindent{\bf{\em Introduction.}}

General Relativity (GR) is currently the best description of the gravitational interaction, and has been successfully tested on different scales~\cite{Will:2014kxa}.
The recent direct detection of gravitational waves (GWs)~\cite{TheLIGOScientific:2016agk,TheLIGOScientific:2016src,Abbott:2020jks} indicates that even dynamical, strong-field regions
are adequately described by GR (up to the precision probed by current detectors).
In spite of its brilliant status, there are a number of conceptual issues with GR, ranging from the large-scale description of the cosmos to the 
fate of classical singularities in gravitational collapse or the incorporation of quantum effects, see e.g.~\cite{Clifton:2011jh,Berti:2015itd,Cardoso:2019rvt} and references therein.
The resolution of some of these challenges most likely requires that GR be superseded by a more sophisticated description.

There is currently no single compelling alternative to GR that solves the above issues without introducing new problems of their own.
However, a variety of modified theories have been proposed, mostly with the view to exploring the mathematical and physics content of possible contenders to GR.
%~\cite{Clifton:2011jh,Berti:2015itd,Cardoso:2019rvt}. 
These frequently include additional degrees of freedom, which might lead to unique observational signatures. 
The simplest modification of GR are scalar-tensor theories, where a new scalar degree of freedom couples to curvature or matter. 
Some of these theories arise naturally as possible alternatives, since they have a well-posed initial value problem and simultaneously evade all known constraints.
Scalars are also a generic prediction of string theory or of extensions of the standard model of particle physics,
and are also natural candidates for dark energy and dark matter~\cite{Arvanitaki:2009fg,Clifton:2011jh,Berti:2015itd,Barack:2018yly,Cardoso:2019rvt}.

\begin{table}[htb]
\centering
\caption{List of some stellar objects which can be prone to the processes described here. We take the radial mode of a neutron star
  (NS) of mass $1.3M_{\odot}$ and radius $L_0=9.7\,{\rm Km}$ described by equation
  of state A in Ref.~\cite{Kokkotas:2000up}. For white dwarfs (WD) we take a degeneracy parameter $1/y_0^2=0.05$ as quoted in
  Refs.~\cite{1961AnAp...24..237S,1991Ap&SS.186..117S,Winget:2008iu}. The Sun's
  description was taken from Refs.~\cite{TurckChieze:1993dw,2012RAA....12.1107T}. The compactness $\mathcal{C}$ is defined in Eq.~\eqref{eq:compactness_def}.
	\label{stars_data}
}
\begin{ruledtabular}
\begin{tabular}{lcccc}
 % \hline
 % \hline
      &    $\mathcal{C}$  &  $L_0$ (km)    &$\omega L_0/c$&  Radial frequency \\
\hline
NS   & $0.3$              & $10$           &0.6           & 3 {\rm kHz} \\%($1.2\times 10^{-11} {\rm eV}$) \\
WD   & $10^{-3}$          & $10^3$         &0.004         & 0.2 {\rm Hz} \\%($8.2\times 10^{-16} {\rm eV}$) \\
Sun  & $10^{-6}$          & $7\times 10^5$ &0.029         & 2 {\rm mHz}% ($8.2\times 10^{-18} {\rm eV}$)
  % \hline\hline
\end{tabular}
\end{ruledtabular}
\end{table}
Depending on the coupling to matter, to curvature and on the self-interactions, such theories and new fundamental fields may give rise to wide array of new effects, such
as the spontaneous ``scalarization'' of objects~\cite{Damour:1993hw,Damour:1996ke,Harada:1998ge,Pani:2010vc,Ramazanoglu:2016kul,Antoniou:2017acq,Doneva:2017bvd,Silva:2017uqg},
screening mechanisms on astrophysical scales~\cite{Khoury:2003aq,Khoury:2003rn,Joyce:2014kja}, etc. Potential astrophysical consequences of these theories
were mostly based on analysis in stationary settings~\cite{Damour:1993hw,Damour:1996ke,Harada:1998ge,Pani:2010vc,Ramazanoglu:2016kul,Cardoso:2020cwo,Khoury:2003aq,Khoury:2003rn,Tsujikawa:2009yf}.
Time-dependent setups include spontaneous scalarization during the inspiral and merger of neutron stars~\cite{Barausse:2012da,Palenzuela:2013hsa,Shibata:2013pra,Khalil:2019wyy},
black holes~\cite{Silva:2020omi,East:2021bqk}, cosmological evolutions
~\cite{Llinares:2013qbh,Llinares:2013qbh,Hagala:2016fks} or situations aimed at understanding well-posedness or other fundamental issues~\cite{Brito:2014ifa,Silvestri:2011ch,Nakamura:2020ihr}.

Here, we wish to understand possible new phenomena induced by time-periodic motion, in particular by vibrating stars such as the ones summarized in Table~\ref{stars_data}.
%Consider for example a star for which the ground-state
%is a trivial scalar. 
Theories for which a constant scalar is a ground state typically have the same stationary solutions as GR, making it challenging to
tell the two apart. However, when such stars are disturbed (stochastically, like our Sun, or via mergers or accretion for a neutron star), they provide a time-periodic background on which a 
scalar fluctuation propagates. Indeed, we show that oscillating stars trigger various instabilities of fundamental fields. 
Such instabilities may facilitate constraints on otherwise hard-to-access parameter space of the theory.

A toy model had been studied in a Minkowski spacetime~\cite{Wang:2009qa,Wang:2013kc}. Our results confirm and extend the general picture, but in a nontrivial way, by dealing also with general relativistic backgrounds and fluctuations and providing realistic timescales for the mechanism (details on the general relativistic case are discussed in the appended Supplemental material).

We use geometrized units $c=G=1$ and parameterize stars by their mass $M$, radius~$L_0$, and compactness
\be
\mathcal{C}\equiv \frac{2M}{L_0}\,.\label{eq:compactness_def}
\ee
%

%%%%%%%%%%%%%%%%%%%%%%%%%%%%%%%%%%%%%%%%%%%%%%%%%%%%%%%%%%%%%%%%%%%%%%%%%%%%%%
\section{Setup}

%%%%%%%%%%%%%%%%%%%%%%%%%%%
% \noindent{\bf{\em Setup.}}
%%%%%%%%%%%%%%%%%%%%%%%%%%%

Our main purpose is to show that oscillating backgrounds can trigger instabilities
of fundamental fields nonminimally coupled to matter. We will show this explicitly for the simplest example of a scalar-tensor theory, but
our results are valid for more general setups.

We focus on theories in which a scalar field $\Phi$ is coupled to matter, described by the Lagrangian density %(written in the so-called ``Einstein frame'')
\begin{equation}
{\cal L}=\frac{R[g]}{16\pi} - \frac{1}{2}g^{\mu\nu}\nabla_{\mu}\Phi\nabla_{\nu}\Phi
-V(\Phi)+{\cal L}_{m}\left[\Psi_{m},\tilde{g}\right]\,,
\end{equation}
where $\tilde{g}_{\mu\nu}=A(\Phi)^{2}g_{\mu\nu}$ is the physical metric and ${\cal L}_{m}$ the Lagrangian describing matter fields $\Psi_m$ (these could be the microscopic, fundamental description leading to notions of density and pressure for example). The equations of motion for the system are
\begin{subequations}
  \label{eq:eoms}
\begin{align}
& R_{\mu\nu}-\frac{1}{2}Rg_{\mu\nu}=8\pi T_{\mu\nu}\,,\\
& T_{\mu\nu}=\nabla_{\mu}\Phi\nabla_{\nu}\Phi-g_{\mu\nu}\left(\frac{1}{2}(\nabla\Phi)^{2}+V(\Phi)\right)+T^{(m)}_{\mu\nu}\,,\nonumber\\
& \square \Phi-V'(\Phi)=-A(\Phi)^{3}\partial_{\Phi}A(\Phi)\tilde{T}^{(m)}\,.\label{KG_effective}  %\\
%
%  & \tilde{\nabla}^{\mu}\tilde{T}_{\mu\nu}^{(m)}=0\,.
\end{align}
\end{subequations}
Here, $\tilde{T}^{(m)}=\tilde{g}^{\mu\nu}\tilde{T}^{(m)}_{\mu\nu}$ while $T^{(m)}_{\mu\nu}$ and $\tilde{T}^{(m)}_{\mu\nu}$ are the energy-momentum tensors of the matter fields with respect to metric $g_{\mu\nu}$ and $\tilde{g}_{\mu\nu}$.

The phenomena we wish to describe are also part of more generic theories: the fundamental ingredient is a position-dependent effective mass function for the scalar~$\Phi$. Once linearized around a specific background, the right-hand side of Eq.~\eqref{KG_effective} is proportional to $\tilde{T}^{(m)}\Phi$. 
In other words, the coupling of the scalar to matter gives rise to an effective scalar mass which depends on the matter content $\tilde{T}^{(m)}\Phi$. This is one key ingredient of our study.
Thus, direct couplings to curvature would also produce similar effects~\cite{Silva:2017uqg,Antoniou:2017acq,Doneva:2017bvd,Ventagli:2020rnx}.

For concreteness, we focus exclusively on the following scalar potential and coupling function,
\begin{equation}
  V(\Phi)=\frac{\mu_0^{2}}{2}\Phi^{2}\,,\qquad A(\Phi)=e^{\frac{\beta}{2} \Phi^{2}}\,,
  \label{eq:potential_coupling}
\end{equation}
% 
% but our results and methods can be generalized in a trivial way.
but our results and methods can be applied to other models. Scalar self-interactions can play an important role in the {\it nonlinear} development of the instability; here we focus exclusively on
the physics at small $\Phi$. Hence our results describe the early-time dynamics of more general self-interacting theories. Here, $\mu_0$ is the (bare) mass parameter of the scalar field (note that the mass $m_{\rm s}$ of the field is related to the mass parameter via $m_{\rm s}=\mu_0\,\hbar$ in these units).
With this choice of scalar potential and coupling function, a vanishing scalar field is a solution to the theory and %.As we will show,
fluctuations around this solution endow the scalar with an effective, time and position-dependent mass.
Our analysis can also be generalized to theories with nontrivial, stable scalar profiles, including popular examples such as spontaneous scalarization~\cite{Damour:1993hw,Damour:1996ke,Harada:1998ge,Silva:2017uqg,Antoniou:2017acq,Doneva:2017bvd} and screening mechanisms~\cite{Khoury:2003aq,Khoury:2003rn,Joyce:2014kja,Hinterbichler:2011ca,Hinterbichler:2010es}.

Let us consider matter fields $\Psi_m$ describing a perfect fluid. Focus on a geometry with vanishing scalar,
$\Phi=0$, describing a static star of (constant, for simplicity) density $\tilde{\rho}_{0}$ and radius $L_0$~\cite{Shapiro:1983du}. The total mass of this solution can be written as $M=\frac{4\pi}{3}\tilde{\rho}_{0}L_0^{3}$.

We ignore the effect of the scalar field on the profile of the fluid. In other words, we deal only with the Klein-Gordon equation~\eqref{KG_effective} on this fixed geometry.
Perturbing around a background solution $\Phi_0(r)$, one finds that the stability of such solutions 
depend dramatically on the value of the coupling $\beta$~\cite{Damour:1993hw,Damour:1996ke,Harada:1998ge,Pani:2010vc}.
For concretness, here we take $\beta<0$ with $\mu^{2}_{0}-\beta\tilde{T}^{(m)}\gg -L_{0}^{-2}$ and such that $\Phi_0=0$ is a stable solution.
The assumption of zero background scalar is a conservative assumption, made to highlight the nontrivial effects of the mechanism we discuss below, which provide a unique tool to constrain
the theory. When the background scalar is not zero other effects become important. For example, an oscillating body with a nontrivial background scalar will radiate monopolar radiation, a topic explored in the past~\cite{Silvestri:2011ch,Sotani:2014tua,Mendes:2018qwo}.

\section{Instabilities of oscillating astrophysical objects}

%%%%%%%%%%%%%%%%%%%%%%%%%%%%%%%%%%%%%%%%%%%%%%%%%%%%%%%%%%%%%%%%%%%%%%%
% \noindent{\bf{\em Instabilities of oscillating astrophysical objects.}}
%%%%%%%%%%%%%%%%%%%%%%%%%%%%%%%%%%%%%%%%%%%%%%%%%%%%%%%%%%%%%%%%%%%%%%%
Let us then consider the dynamical behavior of a (nonminimally coupled) scalar field in a geometry describing a radially oscillating star.
For simplicity, the results shown below assume a Minkowski background and are thus valid only for Newtonian stars.
General Relativity effects change the quantitative but not the qualitative behavior, and are discussed in the apendded supplemental material, including an analysis of general-relativistic fluctuations of compact stars~\cite{1964ApJ...140..417C,1966ApJ...145..505B,Kokkotas:2000up}.
Fluctuations around a background value of the scalar field
are described by the Klein-Gordon equation 
\be
\square\Phi=\mu^{2}\Phi\,,\label{KG_eff_star}
\ee
with an effective mass
\[
\mu^2=\mu_0^2-\beta\tilde{T}^{(m)}\,,
\]
which acts as a position-dependent mass term
that depends on the energy density of the
star. If the star oscillates (with a time-dependent radius $L(t)$), so does the effective mass of the scalar field.
For simplicity we here assume the simple model
\begin{align}
  L(t) & =L_{0}+\delta L\sin\omega t \,,
  \label{eq:radius_oscillation} \\
\mu^{2}(t,r) & = \mu_{0}^{2}+\beta\left(\tilde{\rho}_{0}+\delta\tilde{\rho}(t,r)\sin(\omega t)\right)\,,\label{eq_profile_star}
\end{align}
where $\delta\tilde{\rho}(t,r)$ is the amplitude of the density oscillation, and $\tilde{\rho}_{0}$ is the average of the energy density.
We assume $\beta<0$, in such a way that $\mu_{0}^{2}+\beta \tilde{\rho}_{0}\gtrsim 0$ to avoid tachyonic instabilities.
The competition between different mechanisms is discussed in detail below and in the Supplemental Material, appended at the end.

Here, we assumed $\delta L\ll L_{0}$ and $\delta\tilde{\rho}\ll\tilde{\rho}_{0}$. This dependency has two important aspects:

\noindent $\bullet$ Because the effective mass inside the star is smaller than the exterior bare mass $\mu_0$, cf.\ Eq.~\eqref{eq_profile_star}, long wavelength modes are trapped inside the star. The oscillating star surface therefore provides reflecting boundary conditions at the periodically-varying location $L(t)$. For setups where there is a continuous mass
profile, a radially oscillating star corresponds, nevertheless, to periodically varying field profiles at the surface, possibly changing on scales smaller than a wavelength. 

\noindent $\bullet$ The local density oscillations cause a time-varying effective mass for the scalar, which lends itself to parametric instabilities as we show below.

In astrophysically realistic stars, the time-dependence of the radius and density of the star is quite involved.
Normal modes of oscillation of a star will cause a density profile with
a sinusoidal-like radial profile as well. We studied other profiles for the
perturbation (in particular $\delta\tilde{\rho}\propto \sin\omega r$) and fully general-relativistic settings and find
that they lead to similar results as described below and in the Supplemental material.

We decompose the scalar field into spherical harmonics and evolve each mode, $\Phi_{lm}(t,r)$,
using a fourth-order accurate Runge-Kutta scheme for the time integration where spatial derivatives
are approximated by fourth-order accurate finite difference stencils. 
Radiative boundary conditions are imposed at the boundary of the computational domain, which is not in causal contact with the star during our numerical evolutions.

We use time-symmetric initial data with a profile parameterized by
\begin{align}
\Phi_{lm}=e^{-\left(\frac{r-r_{0}}{\sigma}\right)^{2}}\,, \qquad\dot{\Phi}_{lm}=0\,,\label{eq;momentarily static Gaussian initial data}
\end{align}
where $\sigma$ and $r_{0}$ denoting the width and initial
position of the scalar field pulse.
Since the equation to be solved is linear we set the initial amplitude of the pulse to unity.
We focus our discussion on $l=0$ modes and we fix $r_0=0.5L_0$ and $\delta L=0.1L_0$.
We studied higher-$l$ modes, and found no qualitative difference.
The results discussed below all show fourth-order convergence.

%%%%%%%%%%%%%%%%%%%%%%%%%%%%%%%%%%%%%%%%%%%%%%%%%%%%%%%%%%%%%%%%%%%%%%%%%%%%%%

\section{Results}

%%%%%%%%%%%%%%%%%%%%%%%%%%%%
% \noindent{\bf{\em Results.}}
%%%%%%%%%%%%%%%%%%%%%%%%%%%%
%
Consider setups where the effective mass inside [cf.\ Eq.~\eqref{eq_profile_star}] the star satisfies $\mu_{\rm in}^2<\mu_{0}^2$. Low-energy fluctuations of the scalar are trapped inside the star and subject to conditions at the surface which might be prone to {\it blueshift} instabilities -- the growth of energy in the field can be understood from a simple particle picture as a cumulative Doppler effect~\cite{Dittrich:1993hw}, which is discussed in greater depth in the appended supplemental material. Such an analysis indicates the dominant instability window to be~\cite{Dittrich:1993hw},
\be
\frac{\pi}{L_{0}+\delta L}<\omega<\frac{\pi}{L_{0}-\delta L}\,.\label{blue_inst_window}
\ee
\begin{figure}[th]
\includegraphics[width=0.48\textwidth]{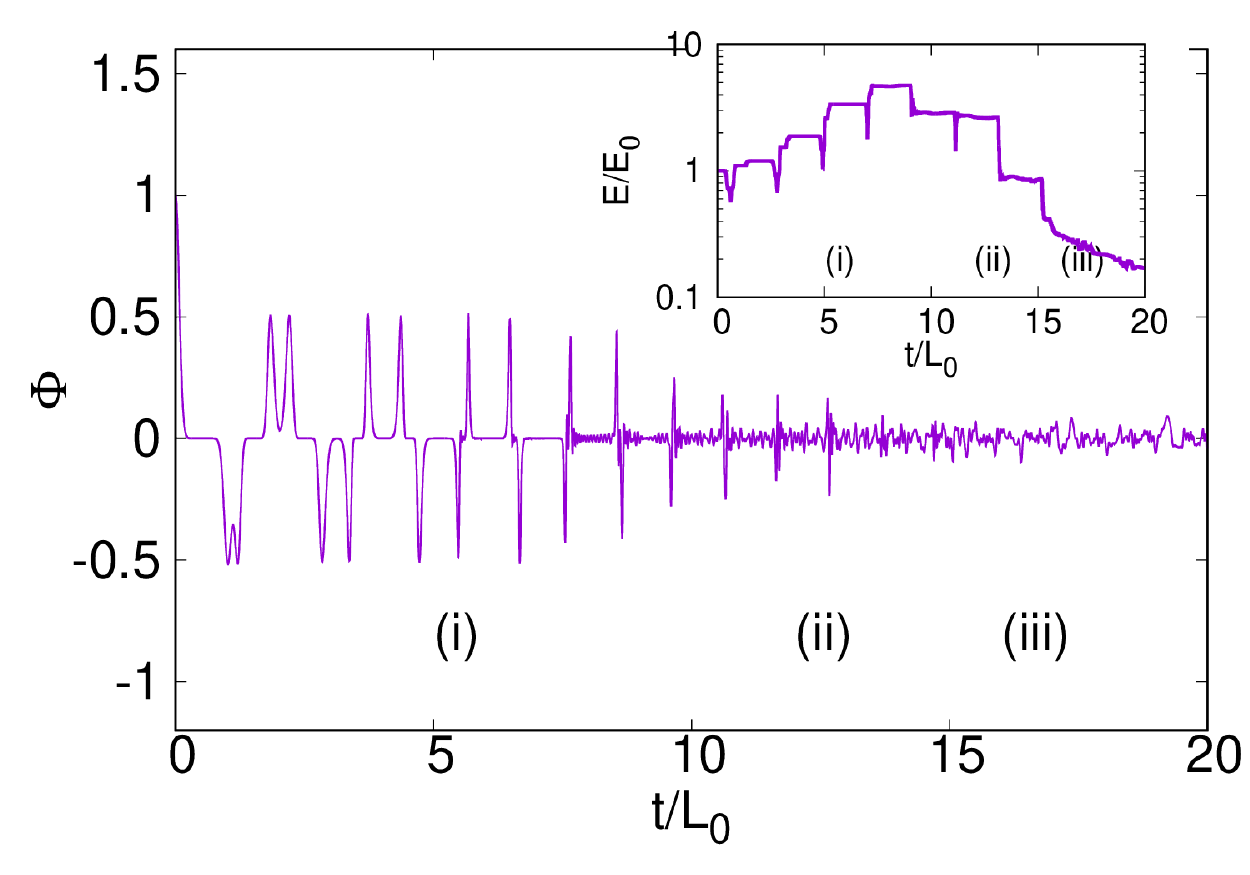}
\caption{Time evolution of a massive scalar field non-minimally coupled to geometry, such that it is effectively massless inside the star ($\mu_{\rm in}=0$)
but with bare mass $\mu_0 L_0=100$ and $\delta\tilde{\rho}=0$. The field ``sees'' only an oscillating boundary with $\omega L_0=3.1$, we froze density variations inside the star.
The field is extracted at $r=0.5L_{0}$ and the corresponding integrated energy inside the star is shown in the inset. 
Here, the initial data is characterized by $\sigma=0.1L_0$. 
 \label{Graph_time_evolution_of_energy_inside_star_ex_ID1_Amp1_wn01_r05_win0_001_bulkmass0_AmpM01_omegaM3_1_L1_l0_outmass100}}
\end{figure}

\begin{figure}[thp]
\includegraphics[width=0.48\textwidth]{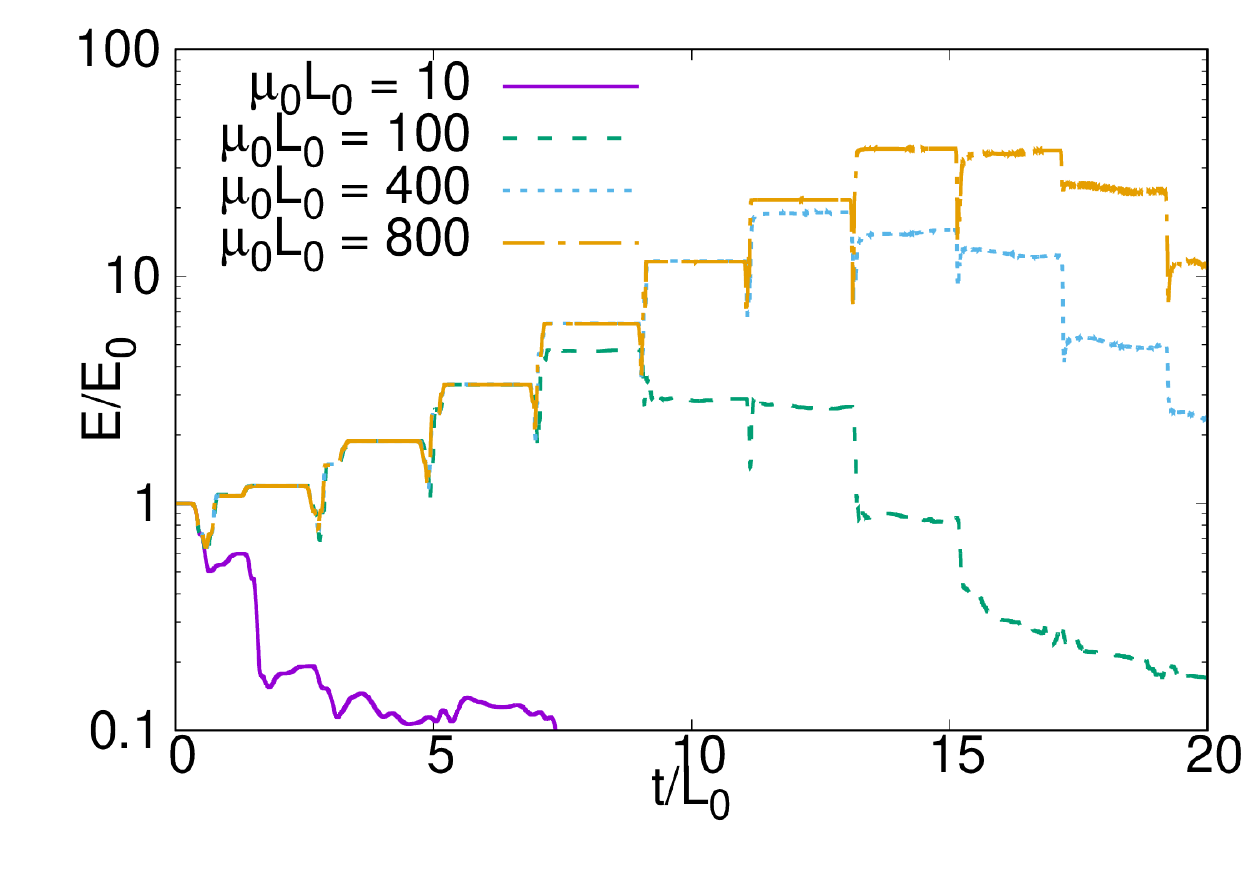}
\caption{Evolution of the integrated energy of a scalar inside an oscillating star, for different bare mass parameters~$\mu_0$.
The coupling to matter is such that the field is massless inside the star ($\mu_{\rm in}=0$). 
The field ``sees'' only an oscillating boundary with $\omega L_0=3.1$, we froze density variations inside the star.
The initial data is characterized by $\sigma=0.1L_0$. 
}
\label{Graph_time_evolution_of_energy_inside_star_Parameter_mass_dependence}
\end{figure}

In such cases, we see a transfer of energy from the oscillating star to the scalar field, increasing the scalar frequency. We can see this behavior in Figs.~\ref{Graph_time_evolution_of_energy_inside_star_ex_ID1_Amp1_wn01_r05_win0_001_bulkmass0_AmpM01_omegaM3_1_L1_l0_outmass100}-\ref{Graph_time_evolution_of_energy_inside_star_Parameter_mass_dependence}.
Upon each reflection at the surface the scalar drifts to higher frequencies, and after a sufficient amount of reflections it is no longer confined: the field is finally able to leak to outside the star (and has frequency $\omega\sim \mu (1+\epsilon),\,\epsilon \ll 1$). This mechanism causes the energy to grow as $E\sim E_0 e^{\lambda_B t}= E_0e^{2n_{\rm ref}L_0\lambda_{B}}$, 
with $n_{\rm ref}$ the number of reflections and $\lambda_{B}$ the instability growth rate. Here, $E_0 \sim \sigma^{-1}$ is the initial dominant spectral content at low energies, the ones that linger long enough to be amplified.
Thus, the field is able to leak away from the star and out to spatial infinity when $e^{2n_{\rm ref}L_0\lambda_{B}}\sigma^{-1}>\mu_0$. This occurs after a number of reflections
\begin{align}
n_{\rm ref}>n_{\ast}\sim \frac{\ln |\mu_0 \sigma|}{2\lambda_{B}L_{0}} \,.\label{eq:number_reflections}
\end{align}
At the critical number of reflections, the total energy confined inside the star as scalar field is
\be
\frac{E}{E_0}\biggr\rvert_{\rm max}=\sigma\mu_0\,,\label{eq_extracted_energy}
\ee
this being also the total amount of energy extracted from the star.

\begin{figure}[th]
\includegraphics[width=0.48\textwidth]{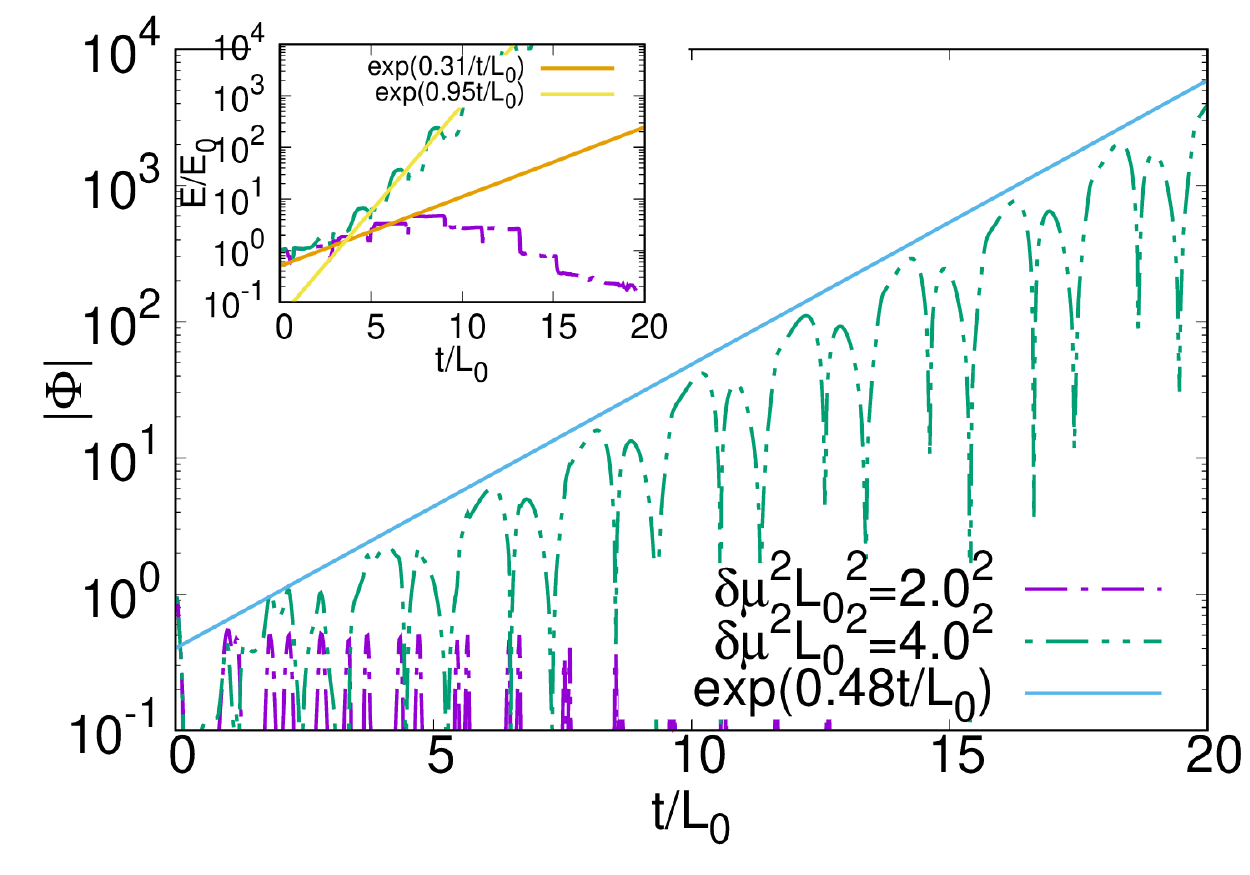}
\caption{Evolution of the energy and scalar field for setups which probe both an oscillating boundary and a time-varying effective mass inside the star.
The initial data has width $\sigma=0.1L_0$, and the star is oscillating with frequency $\omega L_0=3.1$. The (background) effective mass inside the star is zero, $\mu_{\rm in}^2=\mu_0^2+\beta\tilde{\rho}_0\approx 0$, while
the bare mass is $\mu_0L_0=100$. The time-varying component of the density is proportional to $\delta \mu^{2}\equiv \beta \delta \tilde{\rho}$. Even though both the blueshift and parametric instabilities are present,
the parametric mechanism dominates at large $\delta \mu^2$.}
\label{Graph_time_evolution_of_energy_Parameter_mass_dependence_H4_4}
\end{figure}

We have thus far focused only on the effect of oscillating boundaries, while the interior is non-dynamical.
However, real objects will also have an oscillating density. Thus, the effective mass inside the star will have a periodic time variability,
potentially giving rise to ``parametric'' instabilities. Figure~\ref{Graph_time_evolution_of_energy_Parameter_mass_dependence_H4_4}
summarizes our results for non-vanishing $\beta\delta\tilde{\rho}$. Two mechanisms now compete, a blueshift and a parametric mechanism.
For small $L_0^2\delta\tilde{\rho}$, the evolution of the scalar field is almost identical to that with a vanishing $\delta\tilde{\rho}L_{0}^{2}$, and the blueshift mechanism dominates.
For large $\delta\tilde{\rho}L_{0}^{2}$, we observe instead that the amplitude of the scalar field grows while its frequency is barely changing, a clear sign that we are dealing with a different
process. This is in fact a parametric instability.
Contrast this with the blueshift instability cases, where the scalar field pulse becomes narrower (due to the oscillating boundary) as time passes, but its amplitude remains constant. Here, the width of the scalar pulse is roughly constant and the instability is driven by a growth in the amplitude of the field (due to the oscillating effective density).
%

%%%%%%%%%%%%%%%%%%%%%%%%%%%%%%%%%%%%%%%%%%%%%%%%%%%%%%%%%%%%%%%%%%%%%%%%%%%%%%

\section{Application to astrophysical systems}

%%%%%%%%%%%%%%%%%%%%%%%%%%%%%%%%%%%%%%%%%%%%%%%%%%%%%%%%%%
% \noindent{\bf{\em Application to astrophysical systems.}}
%%%%%%%%%%%%%%%%%%%%%%%%%%%%%%%%%%%%%%%%%%%%%%%%%%%%%%%%%%
We have shown that pulsating stars or other objects may excite important instabilities of nonminimally coupled fields.
Compact stars, such as neutron stars, have radial pulsations with frequencies satisfying $\omega L_0\sim 1$ for the lowest overtones~\cite{Kokkotas:2000up}, and are ideal systems
where such instabilities might be relevant.
These change the local density and distribution of the scalar (which could be one dark matter component)
and may even backreact on the star. A precise description of the evolution of the instability requires a precise knowledge 
of stellar oscillations and careful modeling of the evolution of the scalar in such backgrounds. This is a challenging program that requires further study.

Notice first that one can apply our results for stars as long as other dissipation mechanisms are subdominant. Of particular importance are shear viscosity effects, which in neutron stars have a timescale~\cite{1987ApJ...314..234C,2019CQGra..36u5012B} 
\be
\tau_{\eta}\approx 100\,\rho^{-5/4}_{17}T_5^2\left(\frac{L_0}{\rm 10 km}\right)^2\,{\rm s}\,,
\ee
where $\rho_{17}=\rho/(10^{17} {\rm kg/m^3})$, $T_5=T/(10^5\,{\rm K})$ and $T$ is the neutron star temperature.

The blueshift instability is quenched after 
\begin{align}
t_{B} \sim2 L_{0}n_{\ast} %\nonumber\\
& =3\times 10^{-2}{\rm s} \left(\frac{10^{-3}}{\lambda_{B}(\omega)L_{0}}\right)\left(\frac{L_{0}}{10{\rm km}}\right) \notag \\
& \quad {} \times \ln\left|10\frac{\mu}{10^{-10}{\rm eV}}\frac{\sigma}{10{\rm km}}\right|\,,
\label{eq:time scale blue shift}
\end{align}
and on this timescale an energy $E\sim E_0\sigma\mu_0$ is removed from the star.
This result is not very sensitive to the initial conditions. It is, in principle, only weakly affected by backreaction, unless $\sigma \mu_0$ is an extremely large number. 
Our numerical results indicate that $\lambda_{B} L_0\sim 10^{-3}$ is a reasonable estimate for $\delta L>0.01 L_0$.
The instability window for the blueshift mechanism to work is tight, however [cf.\  Eq.~\eqref{blue_inst_window}].
Only large overtones are affected by it, unless the star is oscillating nonlinearly with $\delta L\gtrsim 0.1L_0$.

Consider now the parametric instability. When $\beta\sim-\frac{\mu^{2}}{\tilde{\rho}_{0}}$, with
$\tilde{\rho}_{0}$ the temporal average of the density of the star,
the relevant dynamics is governed by the Mathieu equation~\cite{Abramowitz:1970as,benderorszag}. This particular case provides a test on our results, and allows to estimate analytically the timescales involved.
For small $\delta\mu^{2}$, the instability condition amounts to $L_{0}\omega=4\pi/j$ ($j\in \mathbb{Z}$).
The instability rate time scale for $j=1$, for example, is roughly $t_A\sim 2\omega/\delta\mu^{2}$, or
\[
t_{A}\sim \frac{2\omega}{\delta\mu^{2}}\sim 1{\rm s}
\left(\frac{10 {\rm km}}{L_{0}}\right)\left(\frac{\omega L_{0}}{4\pi}
\right)\left(\frac{10^{-2}\rho_{17}}{\delta\tilde{\rho}}\right)\left(\frac{-10}{\beta}\right)\,,
\]
for small $\delta\mu^{2}/\omega^{2}$. These estimates assume that the field is effectively very light inside the star, which amounts to requiring that
\be
|\beta|\sim 7\left(\frac{0.3}{\mathcal{C}}
\right)\left(\frac{\mu}{10^{-10}{\rm eV}}\right)^{2}
\left(\frac{L_0}{10{\rm km}}\right)^{2}\,,
\ee
but the instabilities discussed here are expected to set in even at small nonzero effective masses.
The star oscillations can be induced by accretion, tidal effects or even mergers~\cite{1994ApJ...426..688R,Chirenti:2016xys,Ma:2020rak}. Note that numerical
relativity simulations show that the amplitude of density perturbations during coalesce, for example, can be significant and of the order of the central density itself~\cite{Perego:2019adq,Bernuzzi:2020txg}. Thus, both mechanisms may act on timescales short enough to be relevant.
In fact, at large couplings $\beta$ -- not yet ruled out for large bare masses -- parametric instabilities will be dominant.
%\ti{I added the following estimate.}
%From the massless condition inside the star, we get typical value of the coupling. 
%\begin{align}
%|\beta|\sim 7\left(
%\frac{0.3}{\mathcal{C}}
%\right)
%\left(
%\frac{m_{\phi}}{10^{-10}{\rm eV}}
%\right)^{2}
%\left(
%\frac{R}{10{\rm km}}
%\right)^{2}
%\end{align}

%%%%%%%%%%%%%%%%%%%%%%%%%%%%%%%%%%%%%%%%%%%%%%%%%%%%%%%%%%%%%%%%%%%%%%%%%%%%%%

\section{Final remarks}

%%%%%%%%%%%%%%%%%%%%%%%%%%%%%%%%%%%%%%%%%%%%%%%%%%%%%%%%%%
% \noindent{\bf{\em Final remarks.}}
%%%%%%%%%%%%%%%%%%%%%%%%%%%%%%%%%%%%%%%%%%%%%%%%%%%%%%%%%%

Nonminimally coupled, massive scalar fields can evade all observational constraints
from the observation of nearly stationary configurations, yet produce distinct signatures when evolving around oscillating backgrounds.
We have shown that there are at least two possible instability mechanisms: one blueshifts 
light scalars inside oscillating stars~\footnote{there were important hints that such an effect could occur.
The classical Fermi acceleration process is one example of energy extraction from a zero-average motion~\cite{Fermi:1949ee}. Similar processes exist for fields trapped inside cavities with periodically-moving boundaries~\cite{247776,PhysRevE.49.3535,Moore:1970,Andreata_2000,Dodonov:2001yb}. We showed that oscillating stars are also prone to such instabilities.};
the second mechanism is of parametric origin, triggered by a periodic oscillation of the star material.
Possible excitation mechanisms for the scalar field could include, for instance, coherently or stochastically oscillating dark matter.

Both instabilities act on short timescales when compared to viscous timescales~\cite{1987ApJ...314..234C}, and are expected
to play a role in neutron star oscillations. They can backreact on the star -- perhaps leading to gravitational collapse, or (in the blueshift mechanism)
simply result in a leakage of high frequency, high amplitude scalar. If a fraction, or all of dark matter
is made of a scalar component, then these mechanisms can act to produce overdensities close to neutron stars, providing one
more route to constraining dark matter. 
Details on observational signatures of these instabilities require further studies, beyond the scope of this work.

Similar instabilities are expected not only in scalar-tensor theories, but also in other theories with vectors or spinors~\cite{Annulli:2019fzq,Minamitsuji:2020hpl,Minamitsuji:2020pak}. Scalarized background solutions are not prone to the type of instabilities discussed here, but star oscillations will lead to radiation emission~\cite{Silvestri:2011ch}.
We expect that the instabilities discussed here could describe and affect other systems where a light degree of freedom is confined to an oscillating background, potentially observable in condensed matter systems.

Finally, other periodic systems include compact binaries; it can be expected that similar blueshift mechanisms act on such binaries.
Their high degree of asymmetry makes it more challenging to model, but their immense gravitational potential energy certainly makes them important candidates.

%%%%%%%%%%%%%%%%%%%%%%%%%%%%%%%%%%%%%%%%%%%%%%%%%%%%%%%

\begin{acknowledgments}
We are grateful to Sebastiano Bernuzzi, Kostas Kokkotas, Shingo Kukita, Tomohiro Nakamura, and Pantelis Pnigouras for useful correspondence and advice.
V. C. is a Villum Investigator supported by VILLUM FONDEN (grant no. 37766).
V.C.\ acknowledges financial support provided under the European Union's H2020 ERC 
Consolidator Grant ``Matter and strong-field gravity: New frontiers in Einstein's 
theory'' grant agreement no. MaGRaTh--646597.
M.Z.\ acknowledges financial support provided by FCT/Portugal through the IF
programme, grant IF/00729/2015.
T.I.\ acknowledges financial support provided under the European Union's H2020 ERC, Starting
Grant agreement no.~DarkGRA--757480.
This project has received funding from the European Union's Horizon 2020 research and innovation programme under the Marie Sklodowska-Curie grant agreement No 690904.
We thank FCT for financial support through Project~No.~UIDB/00099/2020 and through grants PTDC/MAT-APL/30043/2017 and PTDC/FIS-AST/7002/2020.

\end{acknowledgments}

%\appendix

\vspace{5mm}

\begin{center}
{\LARGE \sc Supplemental Material}
\end{center}

We now give more details and expand on the construction outlined in the main text.

%%%%%%%%%%%%%%%%%%%%%%%%%%%%%%%%%%%%%%%%%%%%%%%%%%%%%%%%%%%%%%%%%%
\section{Setup: nonminimal scalar fields}
\label{sec:setup}
\subsection{The background}
%%%%%%%%%%%%%%%%%%%%%%%%%%%%%%%%%%%%%%%%%%%%%%%%%%%%%%%%%%%%%%%%%%

We consider matter fields $\Psi_m$ describing a perfect fluid. Focus on a geometry with vanishing scalar,
$\Phi=0$, describing a static star of (constant) density $\tilde{\rho}_0$ and radius $L_0$. The
ADM mass of this solution can be written as $M=\frac{4\pi}{3}\tilde{\rho}_{0}L_0^{3}$ and its geometry can be expressed
as~\cite{Shapiro:1983du}
\begin{equation}
  \tilde{g}_{\mu\nu}dx^{\mu}dx^{\nu}=-\tilde{\alpha}^{2}dt^{2}+\tilde{a}^{2}dr^{2}
  +r^{2}d\Omega_2\,,
  \label{eq:background}
\end{equation}
where
\begin{align*}
\tilde{\alpha}^2 =
\begin{cases}
\left(\frac{3}{2}\left(1-\frac{2M}{L_0}\right)^{1/2}-\frac{1}{2}\left(1-\frac{2Mr^{2}}{L_0^{3}}\right)^{1/2}\right)^{2}&(r<L_0)\\
1-\frac{2M}{r}&(r>L_0)
\end{cases}
\end{align*}
and
\begin{align*}
\tilde{a}^{2} =
\begin{cases}
\left(1-\frac{2Mr^{2}}{L_0^{3}}\right)^{-1}&(r<L_0)\\
\left(1-\frac{2M}{r}\right)^{-1}&(r>L_0)
\end{cases} \,.
\end{align*}
The profile of the pressure inside the star is
\begin{eqnarray}
\tilde{p}_{0}(r)&=&\tilde{\rho}_{0}\left(
\frac{(1-\mathcal{C}\frac{r^{2}}{L_0^{2}})^{1/2}-(1-\mathcal{C})^{1/2}}
                {3(1-\mathcal{C})^{1/2}-(1-\mathcal{C}\frac{r^{2}}{L_0^{2}})^{1/2}}\right) \,.
                \label{star_profile}
\end{eqnarray}

We can now linearize the equations of motion around this background. Although we focus solely on constant-density stars,
previous results indicate that the details of the model are not important~\cite{Pani:2010vc}. The most relevant feature is the contribution
of the star (trace of) energy tensor $\tilde{T}=3\tilde{p}-\tilde{\rho}$ to the effective scalar mass, and this contribution is qualitatively the same 
for different equations of state.

There are situations when the background scalar is nontrivial, as we explain below. In such circumstances, it backreacts on the geometry.
Although the formalism to handle backreaction is clear, for simplicity here we ignore the effect of the scalar field on the profile of the fluid. In other words, we deal only with the Klein-Gordon equation on a fixed geometry.
As a concrete situation, we consider relativistic stars of constant density as a solution, parameterized by radius $L_0$ and compactness $\mathcal{C}$.
% defined by~Eq.~\eqref{eq:compactness_def}

%%%%%%%%%%%%%%%%%%%%%%%%%%%%%%%%%%%%%%%%%%%%%%%%%%%%%%%%%%%%%%%%%%
\subsection{The dynamical equations}
%%%%%%%%%%%%%%%%%%%%%%%%%%%%%%%%%%%%%%%%%%%%%%%%%%%%%%%%%%%%%%%%%%
In Sec.~\ref{sec:GR}, we deal with full dynamical spacetimes on which the scalar evolves.
For now, we consider simply a time-independent spacetime, since it serves to convey our main message. Let $\Phi_0$ be a static, spherically-symmetric solution of Eq.~(\ref{KG_effective}), with the metric~\eqref{eq:background} and scalar potential and coupling functions given by Eq.~(\ref{eq:potential_coupling}). $\Phi_0$ is then a solution to the equation
\begin{align}
&\partial_{r}^{2}\Phi_{0}+\left(\frac{2}{r}+\frac{\partial_{r}\alpha}{\alpha_{0}}-\frac{\partial_{r}a}{a}
\right)\partial_{r}\Phi_{0} \notag\\
%
%+a^{2}\left(\tilde{T}^{(m)}A^{3}(\Phi_{0})A'(\Phi_{0})-V'(\Phi_{0})\right)&=&0
&\qquad + a^{2}\left(\beta\tilde{T}^{(m)}e^{2\beta\Phi_{0}^{2}} \Phi_{0}
                                                                               -\mu_0^{2}\Phi_{0}\right)=0\,,
\end{align}
where $\alpha$ and $a$ are trivially related to the functions $\tilde a$ and $\tilde \alpha$ defined above via
\[
  (\tilde a, \tilde \alpha) =  (a, \alpha) e^{\frac{\beta}{2}\Phi_0^2} \,.
\]
Let us now consider a general fluctuation around this value,
\begin{eqnarray}
\Phi(t,r,\theta,\phi)=\Phi_{0}(r) + \epsilon\sum_{l,m} \frac{\psi_{lm}(t,r)}{r}Y_{lm}(\theta,\phi)\,,
\label{definition of psi}
\end{eqnarray}
where $\epsilon$ is a bookkeeping parameter and $Y_{lm}(\theta,\phi)$ are the usual scalar spherical harmonics.
Then, keeping only first order terms in $\epsilon$,
we find from Eq.~(\ref{KG_effective}) that $\psi_{lm}$ is governed by
\begin{equation}
-\partial_{t}^{2}\psi-\frac{\alpha^{2}}{a^{2}}\partial_{r}\left(\log{\left(\frac{a}{\alpha}\right)}\right)\partial_{r}\psi+\frac{\alpha^{2}}{a^{2}}\partial_{r}^{2}\psi-V_{\rm eff}\psi=0\,,
\label{eq:eff_potential}
\end{equation}
where we dropped the $lm$ subscripts for clarity and the effective potential $V_{\rm eff}$ is given by
\begin{align}
  V_{\rm eff}& =
               \frac{\alpha^{2}}{ra^{2}}\left(\frac{\partial_{r}\alpha}{\alpha}
               - \frac{\partial_{r}a}{a}\right)+\alpha^{2}\frac{l(l+1)}{r^{2}}\nonumber\\
&\quad+\alpha^{2}\left(\mu_0^{2}-e^{2\beta\Phi_0^{2}}\beta\tilde{T}^{(m)}(1+4\beta\Phi_{0}^{2})\right)\,.\label{eq:Veff}
\end{align}

The stable background solutions $\Phi_0(r)$ depend dramatically on the value of the coupling $\beta$~\cite{Damour:1993hw,Damour:1996ke,Harada:1998ge,Pani:2010vc}.
There is a threshold coupling $\beta_c$ that divides cases where $\Phi_0$ is a stable solutions from those where it is not.
For $\beta>\beta_c (<0)$, $\Phi_0=0$ is a stable solution and the geometry of the star and all its properties are identical to those in GR. 
If $\beta<\beta_{c}$, the trivial scalar profile becomes unstable. In such conditions, a GR star quickly develops a nontrivial profile $\Phi_0$ and is said, for obvious reasons,
to be \emph{scalarized}. In this case, the geometry and properties of the star change, and it would no longer be described by Eq.~\eqref{star_profile}, for example.
All our techniques below can nonetheless be extended to scalarized objects, or directly applied if one's focus is on configurations where $|\Phi_0|\ll 1$ (also known as the decoupling limit).

These scalar-tensor theories are appealing because~\cite{Damour:1993hw,Damour:1996ke,Harada:1998ge,Pani:2010vc,Cardoso:2020cwo}
\be
\beta_c\propto -\left(1+(\mu_0 R)^{2}\right)\mathcal{C}^{-1}\,.\label{crit_beta}
\ee
Thus, they predict scalarization in strong field situations, for example in neutron stars,
but not in stars like our Sun. Thus, these theories satisfy solar-system bounds in a natural way.
On the other hand, several observational predictions regarding neutron stars can be different from GR, and thus interesting constraints can be imposed.
The observation of binary neutron stars can, in principle, be used to constrain scalarization since scalarized stars would radiate more energy
(thus leading to a more rapid inspiral, as compared with the standard general relativistic solutions)~\cite{Damour:1993hw,Damour:1996ke,Harada:1998ge,Pani:2010vc,Antoniadis:2013pzd}. However, low energy excitations are unable to propagate if the field is massive. Thus, for fields with
\begin{eqnarray}
\mu_0 \gg 10^{-16} {\rm eV}\,,
\end{eqnarray}
it is challenging to impose constraints on this mechanism via neutron star observations~\cite{Ramazanoglu:2016kul}.

\begin{figure*}[th]
  \includegraphics[width=0.48\textwidth]{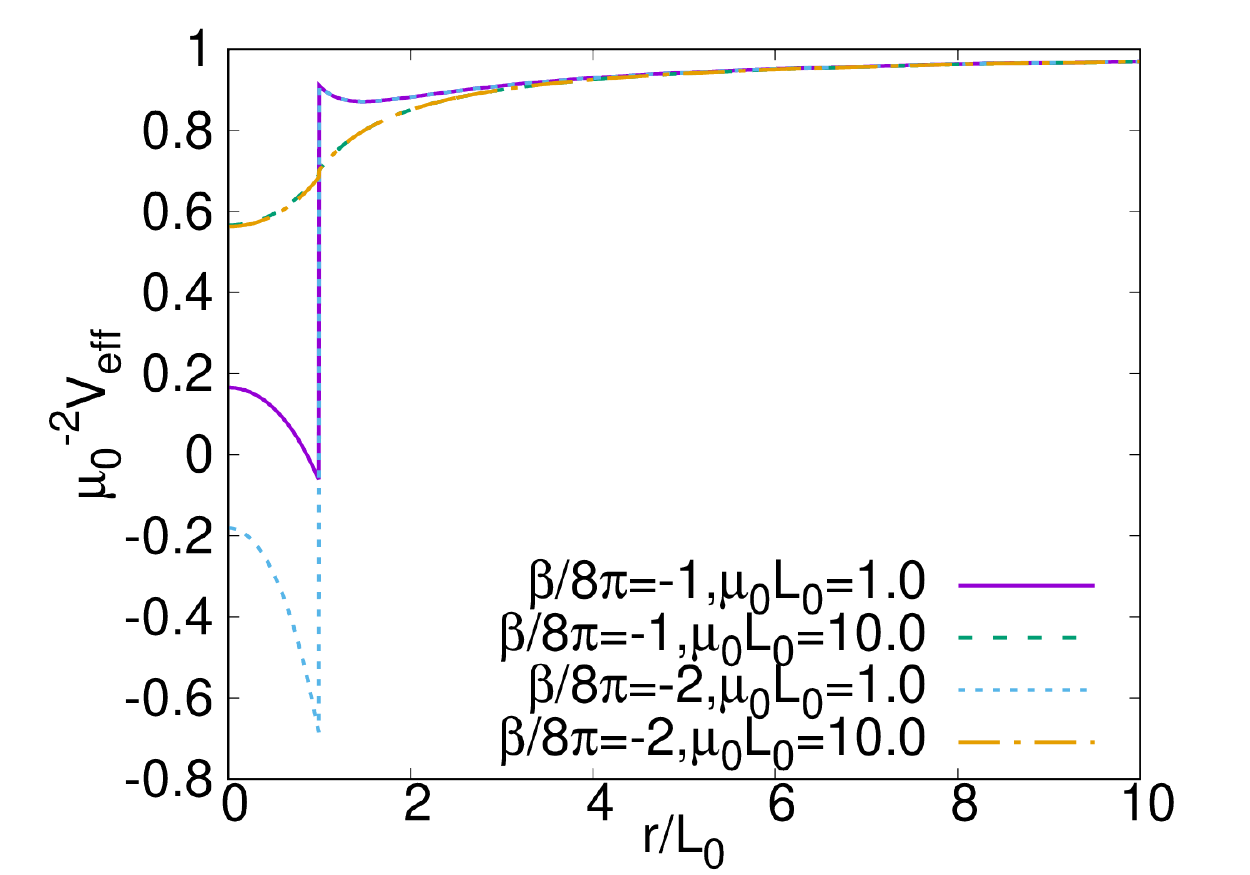}
\hfill
\includegraphics[width=0.48\textwidth]{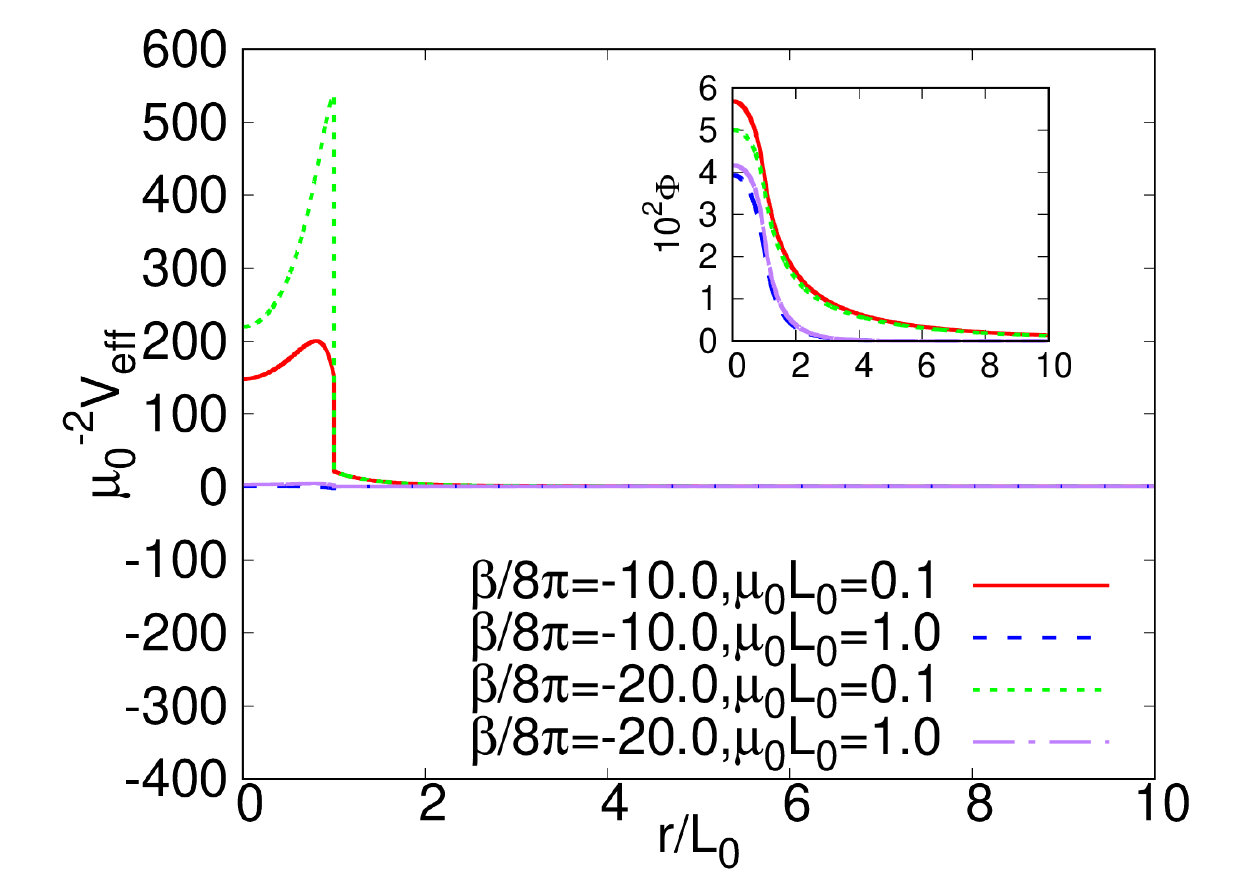}
\caption{(Left panel) Effective potential for spherically-symmetric scalar fluctuations around a relativistic uniform density star with $\mathcal{C}=0.3$
and a trivial background scalar $\Phi_0=0$ (with a coupling $\beta>\beta_c$).
(Right panel) Same as in the left panel, but now for a subcritical coupling $\beta<\beta_{c}$, leading to scalarization of the star. The behavior of the blue and purple lines is hard to see, but their profile is similar to that of the red and green lines.
The corresponding scalar profile is shown in the inset.
}
\label{plot_Veff_case1_1_neutronstar3}
\end{figure*}
To understand the possible dynamics of the scalar field, it is important to know the structure of the effective potential $V_{\rm eff}$. This potential is shown
in Fig.~\ref{plot_Veff_case1_1_neutronstar3}. The importance of the effective potential is most easily seen from Eq.~\eqref{eq:eff_potential}. For fluctuations of energy $\Omega$, 
\be
\psi\sim e^{-i\Omega t}\Upsilon(r)\,, 
\ee
it takes the form of a Schr\"odinger-like equation (notice that $\Omega$ has dimensions of ${\rm mass}^{-1}$ in geometrical units, but is an energy parameter)
\begin{equation}
\frac{\alpha^{2}}{a^{2}}\partial_{r}^{2}\Upsilon-\frac{\alpha^{2}}{a^{2}}\partial_{r}\left(\log{\left(\frac{a}{\alpha}\right)}\right)\partial_{r}\Upsilon+(\Omega^2-V_{\rm eff})\Upsilon=0\,,\label{eq_schr}
\end{equation}
Thus for small enough energies $\Omega^2<V_{\rm eff}$ the fluctuation is unable to tunnel to infinity if $V_{\rm eff}$ asymptotes to a constant positive value.

When $\beta<\beta^{c}$, the object is scalarized with a nontrivial profile $\Phi_{0}\neq 0$ (Fig.~\ref{plot_Veff_case1_1_neutronstar3}, right panel).
Notice from the shape of the effective potential that bound states are hard to achieve inside the object in these cases, since they can easily ``tunnel out'' (in the language of quantum mechanics associated with the Schr\"odinger-like equation \eqref{eq_schr}).

For non-scalarized solutions, with $\Phi_0=0$, Eq.~\eqref{eq:Veff} shows that fluctuations of the scalar are massive with an effective mass
\be
\mu^2=\mu_0^2-\beta\tilde{T}^{(m)}\,,
\ee
which effectively acts as a position-dependent mass term. Consider setups where the effective mass inside the star satisfies $\mu^2<\mu_0^2$ (cf.\ Eq.~\eqref{eq_schr} and the left panel of Fig.~\ref{plot_Veff_case1_1_neutronstar3}). Then, low-energy fluctuations of the scalar are trapped inside the star. This aspect plays a critical role in our analysis and results.
Note that for extremely compact objects the trace of the stress tensor can change sign, and positive couplings may give rise to spontaneous scalarization as well~\cite{Mendes:2014ufa}.
Our results can in principle be applied to all these different scenarios.

%%%%%%%%%%%%%%%%%%%%%%%%%%%%%%%%%%%%%%%%%%%%%%%%%%%%%%%%%%%%%%%%%%
\section{Periodic-motion instabilities in one dimension}
\label{sec:oscillating}
%%%%%%%%%%%%%%%%%%%%%%%%%%%%%%%%%%%%%%%%%%%%%%%%%%%%%%%%%%%%%%%%%%
As we discussed, some classes of scalar-tensor theories are described by a scalar field with an effective mass that depends on the environment. In particular, 
the effective mass $\mu$ of the scalar field inside a star, for example, can be lighter than its vacuum counterpart $\mu_0$.
The potential experienced by the field looks like those depicted in Fig.~\ref{plot_Veff_case1_1_neutronstar3} and bound states inside the star can appear.
Due to this potential barrier, field excitations with wavelength larger than $\mu^{-1}$ are reflected around the surface of the object.
Thus, the field propagates within a cavity whose size is varying periodically, and whose density is also varying periodically. We find that these two aspects
give rise two instabilities of different kinds. A varying cavity size allows for the possibility of energy extraction from the cavity via ``Doppler-like'' exchanges, whereas a varying density gives rise to parametric instabilities.

We describe in details these two types of instability below. Before going into the more complex case of four-dimensional spacetimes, 
we consider a simple 1+1 setup in this section.

%%%%%%%%%%%%%%%%%%%%%%%%%%%%%%%%%%%%%%%%%%%%%%%%%%%%%%%%%%%%%%%%%%
\subsection{The blueshift instability}
%%%%%%%%%%%%%%%%%%%%%%%%%%%%%%%%%%%%%%%%%%%%%%%%%%%%%%%%%%%%%%%%%%
\begin{figure}[th]
\includegraphics[width=0.3\textwidth]{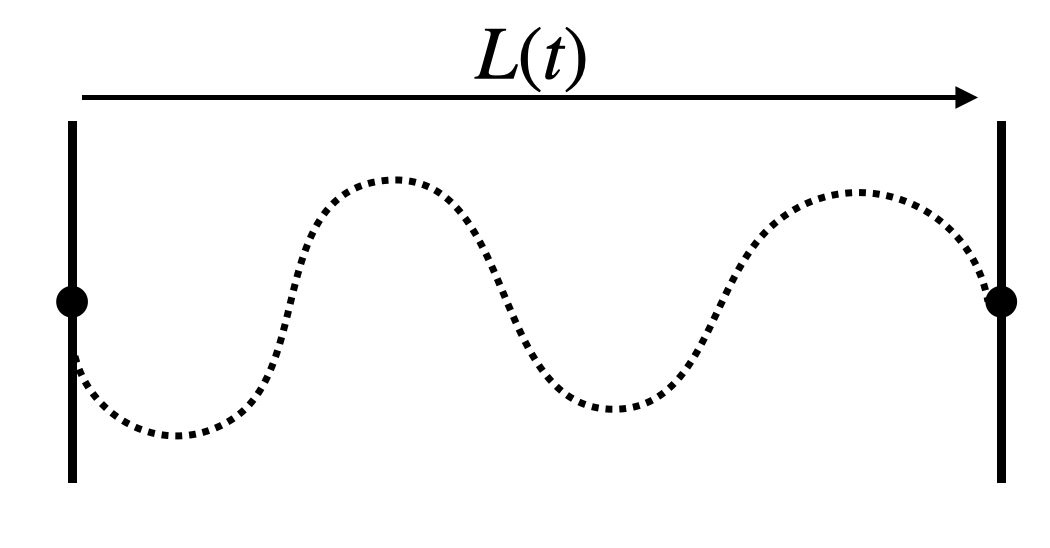}
\caption{A scalar field confined inside a cavity of size $L(t)$.
The left boundary is kept fix and the right boundary (periodically) oscillates in time.
\label{Set_up}}
\end{figure}
We decouple the two effects mentioned above and consider first a scalar field of constant mass,
confined to a one-dimensional oscillating cavity in flat space -- see Fig.~\ref{Set_up}. 
In particular, one cavity boundary is fixed at $x=0$ while the other is located at $L(t)$, as described by Eq.~\eqref{eq:radius_oscillation}.
The scalar field is then governed by the massive Klein-Gordon equation with
Dirichlet boundary conditions imposed at a time-dependent boundary,
\begin{equation}
\begin{aligned}
-\partial_{t}^{2}\Phi+\partial_{x}^{2}\Phi-\mu^{2}\Phi=0\,,\\
\Phi(t,0)=\Phi(t,L(t))=0\,,
\label{eq:KG eq 1+1}
\end{aligned}
\end{equation}
where $\mu$ is the (effective) mass of the scalar field inside the cavity. This models an oscillating star with a
scalar field whose effective mass is small inside the star and large outside.
The system has been investigated previously in other contexts, and shown to give rise to blue or redshifted reflections~\cite{247776,PhysRevE.49.3535}.
After a large enough number of reflections, the energy of the field increases without bound for
\begin{eqnarray}
\frac{N\pi}{L_{0}+\delta L}<\omega<\frac{N\pi}{L_{0}-\delta L}\,,
\label{eq:instability condition}
\end{eqnarray}
where $N$ is an integer~\cite{PhysRevE.49.3535}.
We call this the \textit{blueshift condition} and the associated instability the \textit{blueshift instability}.
In the following we study the mechanism and associated instability in detail.

%%%%%%%%%%%%%%%%%%%%%%%%%%%%%%%%%%%%%%%%%%%%%%%%%%%%%%%%%%%%%%%%%% 
\subsubsection{Estimate from Doppler effect}
\label{sec Doppler effect}
%%%%%%%%%%%%%%%%%%%%%%%%%%%%%%%%%%%%%%%%%%%%%%%%%%%%%%%%%%%%%%%%%%
%
\begin{figure}[t]
\includegraphics[width=0.475\textwidth]{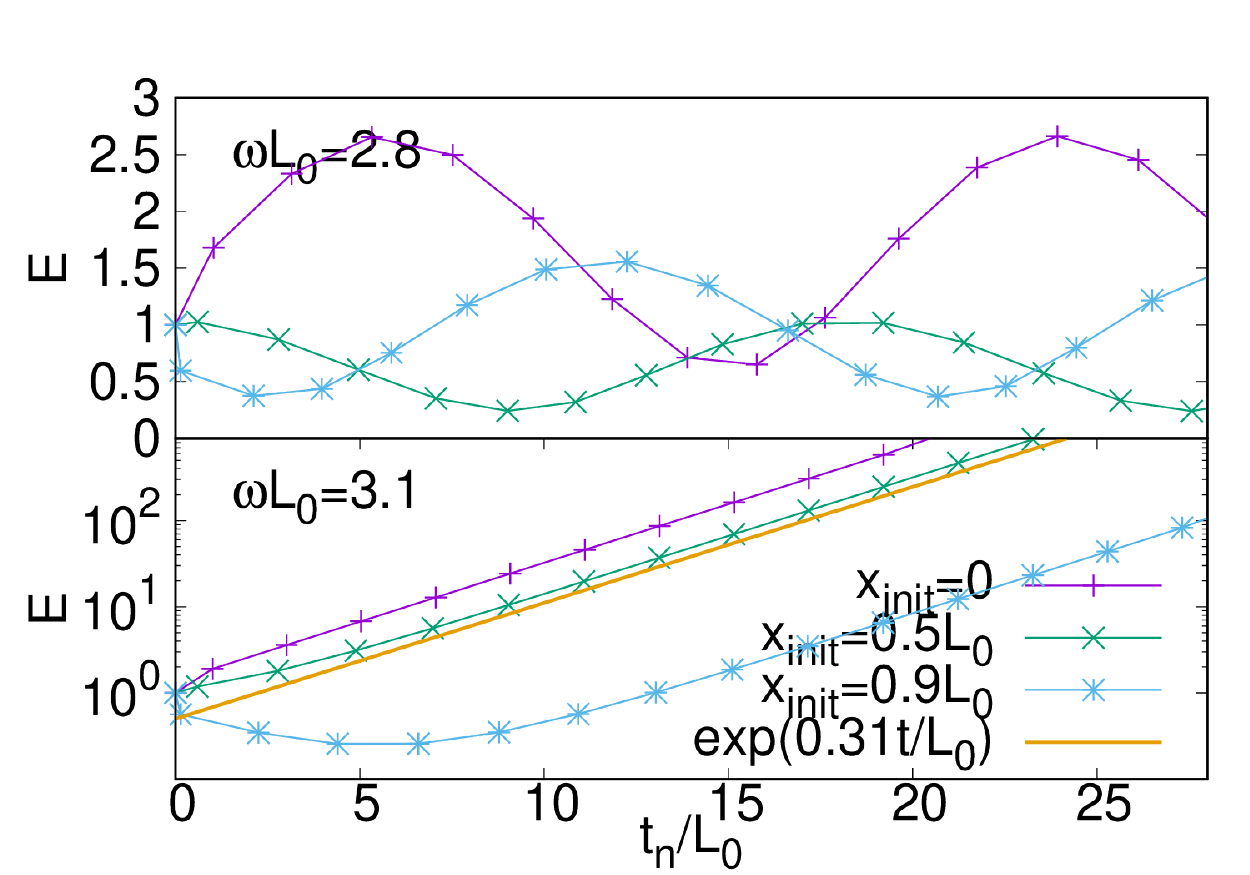}
\caption{
Time evolution of energy of a photon inside a box with an oscillating boundary, solution of recursion relation~\eqref{eq:En-rec}. 
The amplitude of oscillation of the boundary is $\delta L=0.1 L_{0}$, the initial position $x_{\rm init}$ takes different selected values.
For oscillation frequencies $\omega L_0=2.8$ (top panel), the energy oscillates, which means that there's energy transfer between
the photon and the boundary, and they reached equilibrium. This also means that the blueshift instability condition~\eqref{eq:instability condition}
is not satisfied.
For oscillation frequencies that do satisfy that condition, the boundary is constantly giving energy to the photon, resulting in an exponential growth of its energy, as seen in the bottom panel.
\label{plot_example_particle_L0_1_L1_0_1_omega_2_7}
}
\end{figure}

The growth of energy in the field can be understood as a cumulative Doppler effect~\cite{Dittrich:1993hw}. To understand this in more depth and from a different perspective,
we consider the following toy model. A massless particle is inside a perfectly reflecting cavity whose boundary oscillates, and is emitted at $x=x_{\rm init}$ moving to the right with initial (angular) frequency $\Omega_{0}$. After one reflection at the (moving) boundary, the frequency is Doppler-shifted to $\Omega_{1}$,
\be
\Omega_{1}=\frac{1-v(t_{1})}{1+v(t_{1})}\Omega_{0} \,,
\ee
where $v(t)=\dot{L}=\delta L \, \omega \cos\omega t$ is the velocity of the oscillating boundary, and $t_{1}$ the coordinate time when the particle hits the boundary
determined from the equation $t_1+x_{\rm init}=L(t_{1})$.
Using the same argument, one can obtain a recursion relation for the frequency $\Omega_{n_{\rm ref}}$, and therefore for the energy $E_{n_{\rm ref}}$ after $n_{\rm ref}$ reflections
\be
\Omega_{n}=\frac{1-v(t_{n_{\rm ref}})}{1+v(t_{n_{\rm ref}})}\Omega_{n_{\rm ref}-1}\,,\label{eq:En-rec}
\ee
where $t_{n_{\rm ref}}$ is determined by
\be
t_{n_{\rm ref}}-t_{n_{\rm ref}-1}-L(t_{n_{\rm ref}-1})=L(t_{n_{\rm ref}})\,.
\ee

The recursion relation~\eqref{eq:En-rec} can be solved numerically, its solution is shown in Fig.~\ref{plot_example_particle_L0_1_L1_0_1_omega_2_7} for different choices of parameters fixing $\delta L=0.1L_0$.
Note that, for this choice of $\delta L$, condition~\eqref{eq:instability condition} becomes $2.85<\omega L_0 / N <3.49$. We thus choose two representative values for the oscillation frequency $\omega$, one outside the instability interval and one inside. The top panel of the figure shows that when the instability criterion is not satisfied the energy of the particle oscillates in time. In some interactions with the boundary it gains energy, in some it loses energy, the average gain is zero. The bottom panel shows that there are conditions that lead to a constant gain of energy by the particle, and where each reflection is accompanied by a transfer of energy from the boundary to the bouncing massless particle.
In particular, for certain conditions, the energy grows exponentially, 
\be
E\propto e^{\lambda_{B} t}\,,
\ee
and the exponent $\lambda_{B}>0$ is insensitive (or very mildly dependent only) on the initial position $x_{\rm init}$.

\begin{figure}[tb]
  \includegraphics[width=0.48\textwidth]{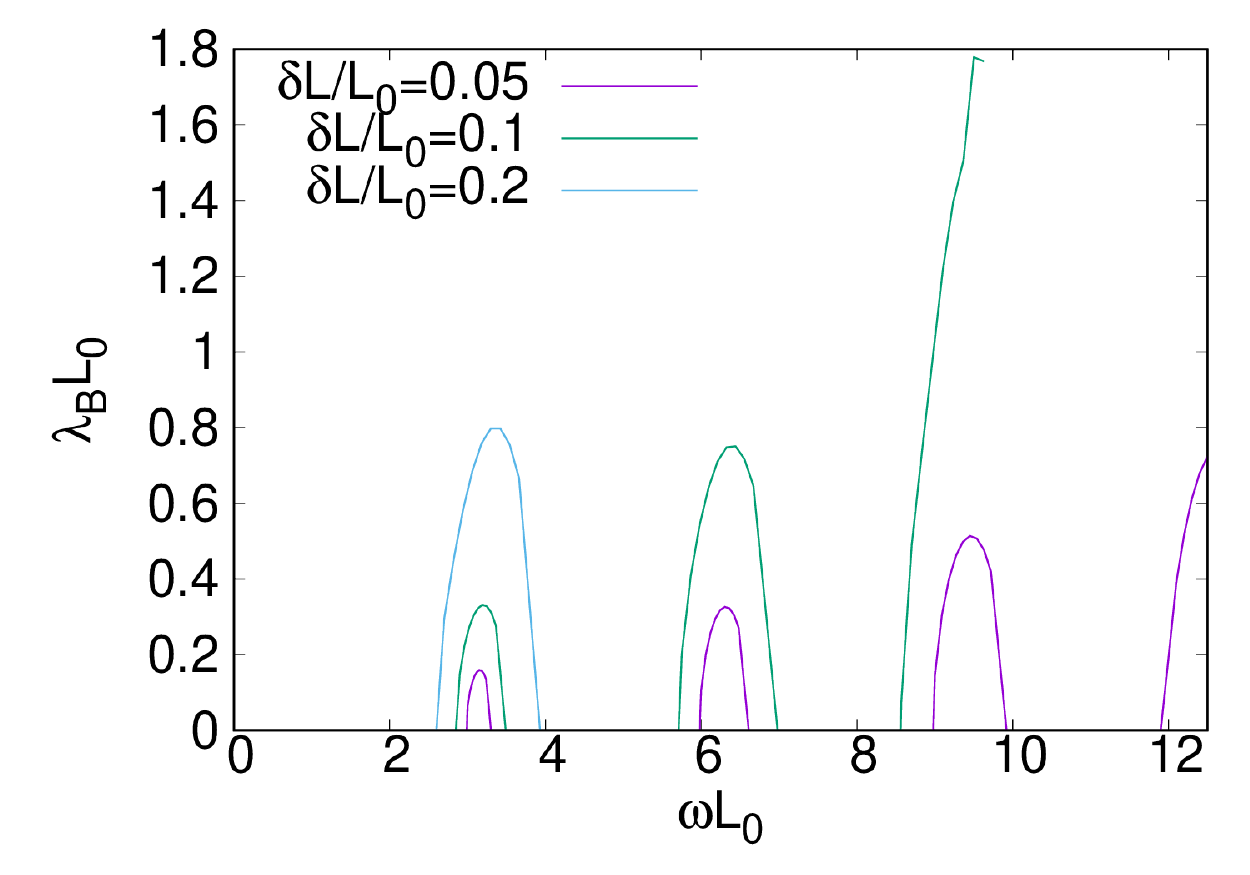}
\caption{Relation between the instability rate $\lambda$ and the frequency of the boundary for each $\delta L$.
   \label{plot_particle_omega_lambda_reoation} }
\end{figure}
Figure~\ref{plot_particle_omega_lambda_reoation} shows the relation between the instability rate, $\lambda_{B}$, and the frequency of oscillation of the boundary $\omega$.
An approximate relation for the lowest-frequency instability window in Fig.~\ref{plot_particle_omega_lambda_reoation} is
\be
\lambda_{B} L_{0} \sim a(\omega)\left(\frac{\delta L}{L_{0}}-\left|\frac{\pi}{\omega L_{0}}-1\right|\right)\,,\label{eq;time scalar blueshift}
\ee
valid in the interval $\pi/(L_{0}+\delta L)<\omega<\pi/(L_{0}-\delta L)$.
The coefficient $a(\omega)$ has a nontrivial dependence on the boundary oscillation frequency and is $a\sim 3.5$ for $\omega L_{0}=\pi$, and $a\sim 4.6$ for $\omega L_{0}=0.9\pi$.

%%%%%%%%%%%%%%%%%%%%%%%%%%%%%%%%%%%%%%%%%%%%%%%%%%%%%%%%%%%%%%%%%%
\subsubsection{An exact solution}
%%%%%%%%%%%%%%%%%%%%%%%%%%%%%%%%%%%%%%%%%%%%%%%%%%%%%%%%%%%%%%%%%%
The results above are based on a simple particle picture, compelling but valid for massless and localized fields only. We now wish to 
show that this behavior arises also when solving the wave equation~\eqref{eq:KG eq 1+1}. In fact, an exact solution for certain prescribed motion can be found for massless scalars subjected to Dirichlet conditions at the boundaries~\cite{Law:1994zza}. The boundaries are located at
\be
x=\{0, L(t)\}\,,
\ee
but -- here and only here -- we will sacrifice the simple boundary motion \eqref{eq:radius_oscillation} in order to find an exact solution to the problem.
In one dimension, the solution of the wave equation takes the form
\be
\Phi=e^{-ik\pi {\cal R}(t+x)}-e^{-ik\pi {\cal R}(t-x)}\,,
\ee
which ensures that the massless KG equation is satisfied. 
Here, $k$ is an arbitrary constant and $k\pi {\cal R}$ is the frequency $\Omega$ of the pulse.
To satisfy the BCs one takes
\be
{\cal R}(t+L(t))={\cal R}(t-L(t))+2\,,
\ee
The solution to this recursion relation solves the problem. For prescribed motion,
\beq
L(t)&=&L_0+\frac{L_0}{2\pi}\left(\arcsin{\left(\sin\theta\cos\frac{2\pi t}{L_0}\right)}-\theta\right)\,,\\
\theta&=&\arctan\frac{\epsilon \pi}{L_0}\,,
\eeq
where $\epsilon$ describes the amplitude of the motion, the solution can be found as
\be
{\cal R}(2n_{\rm ref}L_0+\zeta)=2n_{\rm ref}+\frac{1}{2}-\frac{1}{\pi}\arctan \left(\cot\frac{\zeta\pi}{L_0}-\frac{2n_{\rm ref}\epsilon\pi}{L_0}\right)\,,
\ee
where $\zeta$ is a variable in $[-L_0,L_0]$ and $n\geq 1$ is a positive integer. 
The branch of $\arctan$ must be selected carefully in order to avoid a discontinuity.
Notice that for small $\epsilon$,
\be
\frac{L(t)}{L_0}=1-\frac{\epsilon}{L_0}\sin^2\frac{\pi t}{L_0}+{\cal O}(\epsilon/L_0)^3
\ee
It is easy to show that the system is unstable, precisely because the frequency content is increasing~\cite{Law:1994zza}.
This construction can be generalized to a 3+1 setup.

%%%%%%%%%%%%%%%%%%%%%%%%%%%%%%%%%%%%%%%%%%%%%%%%%%%%%%
\subsubsection{Time evolutions}
\label{sec:oscillating-numerical}
%%%%%%%%%%%%%%%%%%%%%%%%%%%%%%%%%%%%%%%%%%%%%%%%%%%%%%%

The results above are either based on a simple particle picture, or arise within a very special and prescribed boundary motion. We now wish to 
show that this behavior is a generic consequence also when solving the (massive) wave equation~\eqref{eq:KG eq 1+1}. 
To numerically evolve this equation, we first start by introducing new coordinates $(T,X)$ defined by
\be
T(t,x) = t \,, \qquad X(t,x)=\frac{x}{L(t)}\,.
\ee
In these coordinates, the equation of motion for the scalar field $\phi(T,X)$ take the form
\begin{align}
&-\partial_{T}^{2}\Phi+\left(\frac{1}{L^{2}}-\frac{X^{2}\dot{L}^2}{L^2}\right)\partial_{X}^{2}\Phi
+X\left(\frac{\ddot{L}}{L}-\frac{2\dot{L}^2}{L^2}\right)\partial_{X}\Phi\nonumber\\
  & \qquad+2X\frac{\dot{L}}{L}\partial_{T}\partial_{X}\Phi-\mu^{2}\Phi=0\,,
      \label{eq:KG eq 1+1 with X}
\end{align}
Note that we interchangeably use both dots and $\partial_T$ to denote time derivatives with respect to $T$.
The boundary conditions (now at \emph{constant} spatial coordinate $X$) are
\begin{eqnarray}
\Phi(T,X=0)=\Phi(T,X=1)=0 \,.\label{B.C.1+1 with X}
\end{eqnarray}
In the new coordinates, the effect of the moving boundary appears as a shift vector term.
To numerically solve Eq.~\eqref{eq:KG eq 1+1 with X}, we use the characteristic variables $\Phi_{\pm}$ defined by 
\be
\Phi_{\pm}\equiv \frac{1}{2}\left(\dot{\Phi} + \lambda_{\mp} \partial_{X}\Phi\right)\,,\quad\lambda_{\pm}=\pm\frac{1}{L}-X\frac{\dot{L}}{L}\,.
\ee
The equations of motion are then
\begin{subequations}
 \label{eq:1+1eom}
\begin{align}
\dot{\Phi}_{\pm} & = -
\lambda_{\pm}\partial_{X}\Phi_{\pm}
-\frac{\mu^{2}}{2} \Phi\,,\\
\dot{\Phi} & = L \left(\lambda_{+}\Phi_{+}
-  \lambda_{-}\Phi_{-}\right) \,,
\end{align}
\end{subequations}
with the constraint
\begin{equation}
\frac{\partial_{X}\Phi}{L}=-\Phi_{+}+\Phi_{-} \,.
\end{equation}
In these variables, the boundary conditions become
\begin{subequations}
  \label{eq:1+1bc}
\begin{align}
\Phi_{+}(T,0) &= \frac{\lambda_{-}(T)}{\lambda_{+}(T)}\Phi_{-}(T,0) \,,\\
\Phi_{-}(T,1) &= \frac{\lambda_{+}(T)}{\lambda_{-}(T)}\Phi_{+}(T,1) \,.
\end{align}
\end{subequations}
Finally, the energy of the scalar field inside the cavity is given by
\begin{align}
E&=\frac{1}{2}\int_{0}^{L(t)}\left((\partial_{t}\Phi)^{2}+(\partial_{x}\Phi)^{2}+\mu^{2}\Phi^{2}\right)dx \notag\\
&=\int_{0}^{1}LdX\left\{(\Phi_{+}^{2}+\Phi_{-}^{2})+\frac{\mu^{2}}{2}\Phi^{2}\right\} \,.
\end{align}

\begin{figure}[th]
\includegraphics[width=0.48\textwidth]{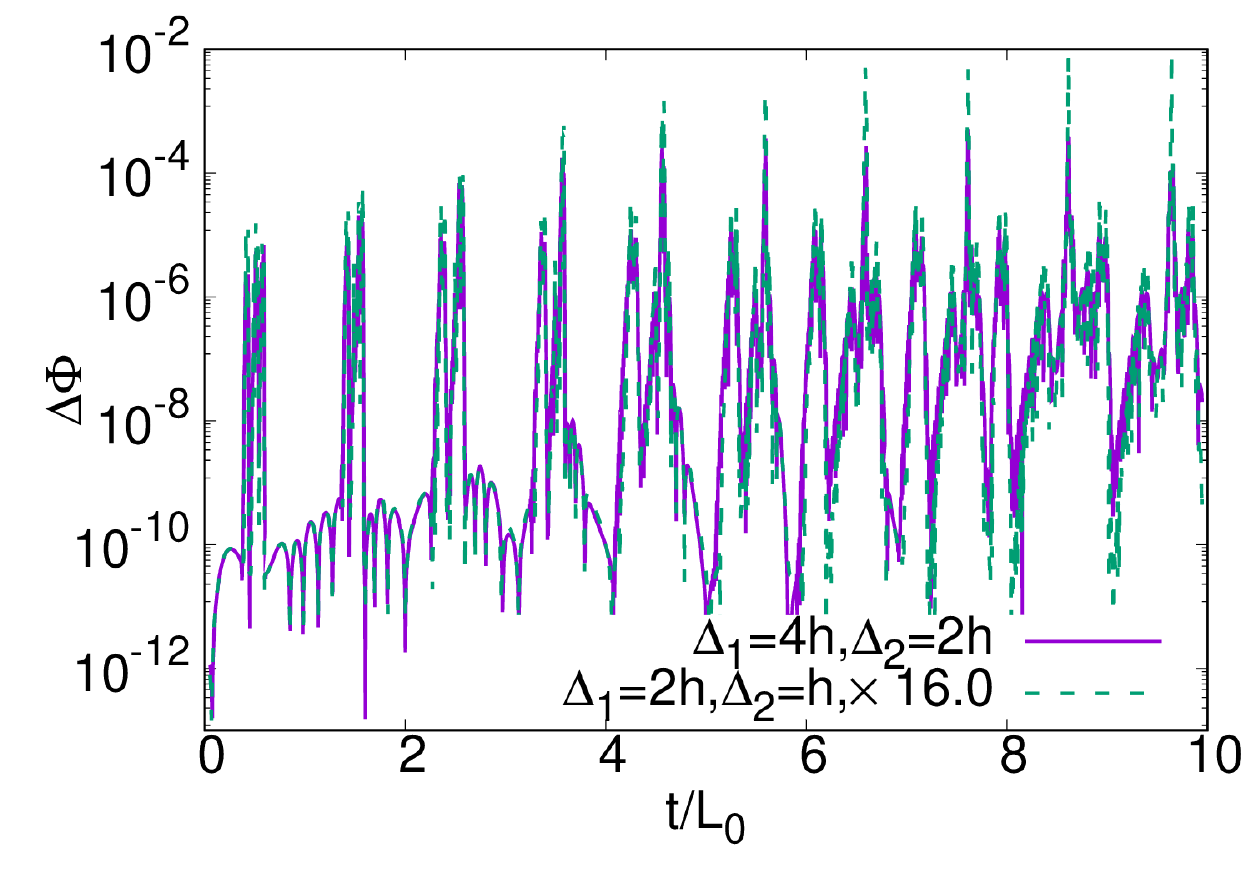}
\caption{Convergence study of the evolution of the scalar field at fixed
radius for the simulation of Fig.~\ref{Graph_time_evolution_of_center_for_draft_ID1_Amp1_wn02_r05_bulkmass0_AmpL01_omegaL_3_1_L1_full2}. The purple line shows the difference between results obtained with low ($4h$) and medium ($2h$) resolutions, while the (dashed) green line shows the difference between results obtained with medium and high ($h$) resolutions multiplied by 16, the expected factor for fourth-order convergence.
\label{convergence_phi_ID1_Amp1_wn01_r05_bulkmass0_AmpL01_omegaL_3_1_L1_full2}
}
\end{figure}

We evolve Eqs.~\eqref{eq:1+1eom} with boundary conditions~\eqref{eq:1+1bc} using
a finite differencing scheme where spatial derivatives are approximated with
4th-order accurate upwind stencils and a 4th-order accurate Runge Kutta scheme
is used for the time integration. %We refer to this code as \texttt{code1}.
We have checked that this code convergences with the expected fourth-order
accuracy, as shown in
Fig.~\ref{convergence_phi_ID1_Amp1_wn01_r05_bulkmass0_AmpL01_omegaL_3_1_L1_full2}.

\begin{figure}[th]
\includegraphics[width=0.48\textwidth]{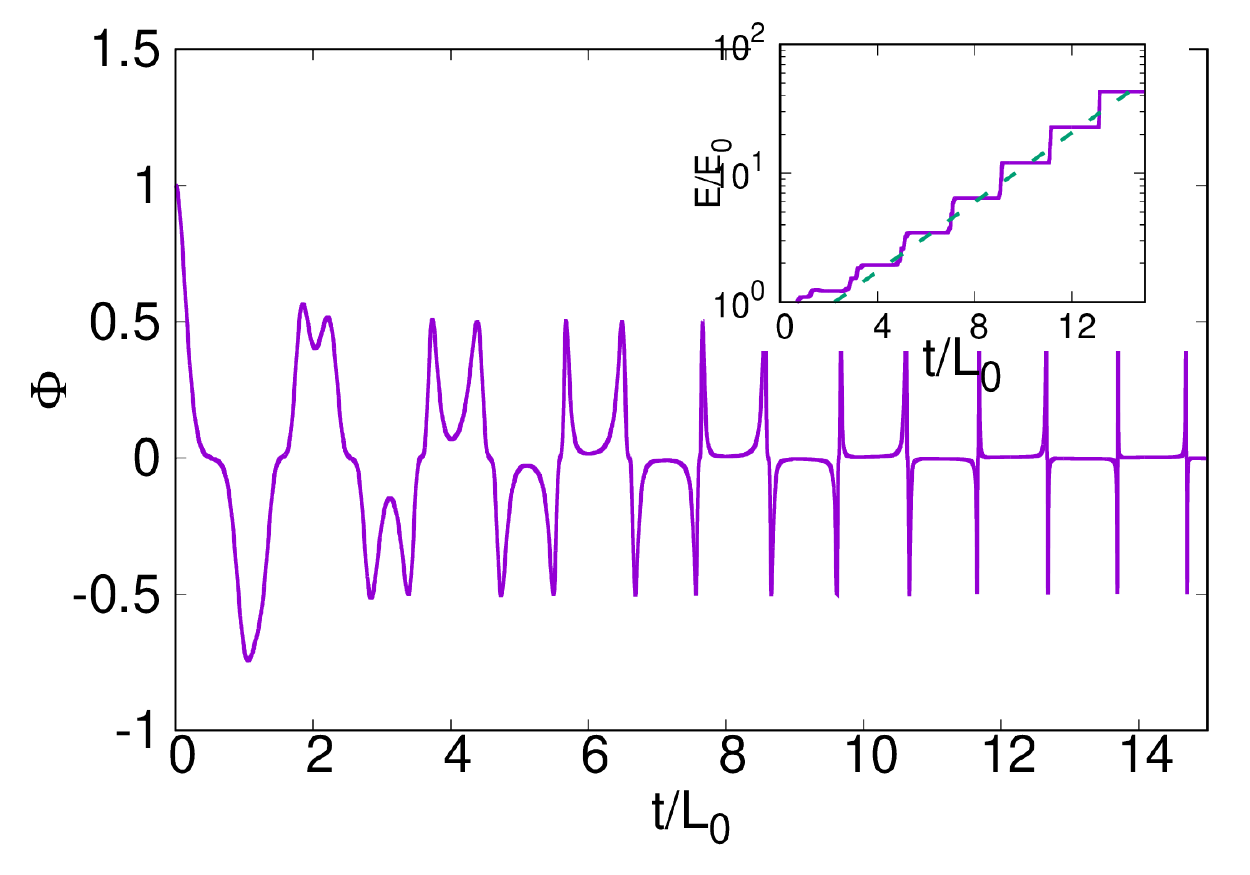}
\caption{Evolution of a gaussian pulse of scalar field and its energy (inset). The field is massless, but confined to a cavity whose right boundary oscillates at $\omega L_0=3.1$
with an amplitude $\delta L=0.1L_0$. The gaussian pulse has width $\sigma=0.2L_0$ and is centered at $r_0=0.5L_0$.
The green dotted line in the inset is the exponential $\exp{(0.31t/L_{0})}$.
\label{Graph_time_evolution_of_center_for_draft_ID1_Amp1_wn02_r05_bulkmass0_AmpL01_omegaL_3_1_L1_full2}
  }
\end{figure}

\begin{figure}[th]
\includegraphics[width=0.48\textwidth]{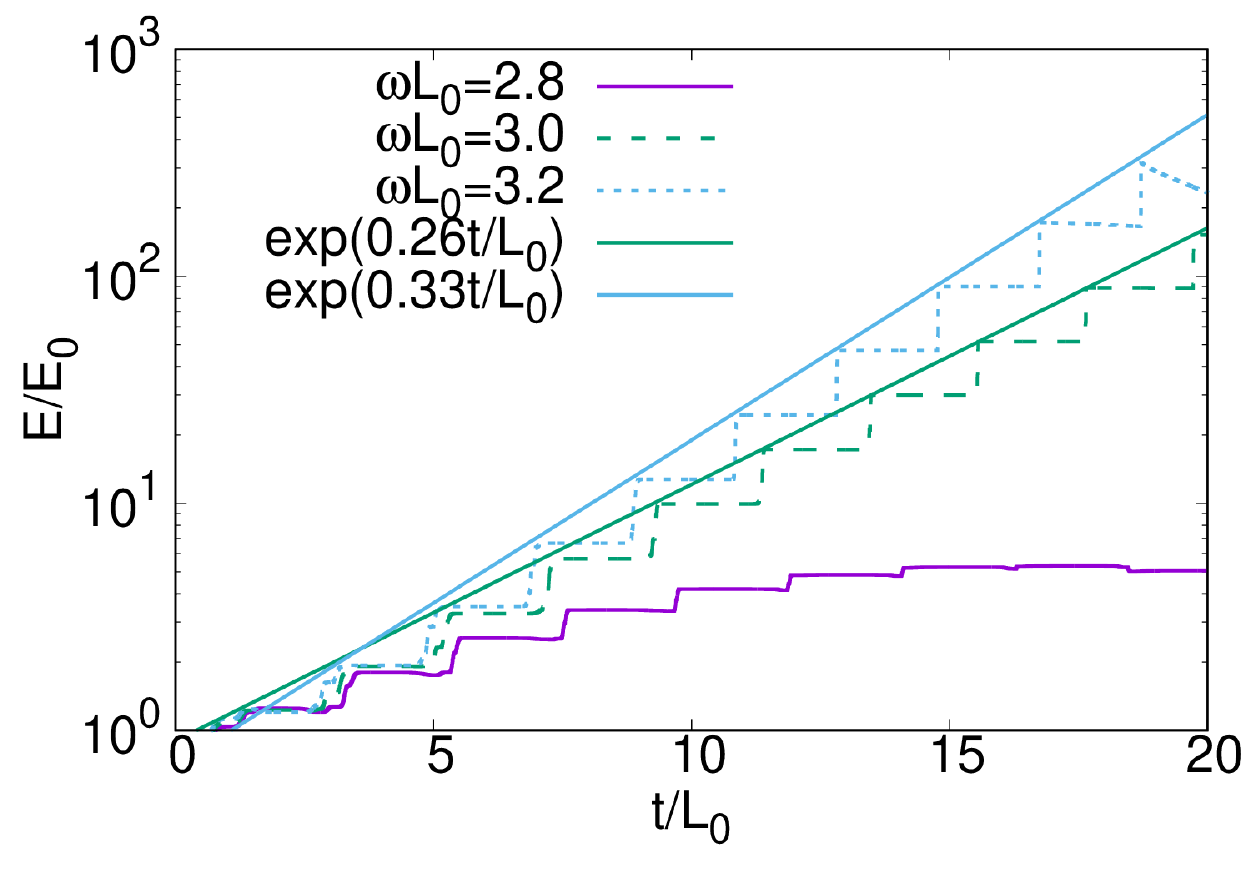}
\caption{Time evolution of the total energy inside the cavity for different boundary oscillation frequencies $\omega L_{0}$. Notice that the purple line, which does not satisfy the instability condition, remains bounded.
\label{Graph_time_evolution_of_energy_Parameter_omega_dependence_H1-3}
}
\end{figure}

\begin{figure}[th]
\includegraphics[width=0.48\textwidth]{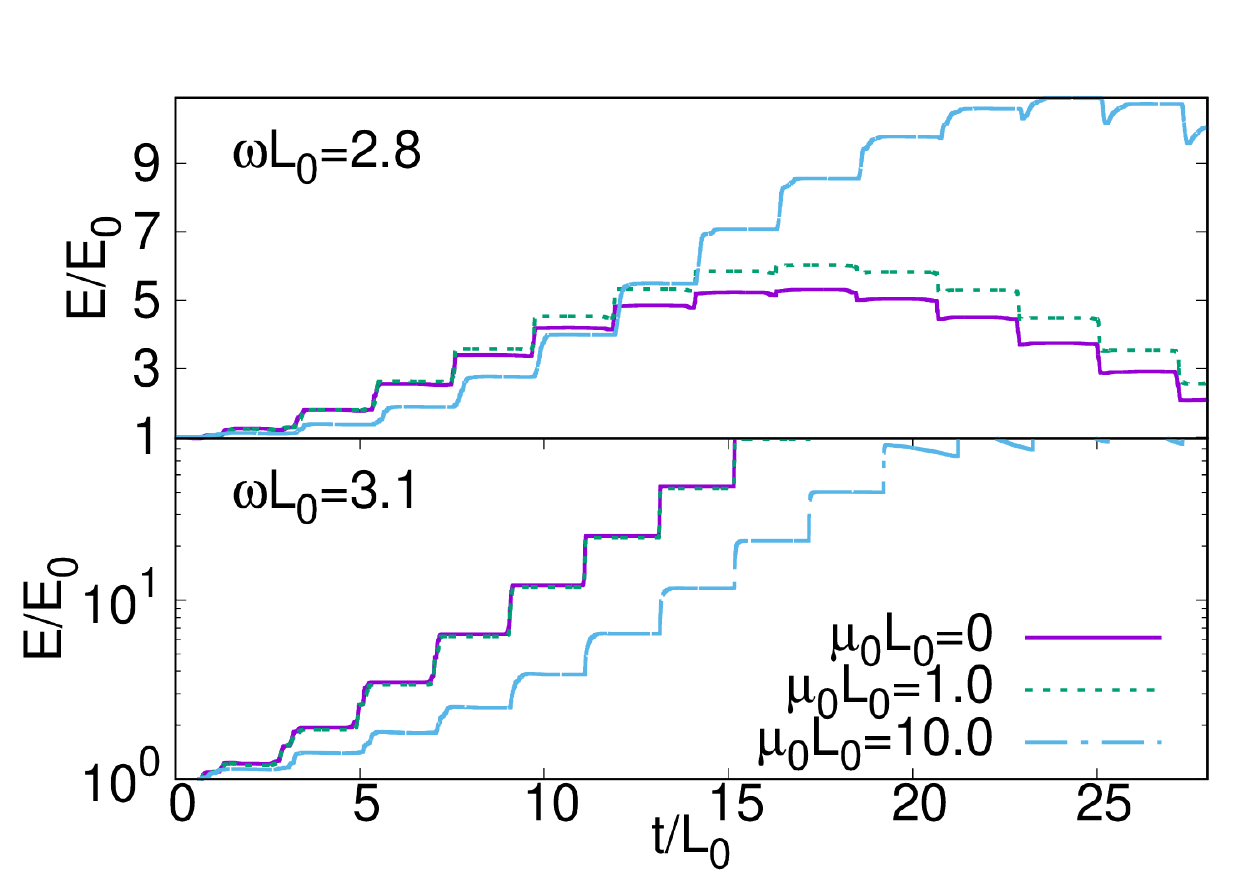}
\caption{Time evolution of the total energy for different values of $\mu L_{0}$.
  Observe that the energy remains bounded in the top panel (where the blueshift
  condition is not satisfied) whereas in the bottom panel it increases without
  bound (where the blueshift condition is satisfied).
\label{Graph_time_evolution_of_energy_Parameter_mass_dependence_H1-3v2_merge}
}
\end{figure}

\begin{figure}[th]
\includegraphics[width=0.48\textwidth]{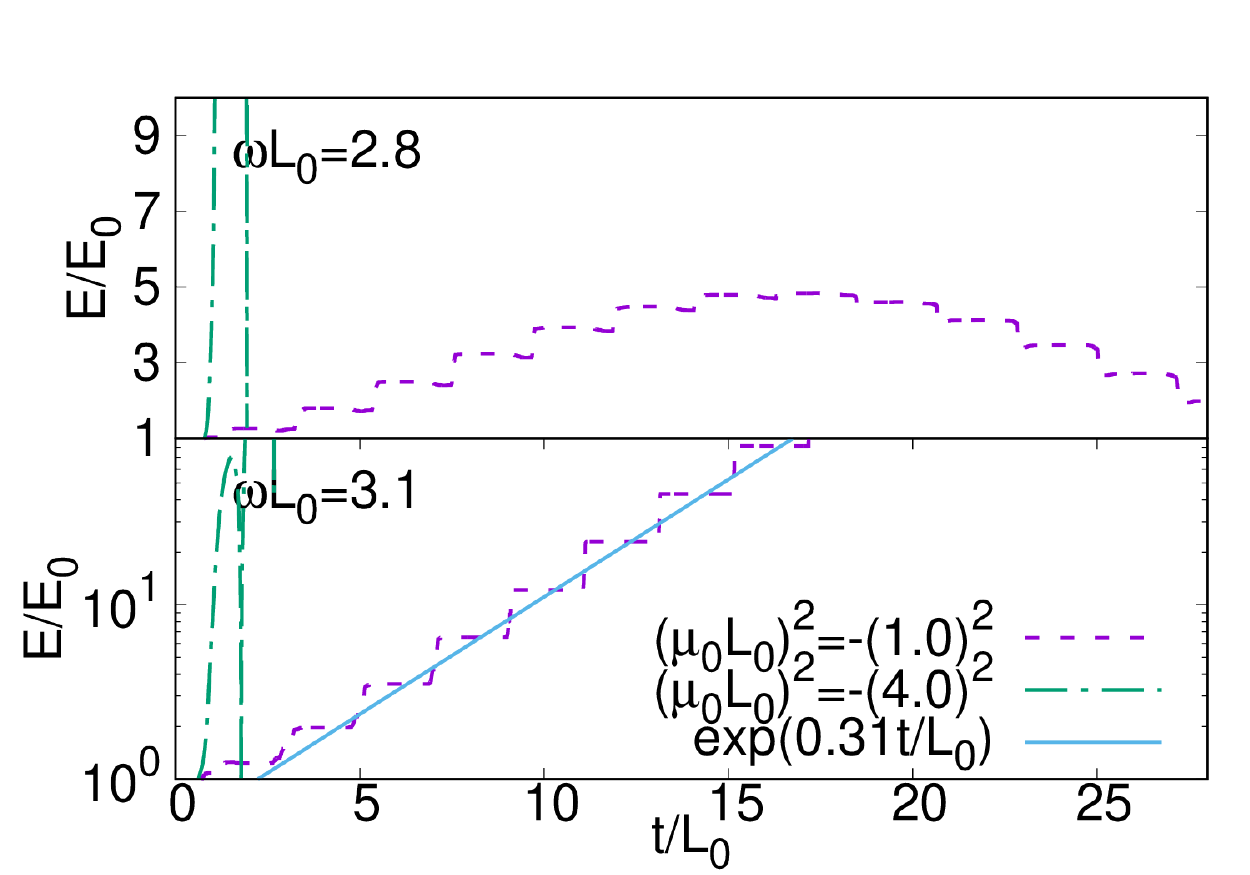}
\caption{Time evolution of the scalar field and energy, for a scalar confined to a cavity of length $L_0$.
The scalar is massive, with $\mu^2<0$, and the cavity boundary oscillates with $\omega L_{0}=2.8$ (top panel) and $\omega L_{0}=3.1$ (bottom panel).
For small $|L_0\mu|$ we recover the previous results for massless fields; in particular a blueshift instability at certain frequency ranges (bottom panel)
and stability outside this window. However, at negative large $(L_0\mu)^{2}$ a new mechanism sets in, that of scalarization governed by a tachyonic mode~\cite{Damour:1993hw,Cardoso:2020cwo}.
(Notice that the tachyonic mode -- the green line -- grows exponentially, the scale just doesn't allow to see any further high-amplitude oscillation).
\label{Graph_time_evolution_of_energy_Parameter_mass_dependence_H1-4v2_merge}
}
\end{figure}

In all our simulations we use time-symmetric initial data, where the scalar field
is parameterized by
\begin{align}
  \Phi(0,X) =e^{-\left(\frac{r-r_{0}}{\sigma}\right)^{2}}\,, \quad
  \dot{\Phi}(0,X) = \frac{\dot{L}}{L}X\partial_{X}\Phi \,,
  \label{eq:ID1}
\end{align}
where $\sigma$ and $r_{0}$ denote the width, and initial position of the scalar field pulse.
Note that the wave equation is linear, hence the amplitude can arbitrarily be set to unity.
In this work we focus on $\sigma=0.2L_0,\,r_0=0.5L_0$ and $\delta L=0.1L_0$.
Our results are summarized in Figs.~\ref{Graph_time_evolution_of_center_for_draft_ID1_Amp1_wn02_r05_bulkmass0_AmpL01_omegaL_3_1_L1_full2}-\ref{Graph_time_evolution_of_energy_Parameter_mass_dependence_H1-4v2_merge}.
We surveyed other regions in parameter space and found no qualitative new features
with respect to those discussed below.

Consider first massless fields, $\mu=0$, inside a cavity with an oscillating boundary. One typical evolution is shown in Fig.~\ref{Graph_time_evolution_of_center_for_draft_ID1_Amp1_wn02_r05_bulkmass0_AmpL01_omegaL_3_1_L1_full2}. This figure is very clear: the scalar field evolves in time with a negligible variation in amplitude, but with a significant evolution of frequency content. Note that each reflection is visible in the figure. 
The scalar field becomes ``sharper'' upon each reflection, signaling a frequency increase. The inset shows the associated increase in the total energy inside the cavity. The energy increase is exponential and the exponent is consistent with the simple estimate from the blueshift condition \eqref{eq:instability condition}.
Thus, the picture outlined in Sec.~\ref{sec Doppler effect} seems to describe evolutions of the wave equation. For these parameters, the energy content inside the cavity grows exponentially, driven by the blueshift conversion. We also verified that the growth timescale $\lambda$ does not strongly depend on the initial position $r_{0}$ of the pulse, 
consistent with the particle picture (cf. Fig.~\ref{plot_example_particle_L0_1_L1_0_1_omega_2_7}).

This interpretation is made stronger with the analysis of different parameters. Figure~\ref{Graph_time_evolution_of_energy_Parameter_omega_dependence_H1-3} shows us that the timescale is sensitive to the boundary oscillation frequency $\omega L_0$ and that at small frequencies the instability disappears. In fact, our numerical results are very consistent with the toy model of Sec.~\ref{sec Doppler effect}, including the threshold frequency prediction. For frequencies outside the instability window, our results show a clear oscillating pattern, again consistent with the previous discussion.
To summarize, there is a clear parallel between our toy model of a massless particle in Sec.~\ref{sec Doppler effect}, and the evolution of the massless wave equation. In particular, there are regions in parameter space where the energy in the scalar field inside the cavity grows exponentially with time. For other parameters, the energy oscillates in time, providing further support to the interpretation in terms of blue and redshift processes.

Thus far, we studied only massless fields.
Figure~\ref{Graph_time_evolution_of_energy_Parameter_mass_dependence_H1-3v2_merge} shows the evolution of the energy in the cavity for different field mass $\mu$ and different frequencies $\omega$.
Notice that one could expect qualitatively new features depending on whether the Compton wavelength of the field is larger or smaller than the cavity size.
Surprisingly, our results show clearly that the instability rate and instability window is only weakly
dependent on $\mu$, even for large $\mu \,L_0$.

Because we are interested also in scalarization, for which the effective mass parameter $\mu^2_{\rm eff}$ can be negative,
we studied this case as well, summarized in Fig.~\ref{Graph_time_evolution_of_energy_Parameter_mass_dependence_H1-4v2_merge}.
Now, another effect can set in, which affects even stationary geometries. Namely, scalarization induced by a tachyonic mode~\cite{Damour:1993hw,Cardoso:2020cwo}. The tachyonic instability is described by a rate $L_0\lambda_{\rm tachyon}\sim \sqrt{|\mu L_{0}|^{2}-1}$; thus at large $|L_0 \mu|$ it can dominate the entire physics.

This is indeed apparent in Fig.~\ref{Graph_time_evolution_of_energy_Parameter_mass_dependence_H1-4v2_merge}. The upper panel summarizes the evolution
outside the blueshift instability window. For small $|L_0\mu|$ our results indicate indeed just an oscillatory behavior for the total energy. At large values
of this coupling however, the tachyonic mode sets in. We observer a similar phenomena inside the blueshift instability window. At large values of $|L_0\mu|$
the tachyon mode takes over the entire process, but for small couplings we recover all the $\mu=0$ results. Our results indicate that there is not relevant change in the scalarization threshold induced by the boundary oscillation.

%%%%%%%%%%%%%%%%%%%%%%%%%%%%%%%%%%%%%%%%
\subsection{Parametric instabilities}
\label{Sec:oscillating density}
%%%%%%%%%%%%%%%%%%%%%%%%%%%%%%%%%%%%%%%%
Up to now, we studied the effects of an oscillating boundary alone. However, whenever an astrophysical object vibrates,
it also changes its local density in a periodic way. To mimic this effect in a one-dimensional setting let us consider the system
\begin{equation}
\begin{aligned}
-\partial_{t}^{2}\Phi+\partial_{x}^{2}\Phi-\mu^{2}\Phi=0\,,\\
\Phi(t,0)=\Phi(t,L_{0})=0\,,
\label{eq:KG eq 1+1 with time dependent mass}
\end{aligned}
\end{equation}
where $\mu^{2}(t)=\bar{\mu}^{2}+\delta\mu^{2}\sin(\omega t)$. Thus, $\bar{\mu}^{2}$ corresponds to the average effective mass in the system.
The oscillating piece $\delta\mu^{2}$ mimics the time-varying density. Later on, when dealing with a realistic setup, a certain condition between $\delta\mu^{2}$
and $\delta L$ will be necessary to ensure mass conservation. For now, we proceed with this idealization and without further constraints.

In Fourier space, that is $\Phi(t,x)=\sum_{k}\phi_{k}(t)e^{ikx}$, Eq.~\eqref{eq:KG eq 1+1 with time dependent mass} can be written as
\begin{align}
\partial_{\tau}^{2}\phi_{k}+\left(\delta + 2 \, \epsilon  \sin(\tau)\right)\phi_{k}=0\,,\label{eq_Mathieu}
\end{align}
where $\phi_{k}$ is the amplitude of the Fourier mode with wavelength $k$, and
\begin{equation}
\tau=\omega t\,,\qquad  \delta=\frac{k^{2}+\bar{\mu}^{2}}{\omega^2}\,,\qquad
\epsilon=\frac{\delta\mu^{2}}{2\,\omega^2}\,.
\end{equation}

Equation \eqref{eq_Mathieu} is the well-known Mathieu equation~\cite{Abramowitz:1970as,benderorszag}, and we refrain from providing further details.
For certain combination of the parameters, Mathieu equation admits unstable solutions for $\delta\sim j^2/4$ with $j\in \mathbb{Z}$~\cite{Boskovic:2018lkj,Chen:1995ena}. In particular,
for small $\epsilon$ and $j=1$,
\be
\delta=\frac{1}{4}+\epsilon \, \delta_{1}\,,\label{condition_parametric1}
\ee
one finds the instability rate $\lambda_A$ (i.e. $\Phi\sim e^{\lambda_A t}$)~\cite{Boskovic:2018lkj,Chen:1995ena},
\be
\lambda_A=\omega \epsilon\sqrt{1-\delta_{1}^{2}}\,.\label{condition_parametric2}
\ee

This instability differs, fundamentally, from the previous ``blueshift'' or Doppler-like mechanism. It is of parametric origin, and it does not lead to no migration to higher frequencies. Instead, modes in the frequency range given by Eq.~\eqref{condition_parametric1} are amplified with the rate \eqref{condition_parametric2}. Thus, this instability can only be quenched via nonlinear effects.

%%%%%%%%%%%%%%%%%%%%%%%%%%%%%%%%%%%%%%%%%%%%%%%%%%%%%%%%%%%%%
\section{Instabilities of oscillating objects in flat space}
\label{sec:oscillating 3dim}
%%%%%%%%%%%%%%%%%%%%%%%%%%%%%%%%%%%%%%%%%%%%%%%%%%%%%%%%%%%%%
% Motivated by the previous results, we now consider a more realistic setup,
% and study the dynamical behavior of a (nonminimally coupled) scalar field in a geometry describing a radially
% oscillating star. The star has total mass $M$ and a time-dependent radius $L(t)$ governed by Eq.~\eqref{eq:radius_oscillation}.

As we saw, fluctuations in some background value of the scalar are described by the Klein-Gordon equation~\eqref{KG_eff_star}
with an effective mass $\mu^2=\mu^{2}(t,r)$ that depends on the energy density of the
star. If the star oscillates, so does the effective mass of the scalar field.
We model this dependence as in Eq.~\eqref{eq_profile_star}, where $\delta\tilde{\rho}(t,r)$ is the amplitude of the density oscillation, and $\tilde{\rho}_{0}$ is the temporal and spatial average of the energy density.
Imposing conservation of the mass $M$ of the star, we find
\be
\tilde{\rho}_{0}=\frac{3}{4\pi}M L_{0}^{-3}\,,\quad \delta L(t)=-\int_{0}^{L_{0}}dr\frac{r^{2}}{L_{0}^{2}}\frac{\delta\tilde{\rho}}{\rho_{0}}\,.
\ee
%
% Here, we assumed $\delta L\ll L_{0}$ and $\delta\tilde{\rho}\ll\tilde{\rho}_{0}$. This dependency has two important aspects:
% %
% \begin{itemize}

% \item The boundary conditions are located at a periodically-varying location,
%   corresponding to the star surface. For setups where there is a continuous mass
%   profile, a radially oscillating star corresponds, nevertheless, to
%   periodically varying field profiles at the surface, possibly changing on
%   scales smaller than a wavelength. Thus, this configuration is prone to a
%   Doppler-like or blueshift instability.

% \item The local density oscillations cause a time-varying effective mass for the
%   scalar, which lends itself to parametric instabilities.
% \end{itemize}

% Note that the profile \eqref{eq_profile_star} is not the most general possible. 
% In astrophysically realistic stars, the time-dependence of the radius of the star will be quite involved and so will its energy density; for simplicity we here assume the same simple model of Eq.~\eqref{eq:radius_oscillation} of Sec.~\ref{sec:oscillating}. In fact, normal modes of oscillation of a star will cause a density profile with a 
% sinusoidal-like radial profile as well. We studied other profiles for the perturbation (in particular $\delta\tilde{\rho}\propto \sin\omega r$) and find that they lead to similar results as described below.

To summarize, we evolve equation~\eqref{KG_eff_star} with an effective mass
\begin{equation}
\mu^{2}(t,r) = 
\begin{cases}
\mu_{\rm in}^{2}+\delta\mu^{2}\sin(\omega t)   &(r<r_-)\\
\mu_{\rm smooth}^{2}(r)&(r_- <r<r_+)\\
\mu_0^{2}     &(r>r_+)\label{eq:deltamu_def2}
\end{cases}
\end{equation}
where $r_\pm = L(t)\pm\frac{1}{2}\Delta{L}$, and $\Delta L$ defines a thickness of the surface of the star, introduced to make $\mu$ a smooth function. Here, $\mu_{\rm smooth}^{2}(r)$ is a smooth fifth-order polynomial connecting the effective mass inside the star, $\mu_{\rm in}^{2}+\delta\mu^{2}\sin(\omega t)$, to its exterior bare value $\mu_0^{2}$. Note that $\mu_{\rm in}^2=\mu_{0}^{2}+\beta \tilde{\rho}_{0}$, from Eq.~\eqref{eq_profile_star}.

% Note that the physics depends sensitively on the coupling between the scalar field and the geometry (through the function $A(\Phi)$ in Eq.~\eqref{KG_effective}).
% In particular, one can distinguish three different regimes. One corresponds to an effective mass inside the star smaller than that outside, $\mu^2<\mu_0^2$.
% In this case, long wavelength modes are reflected at the surface of the star, and effectively confined. These are the modes prone to parametric or Doppler instabilities
% once the star oscillates. 
% It could also happen that $\mu^2<0$, while the field is massless, $\mu_0=0$. For large negative $\mu_0^2$, one expects scalarization instead. 
% Finally, when $\mu^2>\mu_0^2$ we expect none of the instabilities discussed so far, since the field easily leaks to the exterior and disperses to infinity.	

%%%%%%%%%%%%%%%%%%%%%%%%%%%%%%%%%%%%%%%%%%%%%%%%%%%%
\subsection{Numerical setup}
%%%%%%%%%%%%%%%%%%%%%%%%%%%%%%%%%%%%%%%%%%%%%%%%%%%%
Decomposing the scalar field in spherical harmonics, the evolution
of the $(lm)$ mode, $\Phi_{lm}(t,r)$, is governed by
\be
\left(\partial_{r}^{2}+\frac{2}{r}\partial_{r}-\frac{l(l+1)}{r^{2}}-\mu^{2}\right)\Phi_{lm}-\partial_{t}^{2}\Phi_{lm}=0\,.\label{eq:KG equation with spacetime dependent mass}
\ee
Regularity of the field at $r=0$ implies that $\Phi_{lm}\propto r^{l}$ for $l\geq 0$, and we impose radiative boundary conditions at spatial infinity.

As in Sec.~\ref{sec:oscillating-numerical}, we numerically solve
Eq.~\eqref{eq:KG equation with spacetime dependent mass} using a fourth-order
accurate Runge-Kutta scheme for the time integration where spatial derivatives
are approximated by fourth-order accurate finite difference stencils.
We monitor the energy of the scalar field inside the star, defined as
\begin{align}
E_{lm} & =\frac{1}{2}\int_{0}^{L_0}\bigg(|\dot{\Phi}_{lm}|^{2}+|\partial_{r}\Phi_{lm}|^{2} +\frac{l(l+1)}{r^{2}}|\Phi_{lm}|^{2} \notag \\
& \qquad \qquad \quad +\mu^{2}(t,r)|\Phi_{lm}|^{2} \bigg) \, r^2 dr\,,
\end{align}
(although we mostly focus on spherically symmetric modes).

We once again use time-symmetric initial data, with a profile parameterized by \eqref{eq;momentarily static Gaussian initial data},
where $\sigma$ and $r_{0}$ denoting the width and initial
position of the scalar field pulse.
Since the equation to be solved is linear we set the initial amplitude of the pulse to unity.
As we will see below, we will mostly focus on $l=0$ cases and we fix $r_0$ and $\delta L$ to 
\be
r_0=0.5L_0\,,\qquad \delta L=0.1L_0\,.
\ee

\begin{figure}[th]
\includegraphics[width=0.48\textwidth]{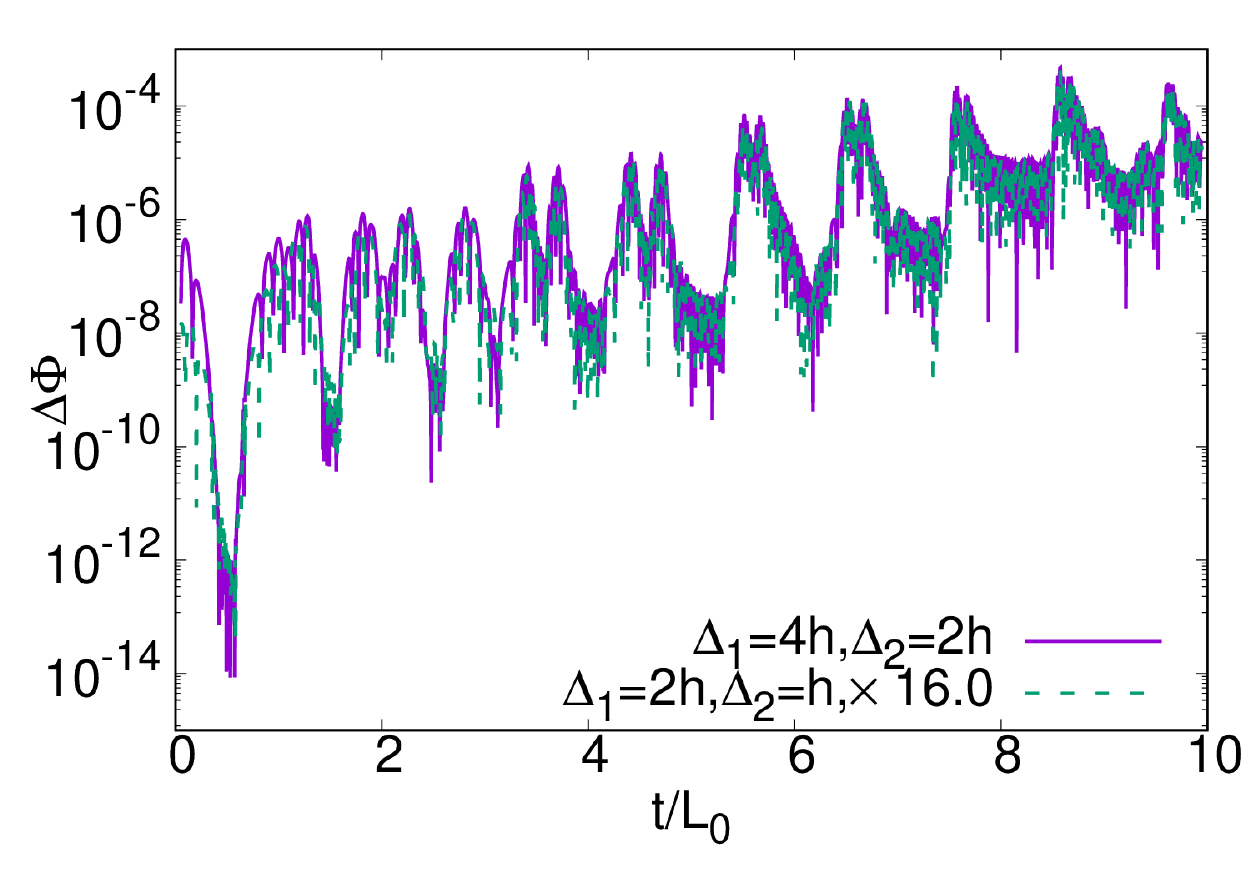}
\caption{%Numerical convergence study.
  % of the evolution of the profile summarized in Fig.~\ref{Graph_time_evolution_of_energy_inside_star_ex_ID1_Amp1_wn01_r05_win0_001_bulkmass0_AmpM01_omegaM3_1_L1_l0_outmass100}.
The purple line shows the difference between results obtained with low ($4h$) and medium ($2h$) resolutions, while the (dashed) green line shows the difference between results obtained with medium and high ($h$) resolutions multiplied by 16, the expected factor for fourth-order convergence.
\label{convergence_phi_ID1_Amp1_wn01_r05_win0_001_bulkmass0_AmpM01_omegaM3_1_L1_l0_outmass100}
}
\end{figure}
All our results are numerically convergent. An example is shown in Fig.~\ref{convergence_phi_ID1_Amp1_wn01_r05_win0_001_bulkmass0_AmpM01_omegaM3_1_L1_l0_outmass100}, showing the expected fourth-order convergence.

%%%%%%%%%%%%%%%%%%%%%%%%%%%%%%%%%%%%%%%%%%%%%%%%%%%%
\subsection{Results}
\subsubsection{Parametric instabilities}
%%%%%%%%%%%%%%%%%%%%%%%%%%%%%%%%%%%%%%%%%%%%%%%%%%%%%%

% % 
% \begin{figure}[th]
% \includegraphics[width=0.48\textwidth]{figs/Graph_time_evolution_of_scalar_energy_merge_Parameter_mass_dependence_H4_4}
% \caption{Evolution of the energy and scalar field for setups which probe both an oscillating boundary and a time-varying effective mass inside the star.
% The initial data has width $\sigma=0.1L_0$, and the star is oscillating with frequency $\omega L_0=3.1$. The effective mass inside the star is zero, $\mu_{\rm in}=0$, while
% the bare mass is $\mu_0L_0=100$.  The time-varying component of the density is proportional to $\delta \mu^{2}$. Notice that both blueshift and parametric instabilities are now observed.}
% \label{Graph_time_evolution_of_energy_Parameter_mass_dependence_H4_4}
% \end{figure}

% 
\begin{figure}[th]
\includegraphics[width=0.48\textwidth]{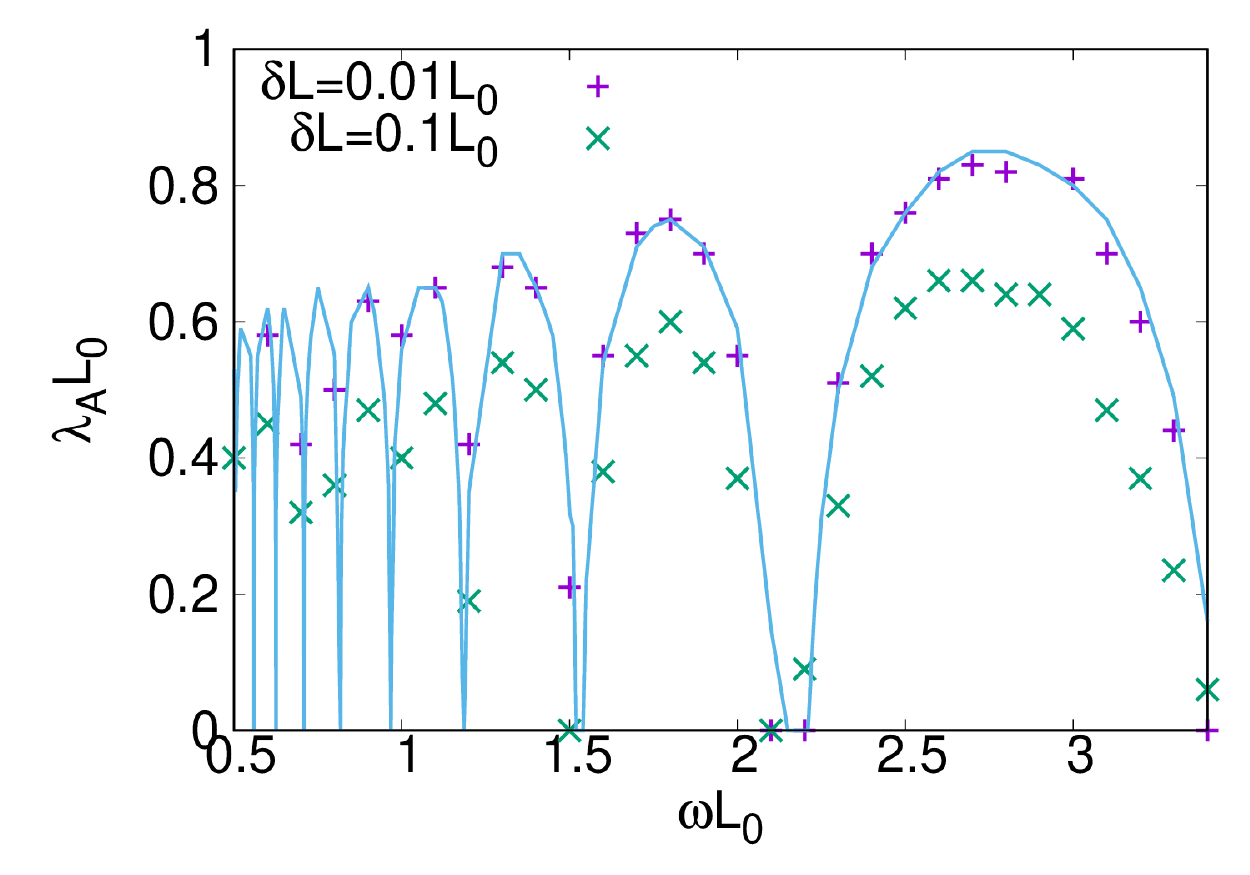}
\caption{Instability rate of oscillating star, when density is allowed to oscillate (for initial condition with width $\sigma=0.1L_0$ but results are insensitive to this choice). Now, two mechanisms compete: a Doppler like or blueshift instability and a parametric instability.
In this example, with $\mu_0L_0=100$, the latter always dominates. Crosses are data points for $\delta\mu^{2}L_{0}^{2}=16$ and different $\delta L/L_{0}$; blue line is prediction from Mathieu equation~[Eq.~\eqref{condition_parametric2}] with $k=2\pi/L_0$. 
}
\label{Relation_omega_lambda_parametric_instability}
\end{figure}

Figure~\ref{Relation_omega_lambda_parametric_instability} shows that the typical growth rate $\lambda_{A}$ of the amplification instability as a function of $\omega L_{0}$ with $\delta\mu^{2}L_{0}^{2}=16$ and different $\delta L/L_{0}$. We used the evolution of the scalar to estimate the rate $\lambda_{A}$ of the instability.
We also numerically evolved Mathieu equation~[Eq.~\eqref{eq_Mathieu}] with $k=\frac{2\pi}{L_{0}}$, $\bar{\mu}=0$ and different $\delta\mu^{2}$, and 
obtained the instability timescale, summarized in Fig.~\ref{Relation_omega_lambda_parametric_instability} (blue line).
Note that $k=\frac{2\pi}{L_{0}}$ is the longest wavelength which can be excited inside the star and the relevant mode governing the instability.
Our results show that the numerical evolution is in good agreement with the simple analysis of Sec.~\ref{Sec:oscillating density}, leading to a Mathieu equation.
For small $\delta L$, the timescale in the numerical simulation is in good agreement with the prediction from Eq.~\eqref{condition_parametric1},
and even for $\delta L=0.1L_{0}$ -- the difference between the growth rate and the ``Mathieu'' prediction is only about 10\%. 
Here, using the well-known properties~\cite{Boskovic:2018lkj,Chen:1995ena} of the Mathieu equation,
we can derive the resonance band in small $\epsilon$, defined in Eq.~\eqref{condition_parametric1}:
\be
\omega\sim \frac{4\pi}{nL_0}\,,
\ee
where $n$ is an integer. This expression and Fig.~\ref{Relation_omega_lambda_parametric_instability} indicate that there are infinite number of instability windows at low frequency. The rate for $n=1$
at small $\epsilon$ reads
\be
L_0\lambda\sim \frac{\beta L_0^2\delta\rho}{8\pi}\,,
\ee
where it is assumed that $L_0\lambda\ll 1$.
These results were derived assuming a simple constant profile for $\delta \rho$. We have studied nontrivial profiles, including sinusoidal shapes (more appropriate for the description
of normal modes) and our results are still compatible with the above description.

%%%%%%%%%%%%%%%%%%%%%%%%%%%%%%%%%%%%%%%%%%%%%%%%%%%%%%%%%%%%%%%%%
\section{Instability of oscillating stars in General Relativity\label{sec:GR}}
%%%%%%%%%%%%%%%%%%%%%%%%%%%%%%%%%%%%%%%%%%%%%%%%%%%%%%%%%%%%%%%%%

We have thus far been considering toy models for instabilities using a flat
space approximation. To study the existence of instabilities in astrophysical
objects we will now consider radial perturbations around a relativistic constant
density star. The corresponding eigenvalue and stability problem was studied previously in classical works~\cite{Chandrasekhar:1964zz,1964ApJ...140..417C,1966ApJ...145..505B,Kokkotas:2000up}.
As we will see, the amplification instability discussed in previous sections also manifests itself in this setting.
This is a highly nontrivial result: the governing equations, as well and the perturbation density profile are complicated functions, which can hardly be related to any
Mathieu-like equation. Nevertheless, our results show that such stars are prone to parametric instabilities.

%%%%%%%%%%%%%%%%%%%%%%%%%%%%%%%%%%%%%%%%%%%%%%%%% 
\subsection{Formulation}
%%%%%%%%%%%%%%%%%%%%%%%%%%%%%%%%%%%%%%%%%%%%%%%%%
\begin{figure}[thb]
  \includegraphics[width=0.48\textwidth]{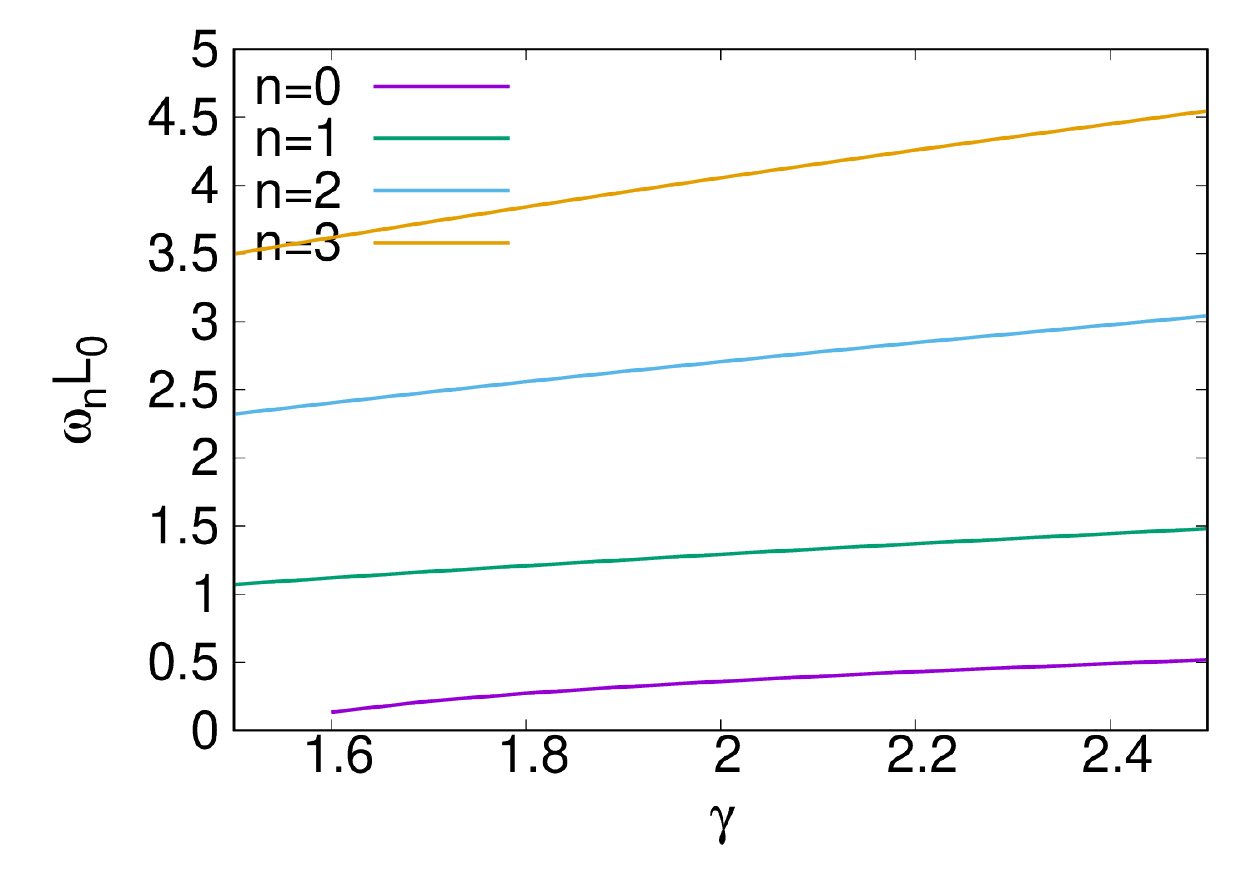}
\caption{The first four normal mode frequencies of a constant-density star with compactness $\mathcal{C}=0.3$, for different index $\gamma$.
We recover well known results in the literature~\cite{1964ApJ...140..417C,1966ApJ...145..505B,Kokkotas:2000up}.
   \label{plot normal mode frequency and gamma} }
\end{figure}
Let us consider perturbations around the metric, density, pressure, and 4-velocity profiles of
the relativistic constant density star given by Eq.~\eqref{eq:background}. Any of these dynamical quantities $X(t,r)=(\alpha,\,a,\,\rho,\,p)$,
is then written as its background, stationary value corresponding to Eq.~\eqref{eq:background}, plus some small linearized fluctuation, 
\be
X(t,r)=X_0(r)+\epsilon \delta X(t,r)\,,
\ee
where $\epsilon$ is a bookkeeping parameter for the perturbation. We also need to consider fluid displacements and we write the radial component of the four-velocity
$u_{r}(t,r)=\epsilon\frac{\tilde{a}_{0}^{2}}{\tilde{\alpha}_{0}}\partial_{t}\xi(t,r)$, with $\xi(t,r)$ the Lagrangian displacement.
From the equation of motion for fluid one obtains the following equation for $\xi$
\beq
&-&\frac{\tilde{a}_{0}^{2}}{\tilde{\alpha}_{0}^{2}}\left(
\tilde{\rho}_{0}+\tilde{p}_{0}\right)\ddot{\xi}=\frac{4}{r}\tilde{p}_{0}'\xi
-\frac{1}{\tilde{a}_{0}\tilde{\alpha}_{0}^{2}}\left(
\tilde{a}_{0}\tilde{\alpha}_{0}^{3}\frac{\gamma \, \tilde{p}_{0}}{r^{2}}\left(
\frac{r^{2}}{\tilde{\alpha}_{0}}\xi\right)'\right)'\nonumber\\
&+&8\pi \tilde{a}_{0}^{2}\tilde{p}_{0}(\tilde{\rho}_{0}+\tilde{p}_{0})\xi-\frac{1}{\tilde{\rho}_{0}+\tilde{p}_{0}}(\tilde{p}_{0}')^{2}\xi\,,
\eeq
where primes stand for radial derivatives and $\gamma$ is the adiabatic index defined as
\begin{align}
\gamma=\frac{\tilde{\rho}+\tilde{p}}{\tilde{p}}\left(
\frac{d\tilde{p}}{d\tilde{\rho}}
\right)\,.
\end{align}

For a constant density star, %the adiabatic index
$\gamma$ is infinite (incompressible fluid);
however, we will assume that it is of (constant) finite value to perform the perturbation~\cite{Chandrasekhar:1964zz,Camilo:2018goy}.
% The stability analysis based on this assumption gives us enough implication for stability of astrophysical relativistic star.
The density, pressure, and metric perturbation can be obtained from the Lagrangian displacement
%\begin{widetext}
%
\begin{subequations}
\begin{align}
\delta\rho&=\frac{\tilde{p}_{0}+\tilde{\rho}_{0}}{\gamma}\left(
4\pi r\tilde{a}_{0}^{2}\tilde{p}_{0}(1+\gamma)+\frac{\tilde{\rho}_{0}}{2r \tilde{p}_{0}}(a^{2}-1)-\gamma \xi'\right)\nonumber\\
&\quad{}+\frac{\tilde{p}_{0}+\tilde{\rho}_{0}}{2r\gamma}
\left(-1-5\gamma+\tilde{a}_{0}^{2}\left(
1+\gamma+8\pi r^{2}\tilde{\rho}_{0}
\right)\right)\xi\,,\\
\delta p&= \frac{\tilde{p}_{0}}{\tilde{\rho}_{0}+\tilde{p}_{0}}\gamma\delta\rho\,,\\
\delta a&=-4\pi r\tilde{a}_{0}(\tilde{p}_{0}+\tilde{\rho}_{0})\xi\,,\\
\delta\alpha'&=-4\pi \tilde{a}_{0}^{4}(\tilde{\rho}_{0}+\tilde{p}_{0})(1+8\pi r^{2}\tilde{p}_{0})\xi\nonumber\\
&\quad{}+\left(\frac{-1+\tilde{a}_{0}^{2}}{r}
+8\pi r\tilde{a}_{0}^{2}\tilde{p}_{0}
-\frac{\tilde{\alpha}_{0}'}{\tilde{\alpha}_{0}}\right)\delta\alpha\,.
\end{align}
\end{subequations}
We focus on the normal mode of the perturbation and assume that $\xi(t,r)=e^{i\omega t}\zeta_{\omega}(r)=e^{i\omega t}\frac{\tilde{\alpha}_{0}(r)}{r^{2}}\zeta_{\omega}$.
The equation for $\zeta_{\omega}$ is
\begin{align*}
\frac{\zeta_{\omega}''}{\zeta_{\omega}}&=
-\frac{\tilde{a}_{0}^{2}}{\gamma\tilde{\alpha}_{0}^{2}}\left(
1+\frac{\tilde{\rho}_{0}}{\tilde{p}_{0}}
\right)\omega^{2}
+8\pi\frac{\tilde{a}_{0}^{2}}{\gamma}(\tilde{p}_{0}+\tilde{\rho}_{0})
+\frac{4\tilde{p}_{0}'}{r\gamma \tilde{p}_{0}}\\
&\quad{}-\frac{\tilde{p}_{0}'^{2}}{\gamma \tilde{p}_{0}(\tilde{\rho}_{0}+\tilde{p}_{0})}
+\left(
\frac{2}{r}
-3\frac{\tilde{\alpha}_{0}'}{\tilde{\alpha}_{0}}
-\frac{\tilde{a}_{0}'}{\tilde{a}_{0}}
-\frac{\tilde{p}_{0}'}{\tilde{p}_{0}}
-\frac{\gamma'}{\gamma}
\right)\frac{\zeta_{\omega}'}{\zeta_{\omega}}\,.
\end{align*}
% \end{widetext}
%
The boundary condition for $\xi_{\omega}(r)$ is
\be
\xi_{\omega}(0)=0\,,\qquad \delta p(L_0)\propto \xi_{\omega}'(L_0)=0\,,
\ee
which corresponds to the following boundary condition for $\zeta_{\omega}(r)$
\begin{equation}
\begin{aligned}
\zeta_{\omega}(0)&=0 \,,\\
\zeta'_{\omega}(L_0)&=-\left(
\frac{\tilde{\alpha}_{0}'(L_0)}{\tilde{\alpha}_{0}(L_0)}
-\frac{1}{L_0}
\right)\zeta_{\omega}(L_0) \,.
\end{aligned}
\end{equation}
We solved the eigenvalue problem for $\omega$ with the above boundary conditions
for different adiabatic index $\gamma$ with a shooting method. Our results give a relation between eigenfrequency $\omega$
and index $\gamma$ for different overtones, labeled by $n$. They are summarized in Fig.~\ref{plot normal mode frequency and gamma}, and are consistent
with classical results in the literature~\cite{1964ApJ...140..417C,1966ApJ...145..505B,Kokkotas:2000up}.
For the simulation of the scalar field, we introduce a 5th
order polynomial smooth transition function $W(r;R_{1},R_{2})$ satisfying
\begin{align*}
W(r;R_{1},R_{2})&=1~~(r\leq R_{1}),\\
W(r;R_{1},R_{2})&=0~~(r\geq R_{2}),\\
W'(R_{1};R_{1},R_{2})&=W'(R_{2};R_{1},R_{2})
                    =W''(R_{1};R_{1},R_{2})\\
  &=W''(R_{2};R_{1},R_{2})=0\,,
\end{align*}
and replace the Lagrangian displacement with the smoothed profile as follows
\begin{align*}
\xi_{\omega}(r)&\to \xi_{\omega}(r)W(r;R_{1},R_{2})\,.
\end{align*}
With this we construct density, pressure, and metric profile from
\begin{align*}
Q(t,r)&=Q_{0}(r+\xi(t,r))+\delta Q(t,r)W(r;R_{1},R_{2}) \\
Q_{0}(r)&=Q_{\rm in}(t,r)W(r;R_{1},R_{2})+Q_{\rm out}(1-W(r;R_{1},R_{2}))
\end{align*}
where $Q(t,r)$ denotes the density, pressure, and the metric, and $Q_{0}(r)$ denotes the background profile for each variable.

% Using constructed metric, density, and pressure,
We are then able to evolve the scalar field $\phi=\frac{\psi}{r}$ through the evolution equation
\begin{widetext}
\begin{align}
\partial_{t}^{2}\phi=\partial_{t}\ln\left|\frac{\alpha}{a}\right|\partial_{t}\phi
+\frac{\alpha^{2}}{a^{2}}\partial_{r}\ln\left|
\frac{\alpha}{a}
\right|\partial_{r}\phi
+\frac{\alpha^{2}}{a^{2}}\left(\frac{2}{r}\partial_{r}\phi
+\partial_{r}^{2}\phi
\right)
-\frac{l(l+1)}{r^{2}}\alpha^{2}\phi
+\beta\tilde{T}\alpha^{2}\phi
-\mu^{2}_0\alpha^{2}\phi \,.
\end{align}
\end{widetext}
For the simulations reported below, we use a momentarily static Gaussian pulse as
initial data parameterized by
\begin{align*}
\phi(0,r)=Ae^{-\left(
\frac{r-r_{0}}{w}
\right)^{2}}\,,\qquad
\dot{\phi}(0,r)=0\,,
\end{align*}
where $A,w,r_{0}$ are the initial amplitude, width, and position of the pulse, respectively. We will always set $r_{0} = 0.5L_0$.

During the evolution we monitor the scalar field at fixed radius, and the
energy of the scalar field inside star, given by
\begin{align}
E=\frac{1}{2}\int_{0}^{L_0} dr\alpha ar^{2}\Big(&\frac{1}{\alpha^{2}}\dot{\phi}^{2}+\frac{1}{a^{2}}\phi'^{2}
+\frac{l(l+1)}{r^{2}}\phi^{2} \nonumber\\
&{}+(\mu_0^{2}-\beta \tilde{T})\phi^{2} \Big)
\end{align}

%%%%%%%%%%%%%%%%%%%%%%%%%%%%%%%%%%%%%%%%%%%%%%%%% 
\subsection{Results}
%%%%%%%%%%%%%%%%%%%%%%%%%%%%%%%%%%%%%%%%%%%%%%%%%
\begin{figure}[thpb]
  \centering
  \includegraphics[width=0.48\textwidth]{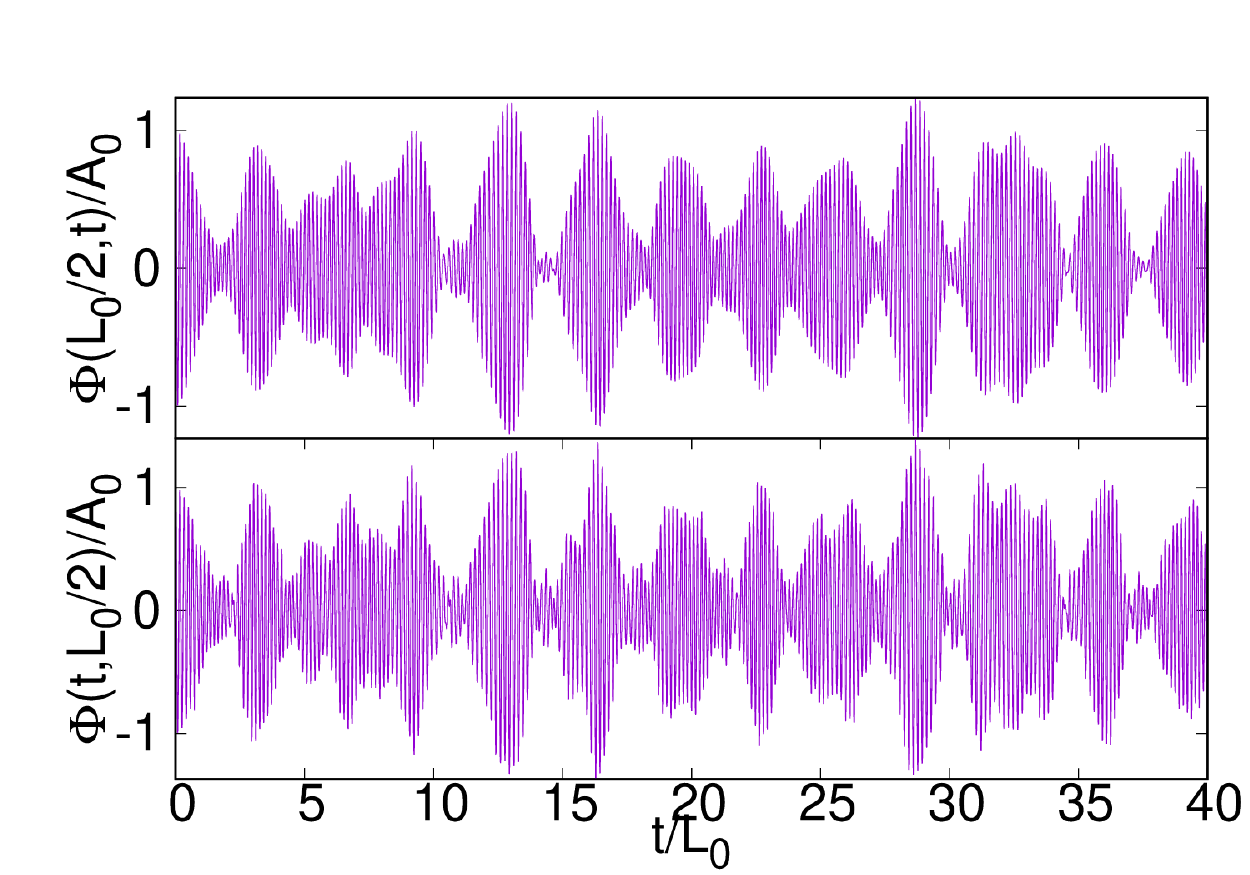}
\caption{
Time evolution with $\mu_0 L_0=100$, $\beta=-280000$, $\delta\rho_{c}\simeq 0.08\tilde{\rho}_{0}$, and different initial width of the pulse.
The initial width $w/L_0$ is $0.2$ (upper panel), and $0.4$ (lower panel).
The background star is perturbed with the fundamental mode ($n=0$).
\label{fig GR no instability}}
\end{figure}
We now take an oscillating compact star. The star oscillations could be caused by accretion
or because it was the product of a merger. The amplitude of such oscillations can in any case be significant~\cite{1994ApJ...426..688R,Chirenti:2016xys,Ma:2020rak,Perego:2019adq,Bernuzzi:2020txg}.
In our simulations we consider a single mode of oscillation, characterized by the number of nodes $n$. The generalization to
accommodate for different modes is straightforward. The control parameters for
the background spacetime are the compactness of the star $\mathcal{C}$, the
adiabatic index $\gamma$, the central value of the density perturbation
$\delta\rho_{c}$, and $n$. Here, the compactness $\mathcal{C}$ and adiabatic
index $\gamma$ are fixed to $0.3$ and $2.0$, which are representative of neutron
stars~\cite{Lattimer:2000nx}. The most drastic astrophysical situations exciting large fluctuations of neutron stars concerns the remnant phase just after a neutron star
merger. According to numerical relativity simulations, the amplitude
of the density perturbations can be significant and of the order of the central density itself~\cite{Perego:2019adq,Bernuzzi:2020txg}. Although such large density perturbations lie beyond the
validity range of the linear approximation for the perturbation, we consider
${\rm max}(\delta\rho(t,0))\simeq 0.28\tilde{\rho}_{0},0.08\tilde{\rho}_{0}$, and use the linear normal
mode derived in previous subsection.

Figure~\ref{fig GR no instability} shows the time evolution of the scalar field at $r=0.5L_0$ with $\mu_0 L_0=100$, $\beta=-280000$, and $\delta\rho_{c}\simeq 0.08\tilde{\rho}(0)$, and the fundamental density perturbation mode $n=0$. Note how the scalar has a time-periodic behavior, but is not growing neither in amplitude nor in frequency.
With these parameters we could not observe any instabilities.

\begin{figure}[thpb]
\includegraphics[width=0.48\textwidth]{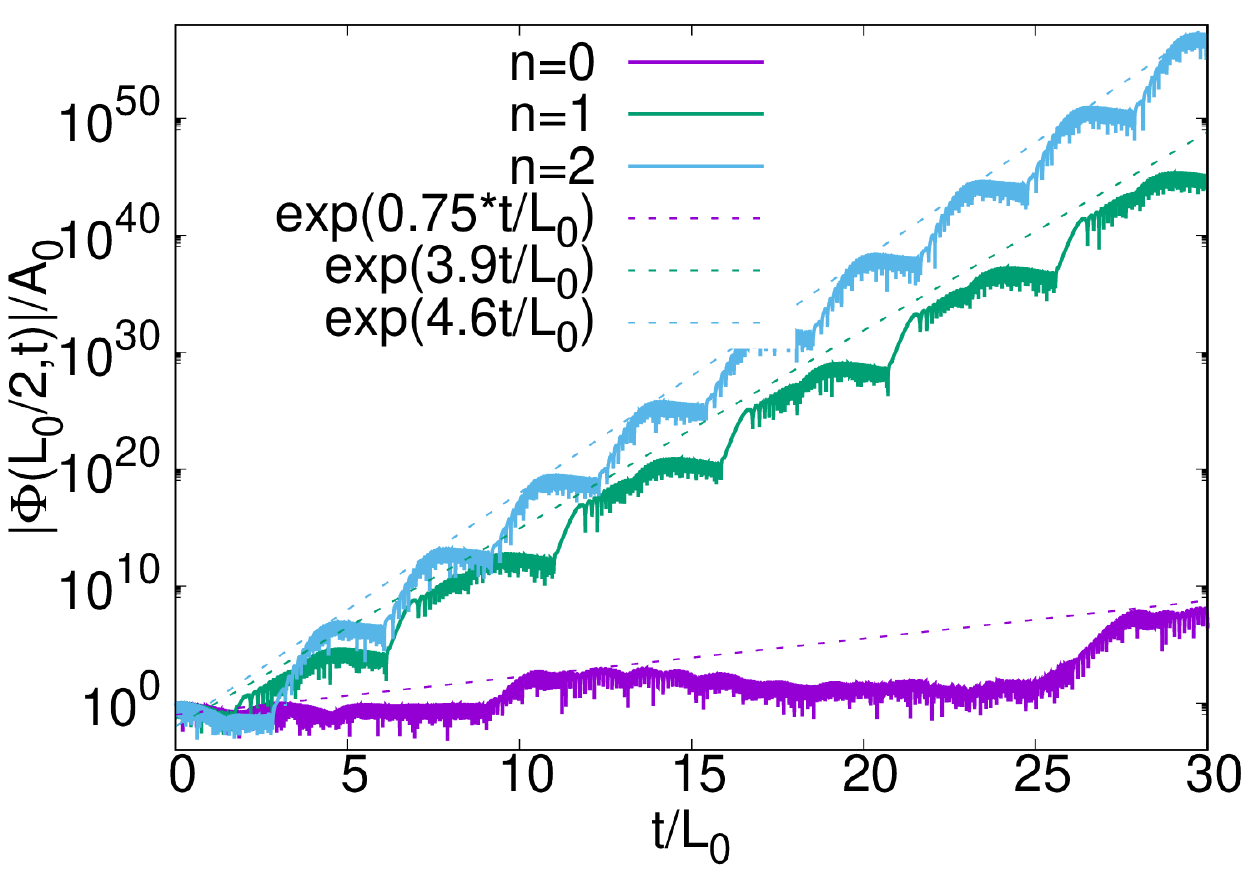}
\caption{
Time evolution with $\mu_0 L_0=100$, $\beta=-280000$, $\delta\rho_{c}\simeq 0.28\tilde{\rho}_{0}$, $w/L_0=0.2$. We plot radial modes of density perturbation with $n=0,1,2$.
Notice that the density fluctuation is large, but still smaller than typical values in seen during the coalescence of two neutron stars~\cite{Perego:2019adq,Bernuzzi:2020txg}. Also, for smaller $\beta$, the instability sets in at small density fluctuations.
\label{fig GR instability}
}
\end{figure}
However, an instability sets in with either larger values of density fluctuation or smaller values of $\beta$ (i.e., larger $\beta^2$).
We note that for massive fields -- our subject here -- observational constraints on $\beta$ are poor or non-existing and any of the values discussed here is compatible with current observations.
For example, Fig.~\ref{fig GR instability} shows the time evolution of the scalar field at $r=0.5L_0$, but now with $\delta\rho_{c}\simeq 0.28\tilde{\rho}_{0}$.
We can compare and contrast the results with those of Fig.~\ref{fig GR no instability}.
For these larger values of density fluctuations, a parametric instability sets in. The frequency of the
fundamental density perturbation mode with $\gamma=2.0$ is about $0.36/L_0$ and
the corresponding period is about $17L_0$. This period is clearly evident in Fig.~\ref{fig GR instability}, in the form of a plateau
in the evolution of the scalar. In other words, the growth proceeds upon interaction with the star with a periodicity
dictated by the star eigenfrequency. As is clear from the evolution equations or the background structure of the star itself,
it is far from trivial that such oscillating stars in General Relativity would trigger parametric instabilities.
Our results confirm it. We note that Fig.~\ref{fig GR instability} refers to a significant density fluctuation. However, our results are mostly sensitive to the combination
$\beta \tilde{\rho}_0$, thus much milder fluctuations also result in instability, provided $\beta^2$ is sufficiently large.

\bibliography{References}

\begin{thebibliography}{77}%
\makeatletter
\providecommand \@ifxundefined [1]{%
 \@ifx{#1\undefined}
}%
\providecommand \@ifnum [1]{%
 \ifnum #1\expandafter \@firstoftwo
 \else \expandafter \@secondoftwo
 \fi
}%
\providecommand \@ifx [1]{%
 \ifx #1\expandafter \@firstoftwo
 \else \expandafter \@secondoftwo
 \fi
}%
\providecommand \natexlab [1]{#1}%
\providecommand \enquote  [1]{``#1''}%
\providecommand \bibnamefont  [1]{#1}%
\providecommand \bibfnamefont [1]{#1}%
\providecommand \citenamefont [1]{#1}%
\providecommand \href@noop [0]{\@secondoftwo}%
\providecommand \href [0]{\begingroup \@sanitize@url \@href}%
\providecommand \@href[1]{\@@startlink{#1}\@@href}%
\providecommand \@@href[1]{\endgroup#1\@@endlink}%
\providecommand \@sanitize@url [0]{\catcode `\\12\catcode `\$12\catcode
  `\&12\catcode `\#12\catcode `\^12\catcode `\_12\catcode `\%12\relax}%
\providecommand \@@startlink[1]{}%
\providecommand \@@endlink[0]{}%
\providecommand \url  [0]{\begingroup\@sanitize@url \@url }%
\providecommand \@url [1]{\endgroup\@href {#1}{\urlprefix }}%
\providecommand \urlprefix  [0]{URL }%
\providecommand \Eprint [0]{\href }%
\providecommand \doibase [0]{http://dx.doi.org/}%
\providecommand \selectlanguage [0]{\@gobble}%
\providecommand \bibinfo  [0]{\@secondoftwo}%
\providecommand \bibfield  [0]{\@secondoftwo}%
\providecommand \translation [1]{[#1]}%
\providecommand \BibitemOpen [0]{}%
\providecommand \bibitemStop [0]{}%
\providecommand \bibitemNoStop [0]{.\EOS\space}%
\providecommand \EOS [0]{\spacefactor3000\relax}%
\providecommand \BibitemShut  [1]{\csname bibitem#1\endcsname}%
\let\auto@bib@innerbib\@empty
%</preamble>
\bibitem [{\citenamefont {Will}(2014)}]{Will:2014kxa}%
  \BibitemOpen
  \bibfield  {author} {\bibinfo {author} {\bibfnamefont {Clifford~M.}\
  \bibnamefont {Will}},\ }\bibfield  {title} {\enquote {\bibinfo {title} {{The
  Confrontation between General Relativity and Experiment}},}\ }\href {\doibase
  10.12942/lrr-2014-4} {\bibfield  {journal} {\bibinfo  {journal} {Living Rev.
  Rel.}\ }\textbf {\bibinfo {volume} {17}},\ \bibinfo {pages} {4} (\bibinfo
  {year} {2014})},\ \Eprint {http://arxiv.org/abs/1403.7377} {arXiv:1403.7377
  [gr-qc]} \BibitemShut {NoStop}%
%%CITATION = ARXIV:1403.7377;%%
\bibitem [{\citenamefont {Abbott}\ \emph
  {et~al.}(2016{\natexlab{a}})\citenamefont {Abbott} \emph
  {et~al.}}]{TheLIGOScientific:2016agk}%
  \BibitemOpen
  \bibfield  {author} {\bibinfo {author} {\bibfnamefont {B.P.}\ \bibnamefont
  {Abbott}} \emph {et~al.} (\bibinfo {collaboration} {LIGO Scientific,
  Virgo}),\ }\bibfield  {title} {\enquote {\bibinfo {title} {{GW150914: The
  Advanced LIGO Detectors in the Era of First Discoveries}},}\ }\href {\doibase
  10.1103/PhysRevLett.116.131103} {\bibfield  {journal} {\bibinfo  {journal}
  {Phys. Rev. Lett.}\ }\textbf {\bibinfo {volume} {116}},\ \bibinfo {pages}
  {131103} (\bibinfo {year} {2016}{\natexlab{a}})},\ \Eprint
  {http://arxiv.org/abs/1602.03838} {arXiv:1602.03838 [gr-qc]} \BibitemShut
  {NoStop}%
\bibitem [{\citenamefont {Abbott}\ \emph
  {et~al.}(2016{\natexlab{b}})\citenamefont {Abbott} \emph
  {et~al.}}]{TheLIGOScientific:2016src}%
  \BibitemOpen
  \bibfield  {author} {\bibinfo {author} {\bibfnamefont {B.~P.}\ \bibnamefont
  {Abbott}} \emph {et~al.} (\bibinfo {collaboration} {LIGO Scientific,
  Virgo}),\ }\bibfield  {title} {\enquote {\bibinfo {title} {{Tests of general
  relativity with GW150914}},}\ }\href {\doibase
  10.1103/PhysRevLett.116.221101, 10.1103/PhysRevLett.121.129902} {\bibfield
  {journal} {\bibinfo  {journal} {Phys. Rev. Lett.}\ }\textbf {\bibinfo
  {volume} {116}},\ \bibinfo {pages} {221101} (\bibinfo {year}
  {2016}{\natexlab{b}})},\ \bibinfo {note} {[Erratum: Phys. Rev.
  Lett.121,no.12,129902(2018)]},\ \Eprint {http://arxiv.org/abs/1602.03841}
  {arXiv:1602.03841 [gr-qc]} \BibitemShut {NoStop}%
%%CITATION = ARXIV:1602.03841;%%
\bibitem [{\citenamefont {Abbott}\ \emph {et~al.}(2020)\citenamefont {Abbott}
  \emph {et~al.}}]{Abbott:2020jks}%
  \BibitemOpen
  \bibfield  {author} {\bibinfo {author} {\bibfnamefont {R.}~\bibnamefont
  {Abbott}} \emph {et~al.} (\bibinfo {collaboration} {LIGO Scientific,
  Virgo}),\ }\href@noop {} {\enquote {\bibinfo {title} {{Tests of General
  Relativity with Binary Black Holes from the second LIGO-Virgo
  Gravitational-Wave Transient Catalog}},}\ } (\bibinfo {year} {2020}),\
  \Eprint {http://arxiv.org/abs/2010.14529} {arXiv:2010.14529 [gr-qc]}
  \BibitemShut {NoStop}%
\bibitem [{\citenamefont {Clifton}\ \emph {et~al.}(2012)\citenamefont
  {Clifton}, \citenamefont {Ferreira}, \citenamefont {Padilla},\ and\
  \citenamefont {Skordis}}]{Clifton:2011jh}%
  \BibitemOpen
  \bibfield  {author} {\bibinfo {author} {\bibfnamefont {Timothy}\ \bibnamefont
  {Clifton}}, \bibinfo {author} {\bibfnamefont {Pedro~G.}\ \bibnamefont
  {Ferreira}}, \bibinfo {author} {\bibfnamefont {Antonio}\ \bibnamefont
  {Padilla}}, \ and\ \bibinfo {author} {\bibfnamefont {Constantinos}\
  \bibnamefont {Skordis}},\ }\bibfield  {title} {\enquote {\bibinfo {title}
  {{Modified Gravity and Cosmology}},}\ }\href {\doibase
  10.1016/j.physrep.2012.01.001} {\bibfield  {journal} {\bibinfo  {journal}
  {Phys. Rept.}\ }\textbf {\bibinfo {volume} {513}},\ \bibinfo {pages} {1--189}
  (\bibinfo {year} {2012})},\ \Eprint {http://arxiv.org/abs/1106.2476}
  {arXiv:1106.2476 [astro-ph.CO]} \BibitemShut {NoStop}%
\bibitem [{\citenamefont {Berti}\ \emph {et~al.}(2015)\citenamefont {Berti}
  \emph {et~al.}}]{Berti:2015itd}%
  \BibitemOpen
  \bibfield  {author} {\bibinfo {author} {\bibfnamefont {Emanuele}\
  \bibnamefont {Berti}} \emph {et~al.},\ }\bibfield  {title} {\enquote
  {\bibinfo {title} {{Testing General Relativity with Present and Future
  Astrophysical Observations}},}\ }\href {\doibase
  10.1088/0264-9381/32/24/243001} {\bibfield  {journal} {\bibinfo  {journal}
  {Class. Quant. Grav.}\ }\textbf {\bibinfo {volume} {32}},\ \bibinfo {pages}
  {243001} (\bibinfo {year} {2015})},\ \Eprint
  {http://arxiv.org/abs/1501.07274} {arXiv:1501.07274 [gr-qc]} \BibitemShut
  {NoStop}%
%%CITATION = ARXIV:1501.07274;%%
\bibitem [{\citenamefont {Cardoso}\ and\ \citenamefont
  {Pani}(2019)}]{Cardoso:2019rvt}%
  \BibitemOpen
  \bibfield  {author} {\bibinfo {author} {\bibfnamefont {Vitor}\ \bibnamefont
  {Cardoso}}\ and\ \bibinfo {author} {\bibfnamefont {Paolo}\ \bibnamefont
  {Pani}},\ }\bibfield  {title} {\enquote {\bibinfo {title} {{Testing the
  nature of dark compact objects: a status report}},}\ }\href {\doibase
  10.1007/s41114-019-0020-4} {\bibfield  {journal} {\bibinfo  {journal} {Living
  Rev. Rel.}\ }\textbf {\bibinfo {volume} {22}},\ \bibinfo {pages} {4}
  (\bibinfo {year} {2019})},\ \Eprint {http://arxiv.org/abs/1904.05363}
  {arXiv:1904.05363 [gr-qc]} \BibitemShut {NoStop}%
\bibitem [{\citenamefont {Arvanitaki}\ \emph {et~al.}(2010)\citenamefont
  {Arvanitaki}, \citenamefont {Dimopoulos}, \citenamefont {Dubovsky},
  \citenamefont {Kaloper},\ and\ \citenamefont
  {March-Russell}}]{Arvanitaki:2009fg}%
  \BibitemOpen
  \bibfield  {author} {\bibinfo {author} {\bibfnamefont {Asimina}\ \bibnamefont
  {Arvanitaki}}, \bibinfo {author} {\bibfnamefont {Savas}\ \bibnamefont
  {Dimopoulos}}, \bibinfo {author} {\bibfnamefont {Sergei}\ \bibnamefont
  {Dubovsky}}, \bibinfo {author} {\bibfnamefont {Nemanja}\ \bibnamefont
  {Kaloper}}, \ and\ \bibinfo {author} {\bibfnamefont {John}\ \bibnamefont
  {March-Russell}},\ }\bibfield  {title} {\enquote {\bibinfo {title} {{String
  Axiverse}},}\ }\href {\doibase 10.1103/PhysRevD.81.123530} {\bibfield
  {journal} {\bibinfo  {journal} {Phys. Rev. D}\ }\textbf {\bibinfo {volume}
  {81}},\ \bibinfo {pages} {123530} (\bibinfo {year} {2010})},\ \Eprint
  {http://arxiv.org/abs/0905.4720} {arXiv:0905.4720 [hep-th]} \BibitemShut
  {NoStop}%
\bibitem [{\citenamefont {Barack}\ \emph {et~al.}(2019)\citenamefont {Barack}
  \emph {et~al.}}]{Barack:2018yly}%
  \BibitemOpen
  \bibfield  {author} {\bibinfo {author} {\bibfnamefont {Leor}\ \bibnamefont
  {Barack}} \emph {et~al.},\ }\bibfield  {title} {\enquote {\bibinfo {title}
  {{Black holes, gravitational waves and fundamental physics: a roadmap}},}\
  }\href {\doibase 10.1088/1361-6382/ab0587} {\bibfield  {journal} {\bibinfo
  {journal} {Class. Quant. Grav.}\ }\textbf {\bibinfo {volume} {36}},\ \bibinfo
  {pages} {143001} (\bibinfo {year} {2019})},\ \Eprint
  {http://arxiv.org/abs/1806.05195} {arXiv:1806.05195 [gr-qc]} \BibitemShut
  {NoStop}%
\bibitem [{\citenamefont {Kokkotas}\ and\ \citenamefont
  {Ruoff}(2001)}]{Kokkotas:2000up}%
  \BibitemOpen
  \bibfield  {author} {\bibinfo {author} {\bibfnamefont {Kostas.~D.}\
  \bibnamefont {Kokkotas}}\ and\ \bibinfo {author} {\bibfnamefont {Johannes}\
  \bibnamefont {Ruoff}},\ }\bibfield  {title} {\enquote {\bibinfo {title}
  {{Radial oscillations of relativistic stars}},}\ }\href {\doibase
  10.1051/0004-6361:20000216} {\bibfield  {journal} {\bibinfo  {journal}
  {Astron. Astrophys.}\ }\textbf {\bibinfo {volume} {366}},\ \bibinfo {pages}
  {565} (\bibinfo {year} {2001})},\ \Eprint
  {http://arxiv.org/abs/gr-qc/0011093} {arXiv:gr-qc/0011093} \BibitemShut
  {NoStop}%
\bibitem [{\citenamefont {{Schatzman}}(1961)}]{1961AnAp...24..237S}%
  \BibitemOpen
  \bibfield  {author} {\bibinfo {author} {\bibfnamefont {E.}~\bibnamefont
  {{Schatzman}}},\ }\bibfield  {title} {\enquote {\bibinfo {title} {{Sur la
  p{\'e}riode de pulsation radiale des naines blanches}},}\ }\href@noop {}
  {\bibfield  {journal} {\bibinfo  {journal} {Annales d'Astrophysique}\
  }\textbf {\bibinfo {volume} {24}},\ \bibinfo {pages} {237} (\bibinfo {year}
  {1961})}\BibitemShut {NoStop}%
\bibitem [{\citenamefont {{Singh}}\ and\ \citenamefont
  {{Das}}(1991)}]{1991Ap&SS.186..117S}%
  \BibitemOpen
  \bibfield  {author} {\bibinfo {author} {\bibfnamefont {Harinder~P.}\
  \bibnamefont {{Singh}}}\ and\ \bibinfo {author} {\bibfnamefont {M.~K.}\
  \bibnamefont {{Das}}},\ }\bibfield  {title} {\enquote {\bibinfo {title}
  {{Radial Oscillations of Finite Temperature White Dwarfs}},}\ }\href
  {\doibase 10.1007/BF00644626} {\bibfield  {journal} {\bibinfo  {journal}
  {Astrophysics and Space Science}\ }\textbf {\bibinfo {volume} {186}},\
  \bibinfo {pages} {117--124} (\bibinfo {year} {1991})}\BibitemShut {NoStop}%
\bibitem [{\citenamefont {Winget}\ and\ \citenamefont
  {Kepler}(2008)}]{Winget:2008iu}%
  \BibitemOpen
  \bibfield  {author} {\bibinfo {author} {\bibfnamefont {D.~E.}\ \bibnamefont
  {Winget}}\ and\ \bibinfo {author} {\bibfnamefont {S.~O.}\ \bibnamefont
  {Kepler}},\ }\bibfield  {title} {\enquote {\bibinfo {title} {{Pulsating White
  Dwarf Stars and Precision Asteroseismology}},}\ }\href {\doibase
  10.1146/annurev.astro.46.060407.145250} {\bibfield  {journal} {\bibinfo
  {journal} {Ann. Rev. Astron. Astrophys.}\ }\textbf {\bibinfo {volume} {46}},\
  \bibinfo {pages} {157} (\bibinfo {year} {2008})},\ \Eprint
  {http://arxiv.org/abs/0806.2573} {arXiv:0806.2573 [astro-ph]} \BibitemShut
  {NoStop}%
\bibitem [{\citenamefont {Turck-Chieze}\ and\ \citenamefont
  {Lopes}(1993)}]{TurckChieze:1993dw}%
  \BibitemOpen
  \bibfield  {author} {\bibinfo {author} {\bibfnamefont {Sylvaine}\
  \bibnamefont {Turck-Chieze}}\ and\ \bibinfo {author} {\bibfnamefont
  {I.}~\bibnamefont {Lopes}},\ }\bibfield  {title} {\enquote {\bibinfo {title}
  {{Toward a unified classical model of the sun: On the sensitivity of
  neutrinos and helioseismology to the microscopic physics}},}\ }\href
  {\doibase 10.1086/172592} {\bibfield  {journal} {\bibinfo  {journal}
  {Astrophys. J.}\ }\textbf {\bibinfo {volume} {408}},\ \bibinfo {pages}
  {347--367} (\bibinfo {year} {1993})}\BibitemShut {NoStop}%
\bibitem [{\citenamefont {{Turck-Chi{\`e}ze}}\ and\ \citenamefont
  {{Lopes}}(2012)}]{2012RAA....12.1107T}%
  \BibitemOpen
  \bibfield  {author} {\bibinfo {author} {\bibfnamefont {Sylvaine}\
  \bibnamefont {{Turck-Chi{\`e}ze}}}\ and\ \bibinfo {author} {\bibfnamefont
  {Il{\'\i}dio}\ \bibnamefont {{Lopes}}},\ }\bibfield  {title} {\enquote
  {\bibinfo {title} {{Solar-stellar astrophysics and dark matter}},}\ }\href
  {\doibase 10.1088/1674-4527/12/8/011} {\bibfield  {journal} {\bibinfo
  {journal} {Research in Astronomy and Astrophysics}\ }\textbf {\bibinfo
  {volume} {12}},\ \bibinfo {pages} {1107--1138} (\bibinfo {year}
  {2012})}\BibitemShut {NoStop}%
\bibitem [{\citenamefont {Damour}\ and\ \citenamefont
  {Esposito-Farese}(1993)}]{Damour:1993hw}%
  \BibitemOpen
  \bibfield  {author} {\bibinfo {author} {\bibfnamefont {Thibault}\
  \bibnamefont {Damour}}\ and\ \bibinfo {author} {\bibfnamefont {Gilles}\
  \bibnamefont {Esposito-Farese}},\ }\bibfield  {title} {\enquote {\bibinfo
  {title} {{Nonperturbative strong field effects in tensor - scalar theories of
  gravitation}},}\ }\href {\doibase 10.1103/PhysRevLett.70.2220} {\bibfield
  {journal} {\bibinfo  {journal} {Phys. Rev. Lett.}\ }\textbf {\bibinfo
  {volume} {70}},\ \bibinfo {pages} {2220--2223} (\bibinfo {year}
  {1993})}\BibitemShut {NoStop}%
\bibitem [{\citenamefont {Damour}\ and\ \citenamefont
  {Esposito-Farese}(1996)}]{Damour:1996ke}%
  \BibitemOpen
  \bibfield  {author} {\bibinfo {author} {\bibfnamefont {Thibault}\
  \bibnamefont {Damour}}\ and\ \bibinfo {author} {\bibfnamefont {Gilles}\
  \bibnamefont {Esposito-Farese}},\ }\bibfield  {title} {\enquote {\bibinfo
  {title} {{Tensor - scalar gravity and binary pulsar experiments}},}\ }\href
  {\doibase 10.1103/PhysRevD.54.1474} {\bibfield  {journal} {\bibinfo
  {journal} {Phys. Rev. D}\ }\textbf {\bibinfo {volume} {54}},\ \bibinfo
  {pages} {1474--1491} (\bibinfo {year} {1996})},\ \Eprint
  {http://arxiv.org/abs/gr-qc/9602056} {arXiv:gr-qc/9602056} \BibitemShut
  {NoStop}%
\bibitem [{\citenamefont {Harada}(1998)}]{Harada:1998ge}%
  \BibitemOpen
  \bibfield  {author} {\bibinfo {author} {\bibfnamefont {Tomohiro}\
  \bibnamefont {Harada}},\ }\bibfield  {title} {\enquote {\bibinfo {title}
  {{Neutron stars in scalar tensor theories of gravity and catastrophe
  theory}},}\ }\href {\doibase 10.1103/PhysRevD.57.4802} {\bibfield  {journal}
  {\bibinfo  {journal} {Phys. Rev. D}\ }\textbf {\bibinfo {volume} {57}},\
  \bibinfo {pages} {4802--4811} (\bibinfo {year} {1998})},\ \Eprint
  {http://arxiv.org/abs/gr-qc/9801049} {arXiv:gr-qc/9801049} \BibitemShut
  {NoStop}%
\bibitem [{\citenamefont {Pani}\ \emph {et~al.}(2011)\citenamefont {Pani},
  \citenamefont {Cardoso}, \citenamefont {Berti}, \citenamefont {Read},\ and\
  \citenamefont {Salgado}}]{Pani:2010vc}%
  \BibitemOpen
  \bibfield  {author} {\bibinfo {author} {\bibfnamefont {Paolo}\ \bibnamefont
  {Pani}}, \bibinfo {author} {\bibfnamefont {Vitor}\ \bibnamefont {Cardoso}},
  \bibinfo {author} {\bibfnamefont {Emanuele}\ \bibnamefont {Berti}}, \bibinfo
  {author} {\bibfnamefont {Jocelyn}\ \bibnamefont {Read}}, \ and\ \bibinfo
  {author} {\bibfnamefont {Marcelo}\ \bibnamefont {Salgado}},\ }\bibfield
  {title} {\enquote {\bibinfo {title} {{The vacuum revealed: the final state of
  vacuum instabilities in compact stars}},}\ }\href {\doibase
  10.1103/PhysRevD.83.081501} {\bibfield  {journal} {\bibinfo  {journal} {Phys.
  Rev. D}\ }\textbf {\bibinfo {volume} {83}},\ \bibinfo {pages} {081501}
  (\bibinfo {year} {2011})},\ \Eprint {http://arxiv.org/abs/1012.1343}
  {arXiv:1012.1343 [gr-qc]} \BibitemShut {NoStop}%
\bibitem [{\citenamefont {Ramazano\u{g}lu}\ and\ \citenamefont
  {Pretorius}(2016)}]{Ramazanoglu:2016kul}%
  \BibitemOpen
  \bibfield  {author} {\bibinfo {author} {\bibfnamefont {Fethi~M.}\
  \bibnamefont {Ramazano\u{g}lu}}\ and\ \bibinfo {author} {\bibfnamefont
  {Frans}\ \bibnamefont {Pretorius}},\ }\bibfield  {title} {\enquote {\bibinfo
  {title} {{Spontaneous Scalarization with Massive Fields}},}\ }\href {\doibase
  10.1103/PhysRevD.93.064005} {\bibfield  {journal} {\bibinfo  {journal} {Phys.
  Rev. D}\ }\textbf {\bibinfo {volume} {93}},\ \bibinfo {pages} {064005}
  (\bibinfo {year} {2016})},\ \Eprint {http://arxiv.org/abs/1601.07475}
  {arXiv:1601.07475 [gr-qc]} \BibitemShut {NoStop}%
\bibitem [{\citenamefont {Antoniou}\ \emph {et~al.}(2018)\citenamefont
  {Antoniou}, \citenamefont {Bakopoulos},\ and\ \citenamefont
  {Kanti}}]{Antoniou:2017acq}%
  \BibitemOpen
  \bibfield  {author} {\bibinfo {author} {\bibfnamefont {G.}~\bibnamefont
  {Antoniou}}, \bibinfo {author} {\bibfnamefont {A.}~\bibnamefont
  {Bakopoulos}}, \ and\ \bibinfo {author} {\bibfnamefont {P.}~\bibnamefont
  {Kanti}},\ }\bibfield  {title} {\enquote {\bibinfo {title} {{Evasion of
  No-Hair Theorems and Novel Black-Hole Solutions in Gauss-Bonnet Theories}},}\
  }\href {\doibase 10.1103/PhysRevLett.120.131102} {\bibfield  {journal}
  {\bibinfo  {journal} {Phys. Rev. Lett.}\ }\textbf {\bibinfo {volume} {120}},\
  \bibinfo {pages} {131102} (\bibinfo {year} {2018})},\ \Eprint
  {http://arxiv.org/abs/1711.03390} {arXiv:1711.03390 [hep-th]} \BibitemShut
  {NoStop}%
\bibitem [{\citenamefont {Doneva}\ and\ \citenamefont
  {Yazadjiev}(2018)}]{Doneva:2017bvd}%
  \BibitemOpen
  \bibfield  {author} {\bibinfo {author} {\bibfnamefont {Daniela~D.}\
  \bibnamefont {Doneva}}\ and\ \bibinfo {author} {\bibfnamefont {Stoytcho~S.}\
  \bibnamefont {Yazadjiev}},\ }\bibfield  {title} {\enquote {\bibinfo {title}
  {{New Gauss-Bonnet Black Holes with Curvature-Induced Scalarization in
  Extended Scalar-Tensor Theories}},}\ }\href {\doibase
  10.1103/PhysRevLett.120.131103} {\bibfield  {journal} {\bibinfo  {journal}
  {Phys. Rev. Lett.}\ }\textbf {\bibinfo {volume} {120}},\ \bibinfo {pages}
  {131103} (\bibinfo {year} {2018})},\ \Eprint
  {http://arxiv.org/abs/1711.01187} {arXiv:1711.01187 [gr-qc]} \BibitemShut
  {NoStop}%
\bibitem [{\citenamefont {Silva}\ \emph {et~al.}(2018)\citenamefont {Silva},
  \citenamefont {Sakstein}, \citenamefont {Gualtieri}, \citenamefont
  {Sotiriou},\ and\ \citenamefont {Berti}}]{Silva:2017uqg}%
  \BibitemOpen
  \bibfield  {author} {\bibinfo {author} {\bibfnamefont {Hector~O.}\
  \bibnamefont {Silva}}, \bibinfo {author} {\bibfnamefont {Jeremy}\
  \bibnamefont {Sakstein}}, \bibinfo {author} {\bibfnamefont {Leonardo}\
  \bibnamefont {Gualtieri}}, \bibinfo {author} {\bibfnamefont {Thomas~P.}\
  \bibnamefont {Sotiriou}}, \ and\ \bibinfo {author} {\bibfnamefont {Emanuele}\
  \bibnamefont {Berti}},\ }\bibfield  {title} {\enquote {\bibinfo {title}
  {{Spontaneous scalarization of black holes and compact stars from a
  Gauss-Bonnet coupling}},}\ }\href {\doibase 10.1103/PhysRevLett.120.131104}
  {\bibfield  {journal} {\bibinfo  {journal} {Phys. Rev. Lett.}\ }\textbf
  {\bibinfo {volume} {120}},\ \bibinfo {pages} {131104} (\bibinfo {year}
  {2018})},\ \Eprint {http://arxiv.org/abs/1711.02080} {arXiv:1711.02080
  [gr-qc]} \BibitemShut {NoStop}%
\bibitem [{\citenamefont {Khoury}\ and\ \citenamefont
  {Weltman}(2004{\natexlab{a}})}]{Khoury:2003aq}%
  \BibitemOpen
  \bibfield  {author} {\bibinfo {author} {\bibfnamefont {Justin}\ \bibnamefont
  {Khoury}}\ and\ \bibinfo {author} {\bibfnamefont {Amanda}\ \bibnamefont
  {Weltman}},\ }\bibfield  {title} {\enquote {\bibinfo {title} {{Chameleon
  fields: Awaiting surprises for tests of gravity in space}},}\ }\href
  {\doibase 10.1103/PhysRevLett.93.171104} {\bibfield  {journal} {\bibinfo
  {journal} {Phys. Rev. Lett.}\ }\textbf {\bibinfo {volume} {93}},\ \bibinfo
  {pages} {171104} (\bibinfo {year} {2004}{\natexlab{a}})},\ \Eprint
  {http://arxiv.org/abs/astro-ph/0309300} {arXiv:astro-ph/0309300} \BibitemShut
  {NoStop}%
\bibitem [{\citenamefont {Khoury}\ and\ \citenamefont
  {Weltman}(2004{\natexlab{b}})}]{Khoury:2003rn}%
  \BibitemOpen
  \bibfield  {author} {\bibinfo {author} {\bibfnamefont {Justin}\ \bibnamefont
  {Khoury}}\ and\ \bibinfo {author} {\bibfnamefont {Amanda}\ \bibnamefont
  {Weltman}},\ }\bibfield  {title} {\enquote {\bibinfo {title} {{Chameleon
  cosmology}},}\ }\href {\doibase 10.1103/PhysRevD.69.044026} {\bibfield
  {journal} {\bibinfo  {journal} {Phys. Rev. D}\ }\textbf {\bibinfo {volume}
  {69}},\ \bibinfo {pages} {044026} (\bibinfo {year} {2004}{\natexlab{b}})},\
  \Eprint {http://arxiv.org/abs/astro-ph/0309411} {arXiv:astro-ph/0309411}
  \BibitemShut {NoStop}%
\bibitem [{\citenamefont {Joyce}\ \emph {et~al.}(2015)\citenamefont {Joyce},
  \citenamefont {Jain}, \citenamefont {Khoury},\ and\ \citenamefont
  {Trodden}}]{Joyce:2014kja}%
  \BibitemOpen
  \bibfield  {author} {\bibinfo {author} {\bibfnamefont {Austin}\ \bibnamefont
  {Joyce}}, \bibinfo {author} {\bibfnamefont {Bhuvnesh}\ \bibnamefont {Jain}},
  \bibinfo {author} {\bibfnamefont {Justin}\ \bibnamefont {Khoury}}, \ and\
  \bibinfo {author} {\bibfnamefont {Mark}\ \bibnamefont {Trodden}},\ }\bibfield
   {title} {\enquote {\bibinfo {title} {{Beyond the Cosmological Standard
  Model}},}\ }\href {\doibase 10.1016/j.physrep.2014.12.002} {\bibfield
  {journal} {\bibinfo  {journal} {Phys. Rept.}\ }\textbf {\bibinfo {volume}
  {568}},\ \bibinfo {pages} {1--98} (\bibinfo {year} {2015})},\ \Eprint
  {http://arxiv.org/abs/1407.0059} {arXiv:1407.0059 [astro-ph.CO]} \BibitemShut
  {NoStop}%
\bibitem [{\citenamefont {Cardoso}\ \emph {et~al.}(2020)\citenamefont
  {Cardoso}, \citenamefont {Foschi},\ and\ \citenamefont
  {Zilhao}}]{Cardoso:2020cwo}%
  \BibitemOpen
  \bibfield  {author} {\bibinfo {author} {\bibfnamefont {Vitor}\ \bibnamefont
  {Cardoso}}, \bibinfo {author} {\bibfnamefont {Arianna}\ \bibnamefont
  {Foschi}}, \ and\ \bibinfo {author} {\bibfnamefont {Miguel}\ \bibnamefont
  {Zilhao}},\ }\bibfield  {title} {\enquote {\bibinfo {title} {{Collective
  scalarization or tachyonization: when averaging fails}},}\ }\href {\doibase
  10.1103/PhysRevLett.124.221104} {\bibfield  {journal} {\bibinfo  {journal}
  {Phys. Rev. Lett.}\ }\textbf {\bibinfo {volume} {124}},\ \bibinfo {pages}
  {221104} (\bibinfo {year} {2020})},\ \Eprint
  {http://arxiv.org/abs/2005.12284} {arXiv:2005.12284 [gr-qc]} \BibitemShut
  {NoStop}%
\bibitem [{\citenamefont {Tsujikawa}\ \emph {et~al.}(2009)\citenamefont
  {Tsujikawa}, \citenamefont {Tamaki},\ and\ \citenamefont
  {Tavakol}}]{Tsujikawa:2009yf}%
  \BibitemOpen
  \bibfield  {author} {\bibinfo {author} {\bibfnamefont {Shinji}\ \bibnamefont
  {Tsujikawa}}, \bibinfo {author} {\bibfnamefont {Takashi}\ \bibnamefont
  {Tamaki}}, \ and\ \bibinfo {author} {\bibfnamefont {Reza}\ \bibnamefont
  {Tavakol}},\ }\bibfield  {title} {\enquote {\bibinfo {title} {{Chameleon
  scalar fields in relativistic gravitational backgrounds}},}\ }\href {\doibase
  10.1088/1475-7516/2009/05/020} {\bibfield  {journal} {\bibinfo  {journal}
  {JCAP}\ }\textbf {\bibinfo {volume} {05}},\ \bibinfo {pages} {020} (\bibinfo
  {year} {2009})},\ \Eprint {http://arxiv.org/abs/0901.3226} {arXiv:0901.3226
  [gr-qc]} \BibitemShut {NoStop}%
\bibitem [{\citenamefont {Barausse}\ \emph {et~al.}(2013)\citenamefont
  {Barausse}, \citenamefont {Palenzuela}, \citenamefont {Ponce},\ and\
  \citenamefont {Lehner}}]{Barausse:2012da}%
  \BibitemOpen
  \bibfield  {author} {\bibinfo {author} {\bibfnamefont {Enrico}\ \bibnamefont
  {Barausse}}, \bibinfo {author} {\bibfnamefont {Carlos}\ \bibnamefont
  {Palenzuela}}, \bibinfo {author} {\bibfnamefont {Marcelo}\ \bibnamefont
  {Ponce}}, \ and\ \bibinfo {author} {\bibfnamefont {Luis}\ \bibnamefont
  {Lehner}},\ }\bibfield  {title} {\enquote {\bibinfo {title} {{Neutron-star
  mergers in scalar-tensor theories of gravity}},}\ }\href {\doibase
  10.1103/PhysRevD.87.081506} {\bibfield  {journal} {\bibinfo  {journal} {Phys.
  Rev. D}\ }\textbf {\bibinfo {volume} {87}},\ \bibinfo {pages} {081506}
  (\bibinfo {year} {2013})},\ \Eprint {http://arxiv.org/abs/1212.5053}
  {arXiv:1212.5053 [gr-qc]} \BibitemShut {NoStop}%
\bibitem [{\citenamefont {Palenzuela}\ \emph {et~al.}(2014)\citenamefont
  {Palenzuela}, \citenamefont {Barausse}, \citenamefont {Ponce},\ and\
  \citenamefont {Lehner}}]{Palenzuela:2013hsa}%
  \BibitemOpen
  \bibfield  {author} {\bibinfo {author} {\bibfnamefont {Carlos}\ \bibnamefont
  {Palenzuela}}, \bibinfo {author} {\bibfnamefont {Enrico}\ \bibnamefont
  {Barausse}}, \bibinfo {author} {\bibfnamefont {Marcelo}\ \bibnamefont
  {Ponce}}, \ and\ \bibinfo {author} {\bibfnamefont {Luis}\ \bibnamefont
  {Lehner}},\ }\bibfield  {title} {\enquote {\bibinfo {title} {{Dynamical
  scalarization of neutron stars in scalar-tensor gravity theories}},}\ }\href
  {\doibase 10.1103/PhysRevD.89.044024} {\bibfield  {journal} {\bibinfo
  {journal} {Phys. Rev. D}\ }\textbf {\bibinfo {volume} {89}},\ \bibinfo
  {pages} {044024} (\bibinfo {year} {2014})},\ \Eprint
  {http://arxiv.org/abs/1310.4481} {arXiv:1310.4481 [gr-qc]} \BibitemShut
  {NoStop}%
\bibitem [{\citenamefont {Shibata}\ \emph {et~al.}(2014)\citenamefont
  {Shibata}, \citenamefont {Taniguchi}, \citenamefont {Okawa},\ and\
  \citenamefont {Buonanno}}]{Shibata:2013pra}%
  \BibitemOpen
  \bibfield  {author} {\bibinfo {author} {\bibfnamefont {Masaru}\ \bibnamefont
  {Shibata}}, \bibinfo {author} {\bibfnamefont {Keisuke}\ \bibnamefont
  {Taniguchi}}, \bibinfo {author} {\bibfnamefont {Hirotada}\ \bibnamefont
  {Okawa}}, \ and\ \bibinfo {author} {\bibfnamefont {Alessandra}\ \bibnamefont
  {Buonanno}},\ }\bibfield  {title} {\enquote {\bibinfo {title} {{Coalescence
  of binary neutron stars in a scalar-tensor theory of gravity}},}\ }\href
  {\doibase 10.1103/PhysRevD.89.084005} {\bibfield  {journal} {\bibinfo
  {journal} {Phys. Rev. D}\ }\textbf {\bibinfo {volume} {89}},\ \bibinfo
  {pages} {084005} (\bibinfo {year} {2014})},\ \Eprint
  {http://arxiv.org/abs/1310.0627} {arXiv:1310.0627 [gr-qc]} \BibitemShut
  {NoStop}%
\bibitem [{\citenamefont {Khalil}\ \emph {et~al.}(2019)\citenamefont {Khalil},
  \citenamefont {Sennett}, \citenamefont {Steinhoff},\ and\ \citenamefont
  {Buonanno}}]{Khalil:2019wyy}%
  \BibitemOpen
  \bibfield  {author} {\bibinfo {author} {\bibfnamefont {Mohammed}\
  \bibnamefont {Khalil}}, \bibinfo {author} {\bibfnamefont {Noah}\ \bibnamefont
  {Sennett}}, \bibinfo {author} {\bibfnamefont {Jan}\ \bibnamefont
  {Steinhoff}}, \ and\ \bibinfo {author} {\bibfnamefont {Alessandra}\
  \bibnamefont {Buonanno}},\ }\bibfield  {title} {\enquote {\bibinfo {title}
  {{Theory-agnostic framework for dynamical scalarization of compact
  binaries}},}\ }\href {\doibase 10.1103/PhysRevD.100.124013} {\bibfield
  {journal} {\bibinfo  {journal} {Phys. Rev. D}\ }\textbf {\bibinfo {volume}
  {100}},\ \bibinfo {pages} {124013} (\bibinfo {year} {2019})},\ \Eprint
  {http://arxiv.org/abs/1906.08161} {arXiv:1906.08161 [gr-qc]} \BibitemShut
  {NoStop}%
\bibitem [{\citenamefont {Silva}\ \emph {et~al.}(2021)\citenamefont {Silva},
  \citenamefont {Witek}, \citenamefont {Elley},\ and\ \citenamefont
  {Yunes}}]{Silva:2020omi}%
  \BibitemOpen
  \bibfield  {author} {\bibinfo {author} {\bibfnamefont {Hector~O.}\
  \bibnamefont {Silva}}, \bibinfo {author} {\bibfnamefont {Helvi}\ \bibnamefont
  {Witek}}, \bibinfo {author} {\bibfnamefont {Matthew}\ \bibnamefont {Elley}},
  \ and\ \bibinfo {author} {\bibfnamefont {Nicol\'as}\ \bibnamefont {Yunes}},\
  }\bibfield  {title} {\enquote {\bibinfo {title} {{Dynamical Descalarization
  in Binary Black Hole Mergers}},}\ }\href {\doibase
  10.1103/PhysRevLett.127.031101} {\bibfield  {journal} {\bibinfo  {journal}
  {Phys. Rev. Lett.}\ }\textbf {\bibinfo {volume} {127}},\ \bibinfo {pages}
  {031101} (\bibinfo {year} {2021})},\ \Eprint
  {http://arxiv.org/abs/2012.10436} {arXiv:2012.10436 [gr-qc]} \BibitemShut
  {NoStop}%
\bibitem [{\citenamefont {East}\ and\ \citenamefont
  {Ripley}(2021)}]{East:2021bqk}%
  \BibitemOpen
  \bibfield  {author} {\bibinfo {author} {\bibfnamefont {William~E.}\
  \bibnamefont {East}}\ and\ \bibinfo {author} {\bibfnamefont {Justin~L.}\
  \bibnamefont {Ripley}},\ }\bibfield  {title} {\enquote {\bibinfo {title}
  {{Dynamics of spontaneous black hole scalarization and mergers in
  Einstein-scalar-Gauss-Bonnet gravity}},}\ }\href@noop {} {\  (\bibinfo {year}
  {2021})},\ \Eprint {http://arxiv.org/abs/2105.08571} {arXiv:2105.08571
  [gr-qc]} \BibitemShut {NoStop}%
\bibitem [{\citenamefont {Llinares}\ and\ \citenamefont
  {Mota}(2013)}]{Llinares:2013qbh}%
  \BibitemOpen
  \bibfield  {author} {\bibinfo {author} {\bibfnamefont {Claudio}\ \bibnamefont
  {Llinares}}\ and\ \bibinfo {author} {\bibfnamefont {David}\ \bibnamefont
  {Mota}},\ }\bibfield  {title} {\enquote {\bibinfo {title} {{Releasing scalar
  fields: cosmological simulations of scalar-tensor theories for gravity beyond
  the static approximation}},}\ }\href {\doibase
  10.1103/PhysRevLett.110.161101} {\bibfield  {journal} {\bibinfo  {journal}
  {Phys. Rev. Lett.}\ }\textbf {\bibinfo {volume} {110}},\ \bibinfo {pages}
  {161101} (\bibinfo {year} {2013})},\ \Eprint {http://arxiv.org/abs/1302.1774}
  {arXiv:1302.1774 [astro-ph.CO]} \BibitemShut {NoStop}%
\bibitem [{\citenamefont {Hagala}\ \emph {et~al.}(2017)\citenamefont {Hagala},
  \citenamefont {Llinares},\ and\ \citenamefont {Mota}}]{Hagala:2016fks}%
  \BibitemOpen
  \bibfield  {author} {\bibinfo {author} {\bibfnamefont {R.}~\bibnamefont
  {Hagala}}, \bibinfo {author} {\bibfnamefont {C.}~\bibnamefont {Llinares}}, \
  and\ \bibinfo {author} {\bibfnamefont {D.F.}\ \bibnamefont {Mota}},\
  }\bibfield  {title} {\enquote {\bibinfo {title} {{Cosmic Tsunamis in Modified
  Gravity: Disruption of Screening Mechanisms from Scalar Waves}},}\ }\href
  {\doibase 10.1103/PhysRevLett.118.101301} {\bibfield  {journal} {\bibinfo
  {journal} {Phys. Rev. Lett.}\ }\textbf {\bibinfo {volume} {118}},\ \bibinfo
  {pages} {101301} (\bibinfo {year} {2017})},\ \Eprint
  {http://arxiv.org/abs/1607.02600} {arXiv:1607.02600 [astro-ph.CO]}
  \BibitemShut {NoStop}%
\bibitem [{\citenamefont {Brito}\ \emph {et~al.}(2014)\citenamefont {Brito},
  \citenamefont {Terrana}, \citenamefont {Johnson},\ and\ \citenamefont
  {Cardoso}}]{Brito:2014ifa}%
  \BibitemOpen
  \bibfield  {author} {\bibinfo {author} {\bibfnamefont {Richard}\ \bibnamefont
  {Brito}}, \bibinfo {author} {\bibfnamefont {Alexandra}\ \bibnamefont
  {Terrana}}, \bibinfo {author} {\bibfnamefont {Matthew}\ \bibnamefont
  {Johnson}}, \ and\ \bibinfo {author} {\bibfnamefont {Vitor}\ \bibnamefont
  {Cardoso}},\ }\bibfield  {title} {\enquote {\bibinfo {title} {{Nonlinear
  dynamical stability of infrared modifications of gravity}},}\ }\href
  {\doibase 10.1103/PhysRevD.90.124035} {\bibfield  {journal} {\bibinfo
  {journal} {Phys. Rev. D}\ }\textbf {\bibinfo {volume} {90}},\ \bibinfo
  {pages} {124035} (\bibinfo {year} {2014})},\ \Eprint
  {http://arxiv.org/abs/1409.0886} {arXiv:1409.0886 [hep-th]} \BibitemShut
  {NoStop}%
\bibitem [{\citenamefont {Silvestri}(2011)}]{Silvestri:2011ch}%
  \BibitemOpen
  \bibfield  {author} {\bibinfo {author} {\bibfnamefont {Alessandra}\
  \bibnamefont {Silvestri}},\ }\bibfield  {title} {\enquote {\bibinfo {title}
  {{Scalar radiation from Chameleon-shielded regions}},}\ }\href {\doibase
  10.1103/PhysRevLett.106.251101} {\bibfield  {journal} {\bibinfo  {journal}
  {Phys. Rev. Lett.}\ }\textbf {\bibinfo {volume} {106}},\ \bibinfo {pages}
  {251101} (\bibinfo {year} {2011})},\ \Eprint {http://arxiv.org/abs/1103.4013}
  {arXiv:1103.4013 [astro-ph.CO]} \BibitemShut {NoStop}%
\bibitem [{\citenamefont {Nakamura}\ \emph {et~al.}(2021)\citenamefont
  {Nakamura}, \citenamefont {Ikeda}, \citenamefont {Saito}, \citenamefont
  {Tanahashi},\ and\ \citenamefont {Yoo}}]{Nakamura:2020ihr}%
  \BibitemOpen
  \bibfield  {author} {\bibinfo {author} {\bibfnamefont {Tomohiro}\
  \bibnamefont {Nakamura}}, \bibinfo {author} {\bibfnamefont {Taishi}\
  \bibnamefont {Ikeda}}, \bibinfo {author} {\bibfnamefont {Ryo}\ \bibnamefont
  {Saito}}, \bibinfo {author} {\bibfnamefont {Norihiro}\ \bibnamefont
  {Tanahashi}}, \ and\ \bibinfo {author} {\bibfnamefont {Chul-Moon}\
  \bibnamefont {Yoo}},\ }\bibfield  {title} {\enquote {\bibinfo {title}
  {{Dynamical Analysis of Screening in Scalar-Tensor Theory}},}\ }\href
  {\doibase 10.1103/PhysRevD.103.024009} {\bibfield  {journal} {\bibinfo
  {journal} {Phys. Rev. D}\ }\textbf {\bibinfo {volume} {103}},\ \bibinfo
  {pages} {024009} (\bibinfo {year} {2021})},\ \Eprint
  {http://arxiv.org/abs/2010.14329} {arXiv:2010.14329 [gr-qc]} \BibitemShut
  {NoStop}%
\bibitem [{\citenamefont {Wang}\ \emph {et~al.}(2011)\citenamefont {Wang},
  \citenamefont {Bonifacio}, \citenamefont {Bingham},\ and\ \citenamefont
  {Tito~Mendonca}}]{Wang:2009qa}%
  \BibitemOpen
  \bibfield  {author} {\bibinfo {author} {\bibfnamefont {Charles H.~T.}\
  \bibnamefont {Wang}}, \bibinfo {author} {\bibfnamefont {Paolo~M.}\
  \bibnamefont {Bonifacio}}, \bibinfo {author} {\bibfnamefont {Robert}\
  \bibnamefont {Bingham}}, \ and\ \bibinfo {author} {\bibfnamefont
  {J.}~\bibnamefont {Tito~Mendonca}},\ }\bibfield  {title} {\enquote {\bibinfo
  {title} {{Parametric instability induced scalar gravitational waves from a
  model pulsating neutron star}},}\ }\href {\doibase
  10.1016/j.physletb.2011.09.111} {\bibfield  {journal} {\bibinfo  {journal}
  {Phys. Lett. B}\ }\textbf {\bibinfo {volume} {705}},\ \bibinfo {pages}
  {148--151} (\bibinfo {year} {2011})},\ \Eprint
  {http://arxiv.org/abs/0905.0706} {arXiv:0905.0706 [gr-qc]} \BibitemShut
  {NoStop}%
\bibitem [{\citenamefont {Wang}\ \emph {et~al.}(2013)\citenamefont {Wang},
  \citenamefont {Hodson}, \citenamefont {Murphy}, \citenamefont {Davies},
  \citenamefont {Bingham},\ and\ \citenamefont {Mendon\c{c}a}}]{Wang:2013kc}%
  \BibitemOpen
  \bibfield  {author} {\bibinfo {author} {\bibfnamefont {C.~H.~T.}\
  \bibnamefont {Wang}}, \bibinfo {author} {\bibfnamefont {A.~O.}\ \bibnamefont
  {Hodson}}, \bibinfo {author} {\bibfnamefont {A.~St.~J.}\ \bibnamefont
  {Murphy}}, \bibinfo {author} {\bibfnamefont {T.~B.}\ \bibnamefont {Davies}},
  \bibinfo {author} {\bibfnamefont {R.}~\bibnamefont {Bingham}}, \ and\
  \bibinfo {author} {\bibfnamefont {J.~T.}\ \bibnamefont {Mendon\c{c}a}},\
  }\bibfield  {title} {\enquote {\bibinfo {title} {{Dynamical trapping and
  relaxation of scalar gravitational fields}},}\ }\href {\doibase
  10.1016/j.physletb.2013.09.002} {\bibfield  {journal} {\bibinfo  {journal}
  {Phys. Lett. B}\ }\textbf {\bibinfo {volume} {726}},\ \bibinfo {pages}
  {791--794} (\bibinfo {year} {2013})},\ \Eprint
  {http://arxiv.org/abs/1301.0269} {arXiv:1301.0269 [gr-qc]} \BibitemShut
  {NoStop}%
\bibitem [{\citenamefont {Ventagli}\ \emph {et~al.}(2020)\citenamefont
  {Ventagli}, \citenamefont {Leh\'ebel},\ and\ \citenamefont
  {Sotiriou}}]{Ventagli:2020rnx}%
  \BibitemOpen
  \bibfield  {author} {\bibinfo {author} {\bibfnamefont {Giulia}\ \bibnamefont
  {Ventagli}}, \bibinfo {author} {\bibfnamefont {Antoine}\ \bibnamefont
  {Leh\'ebel}}, \ and\ \bibinfo {author} {\bibfnamefont {Thomas~P.}\
  \bibnamefont {Sotiriou}},\ }\bibfield  {title} {\enquote {\bibinfo {title}
  {{Onset of spontaneous scalarization in generalized scalar-tensor
  theories}},}\ }\href {\doibase 10.1103/PhysRevD.102.024050} {\bibfield
  {journal} {\bibinfo  {journal} {Phys. Rev. D}\ }\textbf {\bibinfo {volume}
  {102}},\ \bibinfo {pages} {024050} (\bibinfo {year} {2020})},\ \Eprint
  {http://arxiv.org/abs/2006.01153} {arXiv:2006.01153 [gr-qc]} \BibitemShut
  {NoStop}%
\bibitem [{\citenamefont {Hinterbichler}\ \emph {et~al.}(2011)\citenamefont
  {Hinterbichler}, \citenamefont {Khoury}, \citenamefont {Levy},\ and\
  \citenamefont {Matas}}]{Hinterbichler:2011ca}%
  \BibitemOpen
  \bibfield  {author} {\bibinfo {author} {\bibfnamefont {Kurt}\ \bibnamefont
  {Hinterbichler}}, \bibinfo {author} {\bibfnamefont {Justin}\ \bibnamefont
  {Khoury}}, \bibinfo {author} {\bibfnamefont {Aaron}\ \bibnamefont {Levy}}, \
  and\ \bibinfo {author} {\bibfnamefont {Andrew}\ \bibnamefont {Matas}},\
  }\bibfield  {title} {\enquote {\bibinfo {title} {{Symmetron Cosmology}},}\
  }\href {\doibase 10.1103/PhysRevD.84.103521} {\bibfield  {journal} {\bibinfo
  {journal} {Phys. Rev. D}\ }\textbf {\bibinfo {volume} {84}},\ \bibinfo
  {pages} {103521} (\bibinfo {year} {2011})},\ \Eprint
  {http://arxiv.org/abs/1107.2112} {arXiv:1107.2112 [astro-ph.CO]} \BibitemShut
  {NoStop}%
\bibitem [{\citenamefont {Hinterbichler}\ and\ \citenamefont
  {Khoury}(2010)}]{Hinterbichler:2010es}%
  \BibitemOpen
  \bibfield  {author} {\bibinfo {author} {\bibfnamefont {Kurt}\ \bibnamefont
  {Hinterbichler}}\ and\ \bibinfo {author} {\bibfnamefont {Justin}\
  \bibnamefont {Khoury}},\ }\bibfield  {title} {\enquote {\bibinfo {title}
  {{Symmetron Fields: Screening Long-Range Forces Through Local Symmetry
  Restoration}},}\ }\href {\doibase 10.1103/PhysRevLett.104.231301} {\bibfield
  {journal} {\bibinfo  {journal} {Phys. Rev. Lett.}\ }\textbf {\bibinfo
  {volume} {104}},\ \bibinfo {pages} {231301} (\bibinfo {year} {2010})},\
  \Eprint {http://arxiv.org/abs/1001.4525} {arXiv:1001.4525 [hep-th]}
  \BibitemShut {NoStop}%
\bibitem [{\citenamefont {Shapiro}\ and\ \citenamefont
  {Teukolsky}(1983)}]{Shapiro:1983du}%
  \BibitemOpen
  \bibfield  {author} {\bibinfo {author} {\bibfnamefont {S.~L.}\ \bibnamefont
  {Shapiro}}\ and\ \bibinfo {author} {\bibfnamefont {S.~A.}\ \bibnamefont
  {Teukolsky}},\ }\href@noop {} {\emph {\bibinfo {title} {Black holes, white
  dwarfs, and neutron stars: The physics of compact objects}}}\ (\bibinfo
  {publisher} {Wiley},\ \bibinfo {address} {New York},\ \bibinfo {year}
  {1983})\BibitemShut {NoStop}%
\bibitem [{\citenamefont {Sotani}(2014)}]{Sotani:2014tua}%
  \BibitemOpen
  \bibfield  {author} {\bibinfo {author} {\bibfnamefont {Hajime}\ \bibnamefont
  {Sotani}},\ }\bibfield  {title} {\enquote {\bibinfo {title} {{Scalar
  gravitational waves from relativistic stars in scalar-tensor gravity}},}\
  }\href {\doibase 10.1103/PhysRevD.89.064031} {\bibfield  {journal} {\bibinfo
  {journal} {Phys. Rev. D}\ }\textbf {\bibinfo {volume} {89}},\ \bibinfo
  {pages} {064031} (\bibinfo {year} {2014})},\ \Eprint
  {http://arxiv.org/abs/1402.5699} {arXiv:1402.5699 [astro-ph.HE]} \BibitemShut
  {NoStop}%
\bibitem [{\citenamefont {Mendes}\ and\ \citenamefont
  {Ortiz}(2018)}]{Mendes:2018qwo}%
  \BibitemOpen
  \bibfield  {author} {\bibinfo {author} {\bibfnamefont {Raissa F.~P.}\
  \bibnamefont {Mendes}}\ and\ \bibinfo {author} {\bibfnamefont {N\'estor}\
  \bibnamefont {Ortiz}},\ }\bibfield  {title} {\enquote {\bibinfo {title} {{New
  class of quasinormal modes of neutron stars in scalar-tensor gravity}},}\
  }\href {\doibase 10.1103/PhysRevLett.120.201104} {\bibfield  {journal}
  {\bibinfo  {journal} {Phys. Rev. Lett.}\ }\textbf {\bibinfo {volume} {120}},\
  \bibinfo {pages} {201104} (\bibinfo {year} {2018})},\ \Eprint
  {http://arxiv.org/abs/1802.07847} {arXiv:1802.07847 [gr-qc]} \BibitemShut
  {NoStop}%
\bibitem [{\citenamefont {{Chandrasekhar}}(1964)}]{1964ApJ...140..417C}%
  \BibitemOpen
  \bibfield  {author} {\bibinfo {author} {\bibfnamefont {S.}~\bibnamefont
  {{Chandrasekhar}}},\ }\bibfield  {title} {\enquote {\bibinfo {title} {{The
  Dynamical Instability of Gaseous Masses Approaching the Schwarzschild Limit
  in General Relativity.}}}\ }\href {\doibase 10.1086/147938} {\bibfield
  {journal} {\bibinfo  {journal} {\apj}\ }\textbf {\bibinfo {volume} {140}},\
  \bibinfo {pages} {417} (\bibinfo {year} {1964})}\BibitemShut {NoStop}%
\bibitem [{\citenamefont {{Bardeen}}\ \emph {et~al.}(1966)\citenamefont
  {{Bardeen}}, \citenamefont {{Thorne}},\ and\ \citenamefont
  {{Meltzer}}}]{1966ApJ...145..505B}%
  \BibitemOpen
  \bibfield  {author} {\bibinfo {author} {\bibfnamefont {James~M.}\
  \bibnamefont {{Bardeen}}}, \bibinfo {author} {\bibfnamefont {Kip~S.}\
  \bibnamefont {{Thorne}}}, \ and\ \bibinfo {author} {\bibfnamefont {David~W.}\
  \bibnamefont {{Meltzer}}},\ }\bibfield  {title} {\enquote {\bibinfo {title}
  {{A Catalogue of Methods for Studying the Normal Modes of Radial Pulsation of
  General-Relativistic Stellar Models}},}\ }\href {\doibase 10.1086/148791}
  {\bibfield  {journal} {\bibinfo  {journal} {\apj}\ }\textbf {\bibinfo
  {volume} {145}},\ \bibinfo {pages} {505} (\bibinfo {year}
  {1966})}\BibitemShut {NoStop}%
\bibitem [{\citenamefont {Dittrich}\ \emph
  {et~al.}(1994{\natexlab{a}})\citenamefont {Dittrich}, \citenamefont
  {Duclos},\ and\ \citenamefont {{\v{S}}eba}}]{Dittrich:1993hw}%
  \BibitemOpen
  \bibfield  {author} {\bibinfo {author} {\bibfnamefont {J}~\bibnamefont
  {Dittrich}}, \bibinfo {author} {\bibfnamefont {P}~\bibnamefont {Duclos}}, \
  and\ \bibinfo {author} {\bibfnamefont {P}~\bibnamefont {{\v{S}}eba}},\
  }\bibfield  {title} {\enquote {\bibinfo {title} {Instability in a classical
  periodically driven string},}\ }\href@noop {} {\bibfield  {journal} {\bibinfo
   {journal} {Physical Review E}\ }\textbf {\bibinfo {volume} {49}},\ \bibinfo
  {pages} {3535} (\bibinfo {year} {1994}{\natexlab{a}})}\BibitemShut {NoStop}%
\bibitem [{\citenamefont {{Cutler}}\ and\ \citenamefont
  {{Lindblom}}(1987)}]{1987ApJ...314..234C}%
  \BibitemOpen
  \bibfield  {author} {\bibinfo {author} {\bibfnamefont {Curt}\ \bibnamefont
  {{Cutler}}}\ and\ \bibinfo {author} {\bibfnamefont {Lee}\ \bibnamefont
  {{Lindblom}}},\ }\bibfield  {title} {\enquote {\bibinfo {title} {{The Effect
  of Viscosity on Neutron Star Oscillations}},}\ }\href {\doibase
  10.1086/165052} {\bibfield  {journal} {\bibinfo  {journal} {\apj}\ }\textbf
  {\bibinfo {volume} {314}},\ \bibinfo {pages} {234} (\bibinfo {year}
  {1987})}\BibitemShut {NoStop}%
\bibitem [{\citenamefont {{Barta}}(2019)}]{2019CQGra..36u5012B}%
  \BibitemOpen
  \bibfield  {author} {\bibinfo {author} {\bibfnamefont {D{\'a}niel}\
  \bibnamefont {{Barta}}},\ }\bibfield  {title} {\enquote {\bibinfo {title}
  {{Effect of viscosity and thermal conductivity on the radial oscillation and
  relaxation of relativistic stars}},}\ }\href {\doibase
  10.1088/1361-6382/ab472e} {\bibfield  {journal} {\bibinfo  {journal}
  {Classical and Quantum Gravity}\ }\textbf {\bibinfo {volume} {36}},\ \bibinfo
  {eid} {215012} (\bibinfo {year} {2019})},\ \Eprint
  {http://arxiv.org/abs/1904.00907} {arXiv:1904.00907 [gr-qc]} \BibitemShut
  {NoStop}%
\bibitem [{\citenamefont {Abramowitz}\ and\ \citenamefont
  {Stegun}(1972)}]{Abramowitz:1970as}%
  \BibitemOpen
  \bibfield  {author} {\bibinfo {author} {\bibfnamefont {M.}~\bibnamefont
  {Abramowitz}}\ and\ \bibinfo {author} {\bibfnamefont {I.~A.}\ \bibnamefont
  {Stegun}},\ }\href@noop {} {\emph {\bibinfo {title} {Handbook of Mathematical
  Functions with Formulas, Graphs, and Mathematical Tables}}}\ (\bibinfo
  {publisher} {Dover},\ \bibinfo {address} {New York},\ \bibinfo {year}
  {1972})\BibitemShut {NoStop}%
\bibitem [{\citenamefont {Bender}\ and\ \citenamefont
  {Orszag}(1978)}]{benderorszag}%
  \BibitemOpen
  \bibfield  {author} {\bibinfo {author} {\bibfnamefont {C.~M.}\ \bibnamefont
  {Bender}}\ and\ \bibinfo {author} {\bibfnamefont {S.~A.}\ \bibnamefont
  {Orszag}},\ }\href@noop {} {\emph {\bibinfo {title} {Advanced Mathematical
  Methods for Scientists and Engineers}}}\ (\bibinfo  {publisher}
  {McGraw-Hill},\ \bibinfo {address} {New York},\ \bibinfo {year}
  {1978})\BibitemShut {NoStop}%
\bibitem [{\citenamefont {{Reisenegger}}\ and\ \citenamefont
  {{Goldreich}}(1994)}]{1994ApJ...426..688R}%
  \BibitemOpen
  \bibfield  {author} {\bibinfo {author} {\bibfnamefont {Andreas}\ \bibnamefont
  {{Reisenegger}}}\ and\ \bibinfo {author} {\bibfnamefont {Peter}\ \bibnamefont
  {{Goldreich}}},\ }\bibfield  {title} {\enquote {\bibinfo {title} {{Excitation
  of Neutron Star Normal Modes during Binary Inspiral}},}\ }\href {\doibase
  10.1086/174105} {\bibfield  {journal} {\bibinfo  {journal} {\apj}\ }\textbf
  {\bibinfo {volume} {426}},\ \bibinfo {pages} {688} (\bibinfo {year}
  {1994})}\BibitemShut {NoStop}%
\bibitem [{\citenamefont {Chirenti}\ \emph {et~al.}(2017)\citenamefont
  {Chirenti}, \citenamefont {Gold},\ and\ \citenamefont
  {Miller}}]{Chirenti:2016xys}%
  \BibitemOpen
  \bibfield  {author} {\bibinfo {author} {\bibfnamefont {Cecilia}\ \bibnamefont
  {Chirenti}}, \bibinfo {author} {\bibfnamefont {Roman}\ \bibnamefont {Gold}},
  \ and\ \bibinfo {author} {\bibfnamefont {M.~Coleman}\ \bibnamefont
  {Miller}},\ }\bibfield  {title} {\enquote {\bibinfo {title} {{Gravitational
  waves from f-modes excited by the inspiral of highly eccentric neutron star
  binaries}},}\ }\href {\doibase 10.3847/1538-4357/aa5ebb} {\bibfield
  {journal} {\bibinfo  {journal} {Astrophys. J.}\ }\textbf {\bibinfo {volume}
  {837}},\ \bibinfo {pages} {67} (\bibinfo {year} {2017})},\ \Eprint
  {http://arxiv.org/abs/1612.07097} {arXiv:1612.07097 [astro-ph.HE]}
  \BibitemShut {NoStop}%
\bibitem [{\citenamefont {Ma}\ \emph {et~al.}(2020)\citenamefont {Ma},
  \citenamefont {Yu},\ and\ \citenamefont {Chen}}]{Ma:2020rak}%
  \BibitemOpen
  \bibfield  {author} {\bibinfo {author} {\bibfnamefont {Sizheng}\ \bibnamefont
  {Ma}}, \bibinfo {author} {\bibfnamefont {Hang}\ \bibnamefont {Yu}}, \ and\
  \bibinfo {author} {\bibfnamefont {Yanbei}\ \bibnamefont {Chen}},\ }\bibfield
  {title} {\enquote {\bibinfo {title} {{Excitation of f-modes during mergers of
  spinning binary neutron star}},}\ }\href {\doibase
  10.1103/PhysRevD.101.123020} {\bibfield  {journal} {\bibinfo  {journal}
  {Phys. Rev. D}\ }\textbf {\bibinfo {volume} {101}},\ \bibinfo {pages}
  {123020} (\bibinfo {year} {2020})},\ \Eprint
  {http://arxiv.org/abs/2003.02373} {arXiv:2003.02373 [gr-qc]} \BibitemShut
  {NoStop}%
\bibitem [{\citenamefont {Perego}\ \emph {et~al.}(2019)\citenamefont {Perego},
  \citenamefont {Bernuzzi},\ and\ \citenamefont {Radice}}]{Perego:2019adq}%
  \BibitemOpen
  \bibfield  {author} {\bibinfo {author} {\bibfnamefont {Albino}\ \bibnamefont
  {Perego}}, \bibinfo {author} {\bibfnamefont {Sebastiano}\ \bibnamefont
  {Bernuzzi}}, \ and\ \bibinfo {author} {\bibfnamefont {David}\ \bibnamefont
  {Radice}},\ }\bibfield  {title} {\enquote {\bibinfo {title} {{Thermodynamics
  conditions of matter in neutron star mergers}},}\ }\href {\doibase
  10.1140/epja/i2019-12810-7} {\bibfield  {journal} {\bibinfo  {journal} {Eur.
  Phys. J. A}\ }\textbf {\bibinfo {volume} {55}},\ \bibinfo {pages} {124}
  (\bibinfo {year} {2019})},\ \Eprint {http://arxiv.org/abs/1903.07898}
  {arXiv:1903.07898 [gr-qc]} \BibitemShut {NoStop}%
\bibitem [{\citenamefont {Bernuzzi}\ \emph {et~al.}(2020)\citenamefont
  {Bernuzzi} \emph {et~al.}}]{Bernuzzi:2020txg}%
  \BibitemOpen
  \bibfield  {author} {\bibinfo {author} {\bibfnamefont {Sebastiano}\
  \bibnamefont {Bernuzzi}} \emph {et~al.},\ }\bibfield  {title} {\enquote
  {\bibinfo {title} {{Accretion-induced prompt black hole formation in
  asymmetric neutron star mergers, dynamical ejecta and kilonova signals}},}\
  }\href {\doibase 10.1093/mnras/staa1860} {\bibfield  {journal} {\bibinfo
  {journal} {Mon. Not. Roy. Astron. Soc.}\ }\textbf {\bibinfo {volume} {497}},\
  \bibinfo {pages} {1488--1507} (\bibinfo {year} {2020})},\ \Eprint
  {http://arxiv.org/abs/2003.06015} {arXiv:2003.06015 [astro-ph.HE]}
  \BibitemShut {NoStop}%
\bibitem [{Note1()}]{Note1}%
  \BibitemOpen
  \bibinfo {note} {There were important hints that such an effect could occur.
  The classical Fermi acceleration process is one example of energy extraction
  from a zero-average motion~\cite {Fermi:1949ee}. Similar processes exist for
  fields trapped inside cavities with periodically-moving boundaries~\cite
  {247776,PhysRevE.49.3535,Moore:1970,Andreata_2000,Dodonov:2001yb}. We showed
  that oscillating stars are also prone to such instabilities.}\BibitemShut
  {Stop}%
\bibitem [{\citenamefont {Annulli}\ \emph {et~al.}(2019)\citenamefont
  {Annulli}, \citenamefont {Cardoso},\ and\ \citenamefont
  {Gualtieri}}]{Annulli:2019fzq}%
  \BibitemOpen
  \bibfield  {author} {\bibinfo {author} {\bibfnamefont {Lorenzo}\ \bibnamefont
  {Annulli}}, \bibinfo {author} {\bibfnamefont {Vitor}\ \bibnamefont
  {Cardoso}}, \ and\ \bibinfo {author} {\bibfnamefont {Leonardo}\ \bibnamefont
  {Gualtieri}},\ }\bibfield  {title} {\enquote {\bibinfo {title}
  {{Electromagnetism and hidden vector fields in modified gravity theories:
  spontaneous and induced vectorization}},}\ }\href {\doibase
  10.1103/PhysRevD.99.044038} {\bibfield  {journal} {\bibinfo  {journal} {Phys.
  Rev. D}\ }\textbf {\bibinfo {volume} {99}},\ \bibinfo {pages} {044038}
  (\bibinfo {year} {2019})},\ \Eprint {http://arxiv.org/abs/1901.02461}
  {arXiv:1901.02461 [gr-qc]} \BibitemShut {NoStop}%
\bibitem [{\citenamefont
  {Minamitsuji}(2020{\natexlab{a}})}]{Minamitsuji:2020hpl}%
  \BibitemOpen
  \bibfield  {author} {\bibinfo {author} {\bibfnamefont {Masato}\ \bibnamefont
  {Minamitsuji}},\ }\bibfield  {title} {\enquote {\bibinfo {title} {{Stealth
  spontaneous spinorization of relativistic stars}},}\ }\href {\doibase
  10.1103/PhysRevD.102.044048} {\bibfield  {journal} {\bibinfo  {journal}
  {Phys. Rev. D}\ }\textbf {\bibinfo {volume} {102}},\ \bibinfo {pages}
  {044048} (\bibinfo {year} {2020}{\natexlab{a}})},\ \Eprint
  {http://arxiv.org/abs/2008.12758} {arXiv:2008.12758 [gr-qc]} \BibitemShut
  {NoStop}%
\bibitem [{\citenamefont
  {Minamitsuji}(2020{\natexlab{b}})}]{Minamitsuji:2020pak}%
  \BibitemOpen
  \bibfield  {author} {\bibinfo {author} {\bibfnamefont {Masato}\ \bibnamefont
  {Minamitsuji}},\ }\bibfield  {title} {\enquote {\bibinfo {title}
  {{Spontaneous vectorization in the presence of vector field coupling to
  matter}},}\ }\href {\doibase 10.1103/PhysRevD.101.104044} {\bibfield
  {journal} {\bibinfo  {journal} {Phys. Rev. D}\ }\textbf {\bibinfo {volume}
  {101}},\ \bibinfo {pages} {104044} (\bibinfo {year} {2020}{\natexlab{b}})},\
  \Eprint {http://arxiv.org/abs/2003.11885} {arXiv:2003.11885 [gr-qc]}
  \BibitemShut {NoStop}%
\bibitem [{\citenamefont {Antoniadis}\ \emph {et~al.}(2013)\citenamefont
  {Antoniadis} \emph {et~al.}}]{Antoniadis:2013pzd}%
  \BibitemOpen
  \bibfield  {author} {\bibinfo {author} {\bibfnamefont {John}\ \bibnamefont
  {Antoniadis}} \emph {et~al.},\ }\bibfield  {title} {\enquote {\bibinfo
  {title} {{A Massive Pulsar in a Compact Relativistic Binary}},}\ }\href
  {\doibase 10.1126/science.1233232} {\bibfield  {journal} {\bibinfo  {journal}
  {Science}\ }\textbf {\bibinfo {volume} {340}},\ \bibinfo {pages} {6131}
  (\bibinfo {year} {2013})},\ \Eprint {http://arxiv.org/abs/1304.6875}
  {arXiv:1304.6875 [astro-ph.HE]} \BibitemShut {NoStop}%
\bibitem [{\citenamefont {Mendes}(2015)}]{Mendes:2014ufa}%
  \BibitemOpen
  \bibfield  {author} {\bibinfo {author} {\bibfnamefont {Raissa F.~P.}\
  \bibnamefont {Mendes}},\ }\bibfield  {title} {\enquote {\bibinfo {title}
  {{Possibility of setting a new constraint to scalar-tensor theories}},}\
  }\href {\doibase 10.1103/PhysRevD.91.064024} {\bibfield  {journal} {\bibinfo
  {journal} {Phys. Rev. D}\ }\textbf {\bibinfo {volume} {91}},\ \bibinfo
  {pages} {064024} (\bibinfo {year} {2015})},\ \Eprint
  {http://arxiv.org/abs/1412.6789} {arXiv:1412.6789 [gr-qc]} \BibitemShut
  {NoStop}%
\bibitem [{\citenamefont {{Cooper}}(1993)}]{247776}%
  \BibitemOpen
  \bibfield  {author} {\bibinfo {author} {\bibfnamefont {J.}~\bibnamefont
  {{Cooper}}},\ }\bibfield  {title} {\enquote {\bibinfo {title} {Long-time
  behavior and energy growth for electromagnetic waves reflected by a moving
  boundary},}\ }\href {\doibase 10.1109/8.247776} {\bibfield  {journal}
  {\bibinfo  {journal} {IEEE Transactions on Antennas and Propagation}\
  }\textbf {\bibinfo {volume} {41}},\ \bibinfo {pages} {1365--1370} (\bibinfo
  {year} {1993})}\BibitemShut {NoStop}%
\bibitem [{\citenamefont {Dittrich}\ \emph
  {et~al.}(1994{\natexlab{b}})\citenamefont {Dittrich}, \citenamefont
  {Duclos},\ and\ \citenamefont {\ifmmode~\check{S}\else
  \v{S}\fi{}eba}}]{PhysRevE.49.3535}%
  \BibitemOpen
  \bibfield  {author} {\bibinfo {author} {\bibfnamefont {J.}~\bibnamefont
  {Dittrich}}, \bibinfo {author} {\bibfnamefont {P.}~\bibnamefont {Duclos}}, \
  and\ \bibinfo {author} {\bibfnamefont {P.}~\bibnamefont
  {\ifmmode~\check{S}\else \v{S}\fi{}eba}},\ }\bibfield  {title} {\enquote
  {\bibinfo {title} {Instability in a classical periodically driven string},}\
  }\href {\doibase 10.1103/PhysRevE.49.3535} {\bibfield  {journal} {\bibinfo
  {journal} {Phys. Rev. E}\ }\textbf {\bibinfo {volume} {49}},\ \bibinfo
  {pages} {3535--3538} (\bibinfo {year} {1994}{\natexlab{b}})}\BibitemShut
  {NoStop}%
\bibitem [{\citenamefont {Law}(1994)}]{Law:1994zza}%
  \BibitemOpen
  \bibfield  {author} {\bibinfo {author} {\bibfnamefont {C.K.}\ \bibnamefont
  {Law}},\ }\bibfield  {title} {\enquote {\bibinfo {title} {{Resonance Response
  of the Quantum Vacuum to an Oscillating Boundary}},}\ }\href {\doibase
  10.1103/PhysRevLett.73.1931} {\bibfield  {journal} {\bibinfo  {journal}
  {Phys. Rev. Lett.}\ }\textbf {\bibinfo {volume} {73}},\ \bibinfo {pages}
  {1931--1934} (\bibinfo {year} {1994})}\BibitemShut {NoStop}%
\bibitem [{\citenamefont {Boskovic}\ \emph {et~al.}(2019)\citenamefont
  {Boskovic}, \citenamefont {Brito}, \citenamefont {Cardoso}, \citenamefont
  {Ikeda},\ and\ \citenamefont {Witek}}]{Boskovic:2018lkj}%
  \BibitemOpen
  \bibfield  {author} {\bibinfo {author} {\bibfnamefont {Mateja}\ \bibnamefont
  {Boskovic}}, \bibinfo {author} {\bibfnamefont {Richard}\ \bibnamefont
  {Brito}}, \bibinfo {author} {\bibfnamefont {Vitor}\ \bibnamefont {Cardoso}},
  \bibinfo {author} {\bibfnamefont {Taishi}\ \bibnamefont {Ikeda}}, \ and\
  \bibinfo {author} {\bibfnamefont {Helvi}\ \bibnamefont {Witek}},\ }\bibfield
  {title} {\enquote {\bibinfo {title} {{Axionic instabilities and new black
  hole solutions}},}\ }\href {\doibase 10.1103/PhysRevD.99.035006} {\bibfield
  {journal} {\bibinfo  {journal} {Phys. Rev.}\ }\textbf {\bibinfo {volume}
  {D99}},\ \bibinfo {pages} {035006} (\bibinfo {year} {2019})},\ \Eprint
  {http://arxiv.org/abs/1811.04945} {arXiv:1811.04945 [gr-qc]} \BibitemShut
  {NoStop}%
%%CITATION = ARXIV:1811.04945;%%
\bibitem [{\citenamefont {Chen}\ \emph {et~al.}(1996)\citenamefont {Chen},
  \citenamefont {Goldenfeld},\ and\ \citenamefont {Oono}}]{Chen:1995ena}%
  \BibitemOpen
  \bibfield  {author} {\bibinfo {author} {\bibfnamefont {Lin-Yuan}\
  \bibnamefont {Chen}}, \bibinfo {author} {\bibfnamefont {Nigel}\ \bibnamefont
  {Goldenfeld}}, \ and\ \bibinfo {author} {\bibfnamefont {Y.}~\bibnamefont
  {Oono}},\ }\bibfield  {title} {\enquote {\bibinfo {title} {{The
  Renormalization group and singular perturbations: Multiple scales, boundary
  layers and reductive perturbation theory}},}\ }\href {\doibase
  10.1103/PhysRevE.54.376} {\bibfield  {journal} {\bibinfo  {journal} {Phys.
  Rev. E}\ }\textbf {\bibinfo {volume} {54}},\ \bibinfo {pages} {376--394}
  (\bibinfo {year} {1996})},\ \Eprint {http://arxiv.org/abs/hep-th/9506161}
  {arXiv:hep-th/9506161} \BibitemShut {NoStop}%
\bibitem [{\citenamefont {Chandrasekhar}(1964)}]{Chandrasekhar:1964zz}%
  \BibitemOpen
  \bibfield  {author} {\bibinfo {author} {\bibfnamefont {S.}~\bibnamefont
  {Chandrasekhar}},\ }\bibfield  {title} {\enquote {\bibinfo {title} {{The
  Dynamical Instability of Gaseous Masses Approaching the Schwarzschild Limit
  in General Relativity}},}\ }\href {\doibase 10.1086/147938} {\bibfield
  {journal} {\bibinfo  {journal} {Astrophys. J.}\ }\textbf {\bibinfo {volume}
  {140}},\ \bibinfo {pages} {417--433} (\bibinfo {year} {1964})},\ \bibinfo
  {note} {[Erratum: Astrophys.J. 140, 1342 (1964)]}\BibitemShut {NoStop}%
\bibitem [{\citenamefont {Posada}\ and\ \citenamefont
  {Chirenti}(2019)}]{Camilo:2018goy}%
  \BibitemOpen
  \bibfield  {author} {\bibinfo {author} {\bibfnamefont {Camilo}\ \bibnamefont
  {Posada}}\ and\ \bibinfo {author} {\bibfnamefont {Cecilia}\ \bibnamefont
  {Chirenti}},\ }\bibfield  {title} {\enquote {\bibinfo {title} {{On the radial
  stability of ultra compact Schwarzschild stars beyond the Buchdahl limit}},}\
  }\href {\doibase 10.1088/1361-6382/ab0526} {\bibfield  {journal} {\bibinfo
  {journal} {Class. Quant. Grav.}\ }\textbf {\bibinfo {volume} {36}},\ \bibinfo
  {pages} {065004} (\bibinfo {year} {2019})},\ \Eprint
  {http://arxiv.org/abs/1811.09589} {arXiv:1811.09589 [gr-qc]} \BibitemShut
  {NoStop}%
\bibitem [{\citenamefont {Lattimer}\ and\ \citenamefont
  {Prakash}(2001)}]{Lattimer:2000nx}%
  \BibitemOpen
  \bibfield  {author} {\bibinfo {author} {\bibfnamefont {J.~M.}\ \bibnamefont
  {Lattimer}}\ and\ \bibinfo {author} {\bibfnamefont {M.}~\bibnamefont
  {Prakash}},\ }\bibfield  {title} {\enquote {\bibinfo {title} {{Neutron star
  structure and the equation of state}},}\ }\href {\doibase 10.1086/319702}
  {\bibfield  {journal} {\bibinfo  {journal} {Astrophys. J.}\ }\textbf
  {\bibinfo {volume} {550}},\ \bibinfo {pages} {426} (\bibinfo {year}
  {2001})},\ \Eprint {http://arxiv.org/abs/astro-ph/0002232}
  {arXiv:astro-ph/0002232} \BibitemShut {NoStop}%
\bibitem [{\citenamefont {Fermi}(1949)}]{Fermi:1949ee}%
  \BibitemOpen
  \bibfield  {author} {\bibinfo {author} {\bibfnamefont {Enrico}\ \bibnamefont
  {Fermi}},\ }\bibfield  {title} {\enquote {\bibinfo {title} {{On the Origin of
  the Cosmic Radiation}},}\ }\href {\doibase 10.1103/PhysRev.75.1169}
  {\bibfield  {journal} {\bibinfo  {journal} {Phys. Rev.}\ }\textbf {\bibinfo
  {volume} {75}},\ \bibinfo {pages} {1169--1174} (\bibinfo {year}
  {1949})}\BibitemShut {NoStop}%
%%CITATION = PHRVA,75,1169;%%
\bibitem [{\citenamefont {Moore}(1970)}]{Moore:1970}%
  \BibitemOpen
  \bibfield  {author} {\bibinfo {author} {\bibfnamefont {G.~T.}\ \bibnamefont
  {Moore}},\ }\bibfield  {title} {\enquote {\bibinfo {title} {Quantum theory of
  the electromagnetic field in a variable‐length one‐dimensional cavity},}\
  }\href {\doibase 10.1063/1.1665432} {\bibfield  {journal} {\bibinfo
  {journal} {Journal of Mathmatical Physics}\ }\textbf {\bibinfo {volume}
  {11}},\ \bibinfo {pages} {2679} (\bibinfo {year} {1970})}\BibitemShut
  {NoStop}%
\bibitem [{\citenamefont {Andreata}\ and\ \citenamefont
  {Dodonov}(2000)}]{Andreata_2000}%
  \BibitemOpen
  \bibfield  {author} {\bibinfo {author} {\bibfnamefont {M~A}\ \bibnamefont
  {Andreata}}\ and\ \bibinfo {author} {\bibfnamefont {V~V}\ \bibnamefont
  {Dodonov}},\ }\bibfield  {title} {\enquote {\bibinfo {title} {Energy density
  and packet formation in a vibrating cavity},}\ }\href {\doibase
  10.1088/0305-4470/33/16/311} {\bibfield  {journal} {\bibinfo  {journal}
  {Journal of Physics A: Mathematical and General}\ }\textbf {\bibinfo {volume}
  {33}},\ \bibinfo {pages} {3209--3223} (\bibinfo {year} {2000})}\BibitemShut
  {NoStop}%
\bibitem [{\citenamefont {Dodonov}(2001)}]{Dodonov:2001yb}%
  \BibitemOpen
  \bibfield  {author} {\bibinfo {author} {\bibfnamefont {V.~V.}\ \bibnamefont
  {Dodonov}},\ }\bibfield  {title} {\enquote {\bibinfo {title} {{Nonstationary
  Casimir effect and analytical solutions for quantum fields in cavities with
  moving boundaries}},}\ }\href@noop {} {\bibfield  {journal} {\bibinfo
  {journal} {Adv. Chem. Phys.}\ }\textbf {\bibinfo {volume} {119}},\ \bibinfo
  {pages} {309--394} (\bibinfo {year} {2001})},\ \Eprint
  {http://arxiv.org/abs/quant-ph/0106081} {arXiv:quant-ph/0106081} \BibitemShut
  {NoStop}%
\end{thebibliography}%

\end{document}